\documentclass[a4paper,onecolumn,11pt,accepted=2023-11-24]{quantumarticle}
\pdfoutput=1
\usepackage[utf8]{inputenc}
\usepackage[english]{babel}

\usepackage{tikz}
\usepackage{lipsum}

\usepackage[unicode=true]{hyperref}
\usepackage{amsmath}
\usepackage{cite}
\usepackage{amsfonts}
\usepackage{amssymb}
\usepackage{comment}
\usepackage{cite}

\newtheorem{Theorem}{Theorem}

\newtheorem{Cor}{Corollary}

\newtheorem{Lemma}{Lemma} 

\newtheorem{Def}{Definition}

\usepackage{titlesec,amsfonts,braket,csquotes}

\definecolor{dur}{cmyk}{0,1,1,0.3}

\usepackage{soul}

\begin{document}

\title{Measurement-based quantum computation in finite one-dimensional systems: string order implies com\-pu\-ta\-tio\-nal power}

\author{Robert Raussendorf}
\affiliation{Leibniz University Hannover, Hannover, Germany}
\affiliation{Stewart Blusson Quantum Matter Institute, University of British Columbia, Vancouver, Canada}
\orcid{0000-0003-4983-9213}

\author{Wang Yang}
\affiliation{School of Physics, Nankai University, Tianjin, China}

\author{Arnab Adhikary}
\affiliation{Department of Physics and Astronomy, University of British Columbia, Vancouver, Canada}
\affiliation{Stewart Blusson Quantum Matter Institute, University of British Columbia, Vancouver, Canada}

\maketitle

\begin{abstract} We present a new framework for assessing the power of measurement-based quantum computation (MBQC) on short-range entangled symmetric resource states, in spatial dimension one. It requires fewer assumptions than previously known. The formalism can handle finitely extended systems (as opposed to the thermodynamic limit), and does not require translation-invariance. Further, we strengthen the connection between MBQC computational power and string order. Namely, we establish that whenever a suitable set of string order parameters is non-zero, a corresponding set of unitary gates can be realized with fidelity arbitrarily close to unity.
\end{abstract}

\section{Introduction}\label{Intro}

Resource states for measurement-based quantum computation (MBQC) \cite{RB01} are known to be rare in Hilbert space \cite{TooE}. But symmetry adds a twist to this picture. When symmetries are present, in the thermodynamic limit, short-range entangled quantum states group into so-called {\em{computational phases of quantum matter}} \cite{DB1, CBD, M1,Darmawan,Bartl,Bartl3}. From a condensed matter perspective, these phases are symmetry protected topologically (SPT) ordered \cite{GW,Wen1,Wen2,Schuch,Ogata}. From the perspective of quantum computation, these phases are warehouses full of MBQC resource states. Any quantum state in a given SPT phase can be used to realize quantum computations, and, moreover, the {\em{same}} quantum computations. The power of MBQC across SPT phases is uniform \cite{MM2,SPTO1,SPTO2,MScDav,2Duniv, DW, QCA, DAM}. 

The phenomenology of MBQC becomes richer with increasing spatial dimension of the resource states: one dimension (1D) is mostly a test bed for computational methods, 2D reaches quantum computational universality \cite{RB01,Miy,Wei}, and 3D combines universality with fault-tolerance \cite{RHG}. This increase of computational power with dimension is matched in computational phases. The first such phases were identified in 1D \cite{MM2,SPTO1,SPTO2,MScDav}, capable of processing a bounded number of logical qubits. In 2D, examples of universal computational phases are known \cite{2Duniv, DW, QCA, DAM}. In 3D, the fault-tolerance capability of cluster states has been related to SPT order with 1-form symmetry \cite{Bartl2}.
As the phenomenology flourishes with increasing dimension, our understanding diminishes: In spatial dimension one, a classification scheme for computational phases exists \cite{SPTO1, SPTO2, MScDav}; and furthermore a gauge principle underlying MBQC has been identified \cite{gaugeMBQC}. In higher dimensions we have several examples for computational phases, but no classification.

For the reasons just outlined, most current research on the subject of computational phases of quantum matter focuses on higher dimensions. Nonetheless, in the present paper we return to the one-dimensional case, to devise a more versatile formalism for the discussion of MBQC in the presence of symmetry. We do this with the intention of later applying it to 2D and 3D, and beyond that, to identify a unifying framework in which the subjects of foundational interest in MBQC---contextuality, symmetry, temporal order, topological fault-tolerance and gauge principle---can all be discussed. At the beginning of our exploration stands the question: {\em{How is MBQC computational power on symmetric states affected if we transition from infinite to finite systems?}}

The question is well-motivated: quantum computation is about efficiency, hence resource counting. The finite size of an MBQC resource state is thus an essential property. Yet our main interest is conceptual: if we turn to finite systems, the notion of `symmetry protected phase' dissolves. But then, what happens to the cohomological classification of resource states, hence MBQC schemes?

We are  prompted to adopt a novel perspective. Namely, in the discussion of computational phases of quantum matter to date \cite{Bartl, Bartl3, MM2, SPTO1, SPTO2, MScDav, 2Duniv, DW, QCA, DAM}, the resource state is the primary object, the object to classify. The measurement procedure that extracts computational power is almost an afterthought. Now we turn this picture on its head. Phases---symmetry-protected, computational, or otherwise---are not defined in finite systems. This is a priori a detriment, for the classification of SPT order in terms of group cohomology \cite{GW,Wen1,Wen2,Schuch,Ogata} hinges on it. Group cohomology is also the basis for the ``SPT-to-MBQC meat grinder'' \cite{SPTO1,SPTO2}, which converts cohomological data into MBQC schemes. 

As we show in this paper, in the new situation of finite system size, the measurement procedure takes over as the primary object, the object suited to classification. Projective representations, and their cohomological classification, reappear in it. The resource states, in turn,  become the accessory in the formalism. They have to be short-range entangled, symmetric, and possess string order matching the symmetry. And that's all there's to say about them. A first implication of this reversal is that a characterization of MBQC on symmetric resource states in terms of group cohomology can be retained for finite systems.

Advantages of the new formalism---ranging from the conceptual to the more practical---are as follows.
 (I) We strengthen the connection between string order and computational power of MBQC in one dimension. Namely we show that, as long as string order  \cite{denNijs1989,Tasaki1990,SO} is present, however weak, arbitrarily accurate non-trivial computation is possible. (II) We align the MBQC notion of locality (site local) with the SPT notion of local (previously block-local), and (III) We no longer require translation-invariance of the resource state. 
 \medskip

The remainder of this paper is organized as follows. In Section~\ref{AdvExpl} we describe the above-listed advances in greater detail. In Section~\ref{Set} we define our  setting, and introduce the four examples through which we will subsequently illustrate our result, namely the cluster chain, the Kitaev-Gamma chain, a spin chain relating to the output of a Clifford quantum cellular automaton (QCA), and the Ising chain with transverse magnetic field.
In Section~\ref{Res} we state and prove our main result, Theorem~\ref{GT}. It says that multi-particle quantum states can be used as resources for measurement based quantum computation if they (a) are invariant under a suitable group of symmetries, (b) are short-range entangled, and (c) have non-vanishing string order parameters of a form matching the symmetries. We apply the theorem to the examples introduced in the previous section. Section~\ref{BLtoSL} is about block locality vs. site locality. Here we treat the cluster chain and the QCA chain in a refined fashion, leading to blocks of size one. In Section~\ref{StringComp} we discuss the relation between string order parameters and the computational order parameters defined in \cite{SPTO1}. In Section~\ref{COeSO}  we relate string order to quantum contextuality. Section~\ref{Concl} is the conclusion.

\section{Advances of the new formalism}\label{AdvExpl}

We now explain the advances made by the new formalism.

(1) {\em{Computational order equals string order:}} The relevance of string operators for the functioning of MBQC was first recognized in \cite{CBD, DB1}. In \cite{CBD}, quantum correlations describing the fidelity of gate simulations in MBQC were expressed in terms of string operators. In \cite{DB1}, it was shown for ground states of the transverse field cluster model,  the gate fidelity is bounded from below by a constant.

Here, we strengthen the above connection. Namely we show that whenever the string order parameters are {\em{non-zero}}, quantum gates can be realized in MBQC with fidelity arbitrarily close to unity. The higher the fidelity targeted, the larger the section of resource state consumed in the implementation of the gate.

In prior analysis of MBQC on resources states taken from SPT phases \cite{SPTO1,SPTO2}, in the framework of MPS, a computational order parameter $\nu$ was identified that governs the operational overheads of MBQC. It was shown in \cite{SPTO1} how to extract this order parameter from the MPS tensor representing the resource quantum state, but no physical interpretation for it had been found. We now realize that the computational order parameter $\nu$ and the string order parameter are the same.
\medskip

(2) {\em{Block size:}} In the discussion of SPT and MBQC by the MPS formalism, neighbouring spins are grouped into blocks \cite{Bartl, MM2, SPTO1, SPTO2, 2Duniv, DW, QCA, MScDav}, such that the action of the symmetry group on each block is faithful. The block  thereby becomes the natural local unit for the formalism.

In all cases so far considered, the blocks comprise more than a single spin, and this leads to a mismatch from the perspective of MBQC phenomenology. Namely, in standard MBQC, the local unit is a single spin. The measurements driving an MBQC are supposed to be site-local, not just block local. There is thus a gap between the MPS formalism and the phenomenology of interest. In the prior discussions of 1D, the block size is only 2; a gap that was deemed minor. In 2D, however, the block size increases with system size, leading to a very weak result about computational phases of quantum matter if left unaddressed. Therefore, in \cite{2Duniv, DW, QCA}, supplemental arguments have been put on top of the basic formalism to reach block size one. 

The present formalism doesn't require faithfulness of the representations involved, and can therefore handle blocks of any size down to size one. The physically motivated single-site locality of MBQC can be matched by the present formalism in its very algebraic structure, without the need for add-on arguments.

\medskip

(3) {\em{Translation invariance:}} The prior formalism \cite{Bartl, MM2, SPTO1, SPTO2, 2Duniv, DW, QCA, MScDav} requires translation invariance whereas the present formalism doesn't. Translation invariance is tied to the thermodynamic limit: no finite chain is translation-invariant. Therefore, getting rid of the constraint of translation invariance is a precondition for discussing finite systems.

The present formalism achieves this, and in fact permits much greater flexibility than merely permitting the existence of boundaries. For example, the value of the string order parameter may vary with the location of its end point in any fashion. 

\section{The setting}\label{Set}

In this section we define our setting of short-range entangled symmetric states, and introduce the examples that we will subsequently use to illustrate our main theorem.

\subsection{Symmetric short-range entangled states}

As our fundamental notion of ``short-range entangled'', we use that of short-range, bounded depth quantum circuits applied to a product state. Two quantum states are considered equivalent under a given symmetry $G$ if they can be related by a $G$-symmetric such circuit. This is an operationally well-motivated notion in the context of quantum computation.\smallskip

We consider quantum states $|\Phi\rangle$ on open chains of spin 1/2 particles.  The support of the states $|\Phi\rangle$ is grouped into $n$ blocks in the bulk, plus a block $0$ on the left boundary and a block $n+1$ on the right boundary. Graphically, 
$$
\includegraphics[width=10cm]{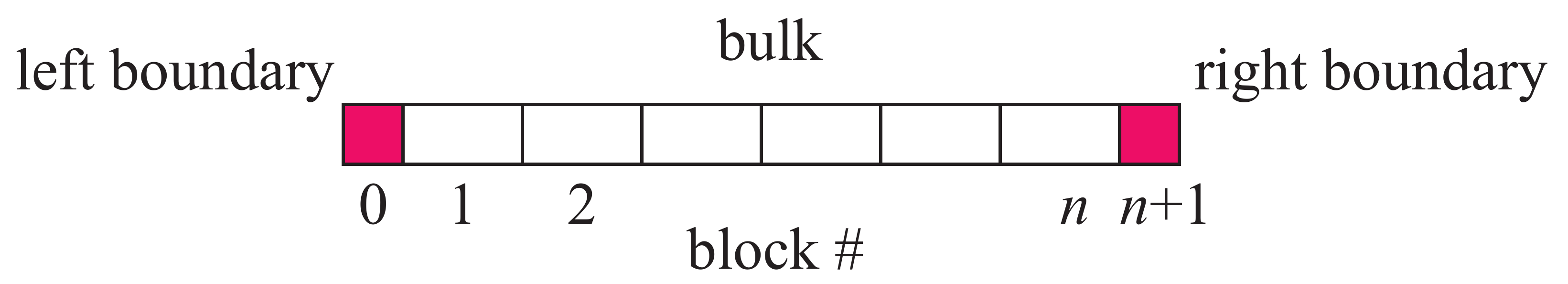}.
$$
The states $|\Phi\rangle$ are short-range entangled and $G$-symmetric.

\paragraph{Symmetry.} The symmetry group $G$ discussed in this paper is of the form $G=(\mathbb{Z}_2)^m$. It acts via a linear representation $U$ on $|\Phi\rangle$,
\begin{equation}\label{PhiSymm}
U(g)|\Phi\rangle = (-1)^{\chi(g)} |\Phi\rangle,\; \chi(g)\in \mathbb{Z}_2,\;\; \forall g\in G.
\end{equation}

\paragraph{Entanglement structure.} 
The resource states $|\Phi\rangle$ we consider are all of the form
\begin{equation}\label{SREstates}
|\Phi\rangle = W_\Phi (|+\rangle |+\rangle..|+\rangle).
\end{equation}
Therein, $W_\Phi$ is a bounded-depth circuit composed of bounded-range gates.  Symmetric such states can arbitrarily closely approximate all ground states in SPT phases \cite{WenBook}. 

We quantify the short-range entangling nature of $W_\Phi$ as follows. We define two subsets of particle block labels on the line, 
$$
\{\leq k\}:= \{0,1,2,..,k\},\; \{>k\} :=\{k+1,k+2,..,n+1\}.
$$
The short-range nature of $W_\Phi$ is specified by an entanglement range $\Delta$. Denoting by $\text{supp}(A)$ the support of a linear operator $A$ on the line segment $\{1,..,n\}$, we make the following definition.
\begin{Def}\label{DefRange}
The entanglement range $\Delta$ of a quantum circuit $W_\Phi$ acting on the spin chain $\{0,..,n+1\}$ is the smallest integer $\Delta\geq 0$ such that, for all $k=0,.., n+1$, it holds that
\begin{equation}\label{DefRangeEq}
\begin{array}{rl}
\text{supp}(W^\dagger_\Phi AW_\Phi) \subset \{\leq (k+\Delta)\},& \forall \,A|\, \text{supp}(A) \subset \{\leq k\},\\
\text{supp}(W^\dagger_\Phi AW_\Phi) \subset \{>(k-\Delta)\},& \forall \,A|\, \text{supp}(A) \subset \{> k\}.
\end{array}
\end{equation}
\end{Def}
The short-range entanglement in resource states $|\Phi\rangle$ enters MBQC through the following lemma.
\begin{Lemma}\label{Prod}
Consider a short-range entangled state $|\Phi\rangle=W_\Phi|+\rangle|+\rangle..|+\rangle$, where the circuit $W_\Phi$ has an entanglement range $\Delta$. Be $A$ and $B$ two linear operators, with their support contained in $\{\leq (k-\Delta)\}$ and $\{>(k+\Delta)\}$, respectively, for any $k=\Delta,..,n+1-\Delta$. Then it holds that
\begin{equation}\label{ProdRel}
\langle  \Phi |AB|\Phi \rangle = \langle \Phi |A|\Phi\rangle  \langle \Phi |B|\Phi\rangle. 
\end{equation}
\end{Lemma}
{\em{Proof of Lemma~\ref{Prod}.}} We define ${\cal{L}}:=\{\leq k\}$,  ${\cal{R}}:=\{>k\}$, and write the product state to which the short-range circuit $W_\Phi$ is applied as $|+\rangle_{\cal{LR}}:=|+\rangle_{\cal{L}} \otimes |+\rangle_{\cal{R}}$, with $|+\rangle_{\cal{L}}=|+\rangle_0\otimes ..\otimes |+\rangle_k$ and $|+\rangle_{\cal{R}}=|+\rangle_{k+1}\otimes ..\otimes |+\rangle_{n+1}$, with all $|+\rangle_i$ reference states on blocks $i$, respectively. Only locality between the left half ${\cal{L}}$ and the right half ${\cal{R}}$ of the chain, split between blocks $k$ and $k+1$, matters. 

We observe that, with the assumptions of the Lemma and Eq.~(\ref{DefRangeEq}) it holds that $\text{supp}(W^\dagger_\Phi A W_\Phi) \subseteq {\cal{L}}$ and $\text{supp}(W^\dagger_\Phi B W_\Phi) \subseteq {\cal{R}}$; hence
\begin{equation}\label{SuppSiz}
W^\dagger_\Phi A W_\Phi = W^\dagger_\Phi A W_\Phi|_{\cal{L}} \otimes I_{\cal{R}},\; \; W^\dagger_\Phi B W_\Phi = I_{\cal{L}} \otimes W^\dagger_\Phi B W_\Phi|_{\cal{R}}.
\end{equation}
We then have 
$$
\begin{array}{rcl}
\langle  \Phi |AB|\Phi \rangle &=& \mbox{}_{\cal{L}}\langle  + |\otimes_{\cal{R}}\!\langle +| \,W^\dagger_\Phi  AB W_\Phi \, |+\rangle_{\cal{L}} \otimes |+\rangle_{\cal{R}}\\
&= & \mbox{}_{\cal{L}}\langle + |\otimes_{\cal{R}}\!\langle +| \left(W^\dagger_\Phi  A W_\Phi\right) \left(W^\dagger_\Phi B W_\Phi \right) |+\rangle_{\cal{L}} \otimes |+\rangle_{\cal{R}}\\
&= & \mbox{}_{\cal{L}} \langle + |  \left.W^\dagger_\Phi  A W_\Phi\right|_{\cal{L}} |+\rangle_{\cal{L}}  \mbox{ }_{\cal{R}}\!\langle +|   \left. W^\dagger_\Phi B W_\Phi \right|_{\cal{R}}  |+\rangle_{\cal{R}}\\
&= & \mbox{}_{\cal{LR}}\langle + |  W^\dagger_\Phi  A W_\Phi |+\rangle_{\cal{LR}}  \mbox{ }_{\cal{LR}} \!\langle +|  W^\dagger_\Phi B W_\Phi   |+\rangle_{\cal{LR}} \\
&=&  \langle \Phi |A|\Phi\rangle  \langle \Phi |B|\Phi\rangle.
\end{array}
$$
Therein, in the third line we have used Eq.~(\ref{SuppSiz}). $\Box$\medskip

\subsection{The role of Hamiltonians in our setting}\label{RGS}

A comment about the role of Hamiltonians and their ground states in  measurement based quantum computation is now in order. From a fundamental point of view, MBQC has nothing to do with Hamiltonians at all; it is only about states and measurements. Yet, all examples in this paper consider ground states of Hamiltonians; see Section~\ref{Exa} below. Here we explain this dichotomy.

First, Hamiltonians do find a role to play in MBQC, in the following way. It was observed in \cite{TooE} that, when sampled uniformly from Hilbert space, computationally useful resource states are extremely rare. This prompted the question: How frequent are computational resources among quantum states that naturally occur? A common notion of `naturally occurring' is ground states of simple Hamiltonians. In this regard it has been established, for example, that AKLT states in dimension two are universal for MBQC \cite{Miy,Wei}. 

The idea of ground states as computational resources fully came into its own with the discovery of computational phases of quantum matter \cite{DB1, CBD, M1,Darmawan,Bartl}, when it was understood that entire symmetry protected topological phases have computational power \cite{MM2,SPTO1,SPTO2,MScDav} and can even be universal \cite{2Duniv, DW, QCA}. A counterpoint to the above scarcity of resource states argument \cite{TooE} is thereby made: in the presence of symmetry, computational resources are no longer rare. The ground state manifold splits into extended phases, some of which have computational power and others don't. Computational phases of quantum matter represent the strong case for invoking Hamiltonians in the discussion of MBQC.

In the present paper, we consider finite systems. The notion of `phase' does therefore no longer apply; and with it disappears the most enticing motivation for considering Hamiltonians. However, the earlier motivation remains: Ground states model naturally occurring states---this applies to finite systems just as well as to infinite ones. There's still a case for invoking Hamiltonians.

A shift occurs with the formal criterion for `short-range entangled'  we impose, Eq.~(\ref{DefRangeEq}). It is based on bounded-depth quantum circuits composed of short-range gates. The manifold of quantum states described in this fashion has an operational motivation in its own right: those states are all equally hard to create. On the other hand, ground states of gapped local Hamiltonians, such as those we use as examples, have exponential decay of correlations \cite{Hast}. Thus they only approximately realize our notion of `short-range entangled'. This approximation notwithstanding, we use examples based on ground states of Hamiltonians, to connect with familiar physics of spin chains.
 
\subsection{Examples}\label{Exa}

Here we introduce four examples of ground state families. We will subsequently use them to illustrate the corresponding MBQC quantum computational power. The examples are (i) the cluster chain, (ii) the Kitaev-Gamma chain, (iii) a spin chain related to quantum cellular automata, and (iv) the Ising chain.

\subsubsection{The cluster chain}

The cluster state is the ``standard'' resource in measurement based quantum computation. Cluster states in spatial dimension two are computationally universal \cite{RB01}. In the simpler one-dimensional case that we discuss here, a single logical qubit can be simulated. The 1D cluster state lies inside a symmetry protected topological (SPT) phase with symmetry group $\mathbb{Z}_2\times \mathbb{Z}_2$. It was demonstrated in \cite{Bartl} that the ability to perform measurement based quantum computational wire extends from the cluster state to the entire SPT phase surrounding it. Subsequently, the same was shown for computational capability; it too is uniform across the $\mathbb{Z}_2\times \mathbb{Z}_2$ cluster phase \cite{SPTO1,SPTO2}. 
The cluster chain is the standard example for computational phases of quantum matter, and arguably the most thoroughly studied. We include it here to illustrate the new formalism in a familiar scenario.

\paragraph{Model.} We now define the 1D cluster state and its surrounding phase. We consider a chain of $N$ spins 1/2. W.l.o.g. we choose $N$ odd. The cluster state $|{\cal{C}}\rangle$ is a stabilizer state, constrained by the eigenvalue equations 
$$
Z_{i-1}X_iZ_{i+1} |{\cal{C}}\rangle = |{\cal{C}}\rangle,\;\;i=2,..,N-1
$$
in the bulk, and
$$
X_1Z_2 |{\cal{C}}\rangle = Z_{N-1}X_N |{\cal{C}}\rangle = |{\cal{C}}\rangle
$$
at the boundary. The above stabilizer constraints specify the cluster state uniquely, up to a global phase.

\paragraph{Symmetry.} The stabilizer is an Abelian group, with a subgroup
\begin{equation}\label{symm}
G= \mathbb{Z}_2\times \mathbb{Z}_2 \cong \langle ZXIXIXIX...IXZ, XIXIXIXI...IX\rangle.
\end{equation}
$G$ is the symmetry group of interest. The cluster phase is the phase of $G$-symmetric states that contains the cluster state.

To assess computational power, we consider the order parameters
\begin{equation}\label{CSO}
\begin{array}{rcl}
\sigma_e &=& \langle I...IZXIXI..IX  \rangle,\\
\sigma_o &=& \langle I...IZXIXI..IXZ  \rangle,\\
\sigma_{o+e} &=& \langle I...IZYXX..XXY\rangle.
\end{array}
\end{equation}
For $\sigma_e$ and $\sigma_{o+e}$, the left-most Pauli operator $Z$ is located on an even-numbered qubit, and for $\sigma_o$ on an odd-numbered qubit. 

We show later that the expectation values $\sigma_o$, $\sigma_e$ and $\sigma_{o+e}$ are associated with logical rotations generated by $\sigma_z$, $\sigma_x$ and $\sigma_y$, respectively. To implement such rotations, the expectation values of Eq.~(\ref{CSO}) must be non-zero.\medskip

\begin{figure}
\begin{center}
\includegraphics[width=9cm]{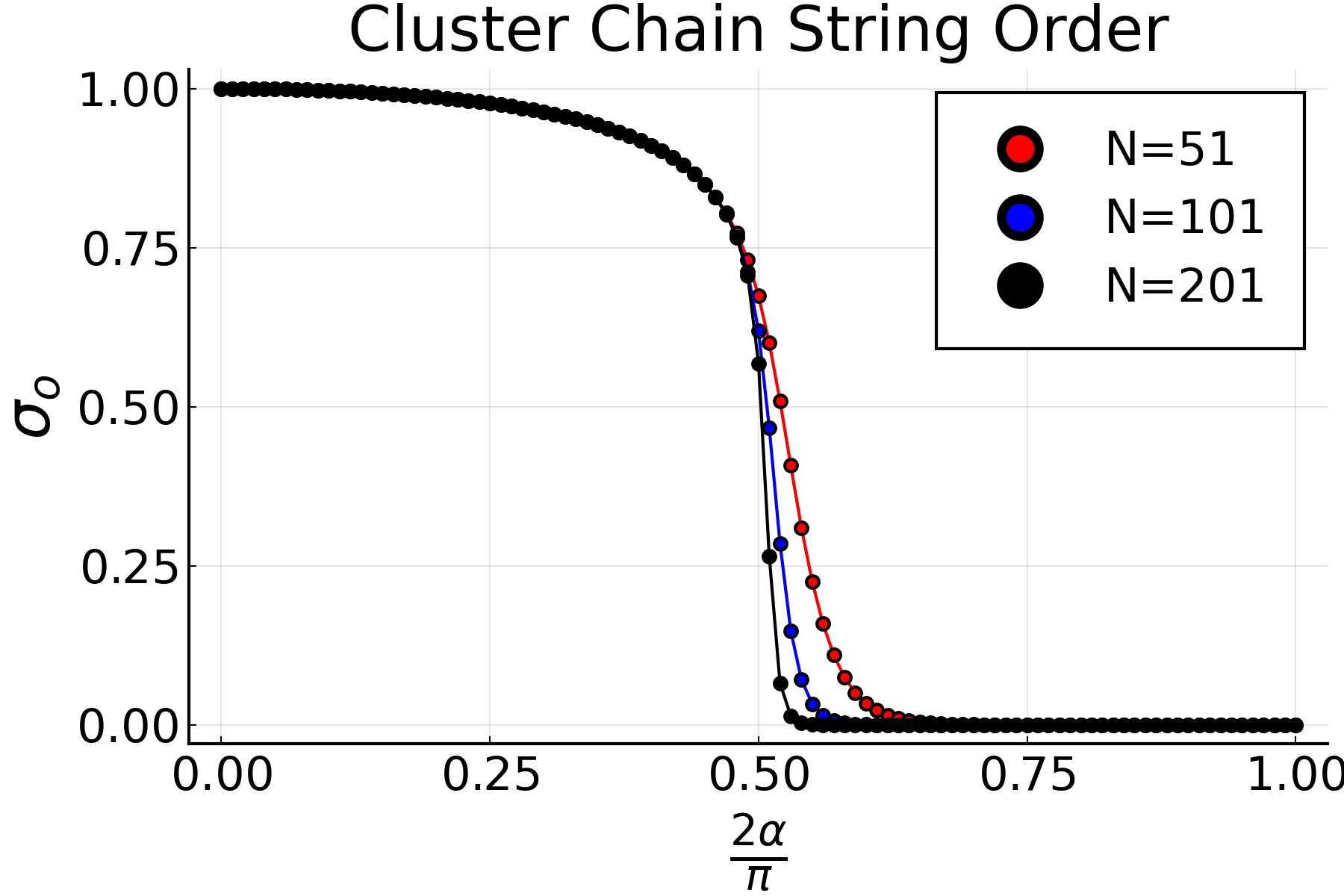}
\caption{\label{clusterNum}Order parameter $\sigma_{o}$ as a function of the sweep parameter $\alpha$ for the transverse field cluster Hamiltonian. The starting point of the string orders is deep in the bulk around $N/4$. The multiplicity of the curves is due to differing values of the chain length $N$. The plot for $\sigma_{e}$ is indistinguishable to the naked eye from the above curves and is therefore omitted. }
\end{center}
\end{figure}

For illustration, we consider a one-dimensional line in the phase diagram of $\mathbb{Z}_2 \times \mathbb{Z}_2$-symmetric states. Namely, we consider the ground states
 of the cluster Hamiltonian with magnetic field, 
\begin{equation}\label{CluHa}
H (\alpha)= -\cos \alpha \left( X_1Z_2 + Z_{N-1}X_N +\sum_{i=2}^{N-1} Z_{i-1}X_iZ_{i+1} \right) -  \sin \alpha \sum_{i=2}^{N-1} X_i,
\end{equation}
parametrized by an interpolation parameter $\alpha$. 

\paragraph{Phase diagram.} When $\alpha = 0$, the ground state is a 1D cluster state. When $\alpha=\pi/2$, then the ground state is fourfold degenerate , $|g(\pi/2)\rangle = |\pm\rangle_1 |+\rangle_2..|+\rangle_{N-1} |\pm\rangle_N$. At $\alpha=\pi/4$ occurs a change-over from cluster-like states to trivial (unentangled) states. This change-over is marked by the string order parameters $\sigma$ changing from non-zero to zero. The larger the chain length $N$, the sharper the drop. In the thermodynamic limit, the change-over becomes a phase transition. See Fig.~\ref{clusterNum} for a plot of the order parameters as a function of $\alpha$.

\subsubsection{The Kitaev-Gamma chain}
\label{subsec:KG}

One-dimensional Kitaev spin models \cite{Agrapidis2018,Yang2019,Yang2020,Luo2021,Yang2022a,Yang2022e} are 1D versions of the generalized Kitaev spin-1/2 models on the honeycomb lattice \cite{Kitaev2006,Nayak2008} used to describe real Kitaev materials \cite{Jackeli2009,Rau2014,Rau2016,Winter2017,Hermanns2018}.
Besides providing useful information for the 2D Kitaev physics \cite{Yang2022e}, 1D Kitaev models
have intricate nonsymmorphic symmetry group structures \cite{Yang2019,Yang2020,Yang2022a},
and contain rich strongly correlated physics, including emergent conformal symmetries \cite{Yang2019}, nonlocal string order parameters \cite{Luo2021} and exotic symmetry breaking phases \cite{Yang2019,Yang2020,Luo2021}, which make such 1D studies intriguing on their own. 

The purpose the Kitaev-Gamma chain example is two-fold. First, the Kitaev-Gamma chain is, more than the other examples, at home in condensed matter physics. Thus, it best represents the overlap area between condensed matter physics and quantum computation explored in this paper. 

Second, it illustrates the interplay between symmetry action and locality. The cluster and the Kitaev-Gamma chain are both invariant under $\mathbb{Z}_2\times \mathbb{Z}_2$ symmetry, and live in the unique topologically non-trivial phase. But only MBQC on the cluster chain can be made site-local; see Section~\ref{BLtoSL}. The reason is the difference in the representation of symmetry on the physical spins.

\begin{figure}
\begin{center}
\includegraphics[width=11cm]{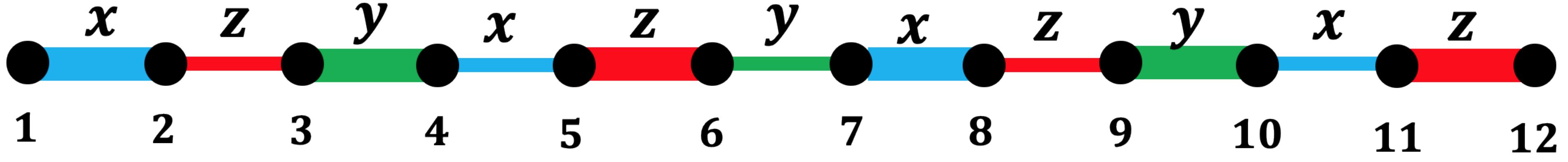} 
\caption{
Bond pattern for the 1D bond-alternating Kitaev-Gamma model in the rotated frame.
The thick and thin lines represent the alternating pattern of the bond strengths. 
\label{fig:bonds} 
}
\end{center}
\end{figure}

\paragraph{Model.}
The model that we consider is the 1D spin-1/2 bond-alternating Kitaev-Gamma model \cite{Luo2021,Yang2022a}.
After applying a unitary transformation $U_6$,
the system is called in the rotated frame
and the Hamiltonian acquires the form \cite{Yang2019}
\begin{eqnarray}
H_{K\Gamma}^\prime=\sum_{\gamma=<ij>} g_\gamma [-K S_i^\gamma S_j^\gamma-\Gamma (S_i^\alpha S_j^\alpha+S_i^\beta S_j^\beta)],
\label{eq:KG_U6}
\end{eqnarray}
in which:
$S_i^x=\frac{1}{2}X_i$, $S_i^y=\frac{1}{2}Y_i$, $S_i^z=\frac{1}{2}Z_i$ are the spin-$1/2$ operators on site $i$;
$\gamma\in\{x,y,z\}$ is the spin direction associated with the bond connecting the nearest neighboring sites $i$ and $j$ as shown in Fig. \ref{fig:bonds}; 
$(\gamma,\alpha,\beta)$ (all belonging to $\{x,y,z\}$) form a local right-handed coordinate system in spin space for sites $i$ and $j$ connected by the bond $\gamma$;
$K$ and $\Gamma$ are the Kitaev and Gamma interactions, respectively;
and $g_\gamma>0$ ($\gamma=x,y$) is the parameter for the bond strength on $\gamma$-bond.
The Hamiltonian $H_{K\Gamma}$ before the $U_6$ transformation
and the definition of $U_6$  are included in Appendix \ref{app:KG}. 

\paragraph{Symmetry.}
The Hamiltonian $H^\prime_{K\Gamma}$ 
has an intricate symmetry group structure \cite{Yang2022a}. 
Namely, $H^\prime_{K\Gamma}$ is invariant under $T$, $U(R_{2a})T_{2a}$, $U(R_M)M$, $U(R(\hat{x},\pi))$, $U(R(\hat{y},\pi))$, and $U(R(\hat{z},\pi))$,
where $T:S_i^\alpha\rightarrow -S_i^\alpha$ ($\alpha=x,y,z$) is the time reversal operation;
$T_{ma}:S_i^\alpha\rightarrow S_{i+m}^\alpha$ is the translation operation  by $m\in \mathbb{Z}$ lattice sites;
$M:S_i^\alpha\rightarrow S_{7-i}^\alpha$ is the spatial inversion operation with the inversion center located at the middle point between sites $3$ and $4$;
$R(\hat{n},\phi)$ is a global SU(2) spin rotation around $\hat{n}$-direction by an angle $\phi$; 
$R_{2a}=R(\frac{1}{\sqrt{3}}(1,1,1),\frac{2\pi}{3})$; 
$R_M=R(\frac{1}{\sqrt{2}}(1,0,-1),\pi)$;
and $U$ is the representation of the SU(2) group on the Hilbert space of the  whole spin chain. 
Clearly, the symmetry group $G_{K\Gamma}$ of $H^\prime_{K\Gamma}$ contains a $\mathbb{Z}_2\times \mathbb{Z}_2$ subgroup generated by $\{U(R(\hat{z},\pi)),U(R(\hat{x},\pi))\}$, 
where the explicit expression of $U(R(\hat{\alpha},\pi))$ ($\alpha\in\{x,y,z\}$) is
\begin{eqnarray}
U(R(\hat{\alpha},\pi))=\Pi_{i=1}^N 2S_i^\alpha
\label{eq:sym_Z2_Z2_KG}
\end{eqnarray}
in which $N\in 2\mathbb{Z}$ is the length of the chain.
More generally, it has been proved in Ref. \cite{Yang2022a} that $G_{K\Gamma}$ satisfies the following short exact sequence,
\begin{eqnarray}
1\rightarrow \langle T_{6a}\rangle\rightarrow G_{K\Gamma} \rightarrow O_h \rightarrow 1,
\label{eq:short_exact}
\end{eqnarray}
in which $\langle T_{6a}\rangle$ denotes the group generated by $T_{6a}$,
and $O_h\cong S_4 \times \mathbb{Z}_2$ is the full octahedral group  where $S_4$ ($\supseteq \mathbb{Z}_2\times \mathbb{Z}_2$) is the permutation group of order $24$. 
We note that $G_{K\Gamma}$ is nonsymmorphic in the sense that  Eq. (\ref{eq:short_exact}) is a non-split short exact sequence. 
For the purpose of MBQC in this work, we will only use the $\mathbb{Z}_2\times\mathbb{Z}_2$ subgroup in $G_{K\Gamma}$. 
How other nonsymmorphic symmetry operations beyond the $\mathbb{Z}_2\times\mathbb{Z}_2$ subgroup play a role in MBQC is an interesting and open question, which will be left for future investigations. 

\begin{figure}
\begin{center}
\includegraphics[width=7cm]{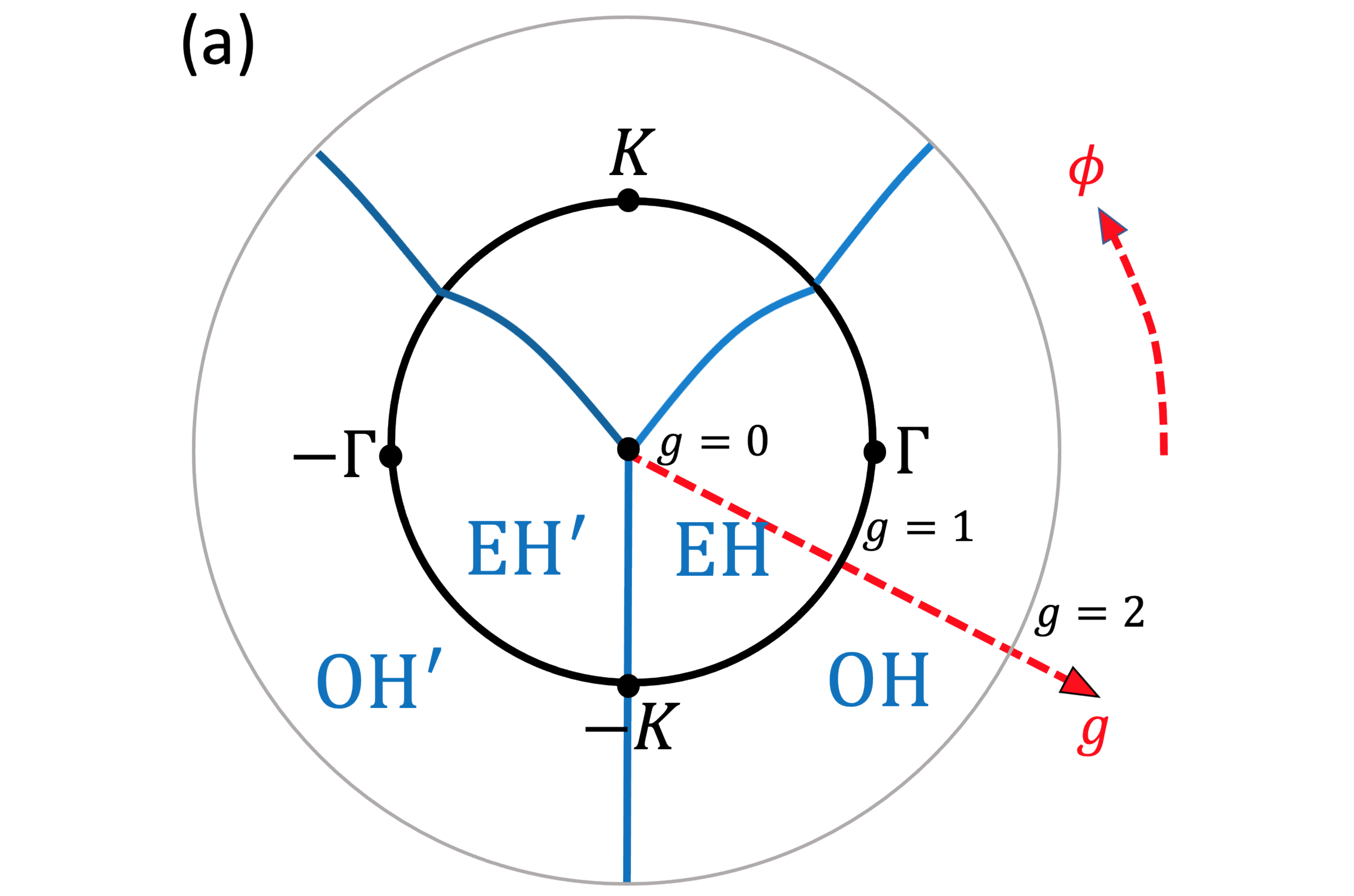} 
\includegraphics[width=7cm]{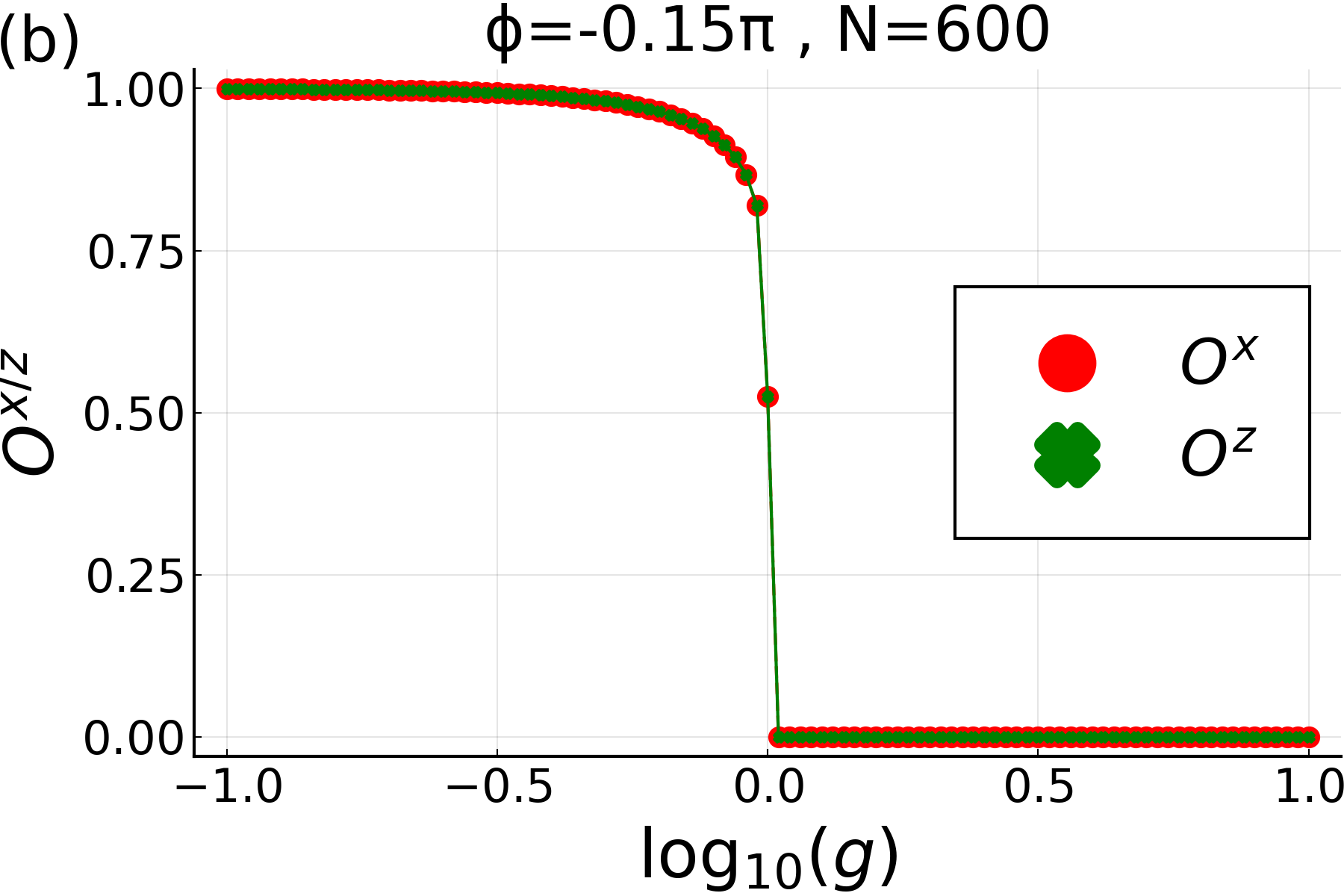}
\caption{
(a) Schematic plot of the phase diagram of the spin-1/2 bond-alternating Kitaev-Gamma chain, 
(b) string order parameters $O^\alpha(\frac{N}{2}+1,N)$ ($\alpha=x,z$) as functions of $\log(g)$ at $\phi=-0.15\pi$.
In (a), $K=\sin(\phi)$, $\Gamma=\cos(\phi)$, and $g=g_y/g_x$;
$\phi$ and $g$ are represented as azimuthal and radial coordinates, respectively;
the center point, middle circle, and outer circle  correspond to $g=0$, $g=1$, and $g=2$, respectively;
the red dashed line is along the $\phi=-0.15\pi$ radial direction;
only the EH, EH$^\prime$, OH, OH$^\prime$ phases are shown,
and the full phase diagram can be found in Ref. \cite{Luo2021}.
In (b), DMRG simulations are performed on open chains with system size $N=600$ ending with an $x$-bond on the right boundary.
\label{fig:phase_KG}
}
\end{center}
\end{figure}

\paragraph{Phase diagram.}
We briefly describe the phase diagram of the model defined in Eq. (\ref{eq:KG_U6}) \cite{Luo2021}, 
using the parametrization 
$K=\sin(\phi), ~\Gamma=\cos(\phi), ~ g=g_y/g_x$. 
There are four SPT phases in the phase diagram shown in Fig. \ref{fig:phase_KG} (a) \cite{Luo2021}, i.e., the EH, EH$^\prime$, OH, and OH$^\prime$ phases.
Since the other three SPT phases are related to the EH phase by unitary transformations (for details, see Appendix \ref{app:KG}),
it is enough to consider the EH phase,
which is characterized by the following non-vanishing string order parameter
in the $|j-i|\rightarrow \infty$ limit in the rotated frame ($\alpha\in\{x,y,z\}$)  \cite{Luo2021},
\begin{eqnarray}
O^\alpha(2i+1,2j) =   \langle \Pi_{k=2i+1}^{2j} 2S_k^\alpha  \rangle.
\label{eq:def_Oe}
\end{eqnarray}
We note that ``EH" is ``even-Haldane" for short, 
the name of which is chosen because of the fact that
the phase is in the same SPT phase as the Haldane phase of the bilinear-biquadratic spin-1 chain \cite{Haldane1983,Affleck1987,Chen2011}.

Fig. \ref{fig:phase_KG} (b) shows the numerical values of the string order parameters $O^\alpha(\frac{N}{2}+1,N)$ ($\alpha=x,z$) in the rotated frame as a function of $\log (g)$ 
for an even-length chain ending with an $x$-bond on the right boundary,
where $\phi$  is fixed to be $-0.15\pi$
corresponding to the red dashed line in Fig. \ref{fig:phase_KG}.
Clearly, as can be seen from Fig. \ref{fig:phase_KG} (b), there is a phase transition at $g=1$, separating the EH phase  in the $g<1$ region from another phase (in fact, the OH phase) in the $g>1$ region.
Discussions of the OH phase are included in Appendix \ref{app:KG}.
The system in the EH phase has a non-degenerate ground state $|\Phi_{K\Gamma}\rangle$, with a spectral gap separating the ground state from the excited states. 
The state $|\Phi_{K\Gamma}\rangle$ is short-range entangled with the $\mathbb{Z}_2\times \mathbb{Z}_2$ symmetries in Eq. (\ref{eq:sym_Z2_Z2_KG}), and can be used for MBQC purposes to be discussed in Sec. \ref{subsec:MBQC_KG}.  

\subsubsection{Cellular automaton states}

In this section, we study MBQC resource states that are more entangled siblings of the 1D cluster state, and which have a larger symmetry group than the two previous examples. Indeed the purpose of this example is to illustrate that our main theorem applies beyond $G=\mathbb{Z}_2\times \mathbb{Z}_2$.

The resource states discussed here, at the renormalization group fixed point, are generated by Clifford cellular automata, in  $\tau$ rounds of applying the transition function. The cluster state arises in this fashion, for $\tau=1$. The larger $\tau$ the larger the number of logical qubits that can be processed. It has recently been shown that universal MBQC resource states can be created in 1D in this fashion, when $\tau$ is allowed to scale \cite{DTS}. However, here we are content with a fixed value of $\tau$, $\tau=2$, yielding a model with two logical qubits. The symmetry group is $\left(\mathbb{Z}_2\right)^4$.

Specifically, we consider a quantum circuit with nearest neighbour interactions that takes the product state to the 1D cluster state of $N$ qubits i.e.
\begin{equation}\label{QCA}
   U_N =  \prod_{i=1}^{N-1} CZ_{i,i+1} \prod_{i=1}^{N} H_i.
\end{equation}
where $H= \frac{X+Z}{\sqrt{2}}$, and $CZ= |0 \rangle \langle 0|    \otimes I + |1\rangle\langle 1| \otimes Z$. Now, this circuit is applied $\tau$ times to the product state $\otimes_i |0\rangle_i$, arriving at the resource states
\begin{equation}
|\Phi_{\tau}\rangle= \left(U_N\right)^{\tau} \left(\bigotimes_{i=1}^{N} | 0\rangle_i\right) . \end{equation}
The resources states $|\Phi_{\tau}\rangle$ are capable of encoding $\tau$ logical qubits on which MBQC can be performed. Such states can also be seen as fixed points belonging to different quantum phases with non-trivial SPT order. 

In the remainder of this section, we focus on the case of $\tau=2$, which suffices for our present purpose.
We first note that $|\Phi_{2}\rangle$ is a stabilizer state, constrained by the eigenvalue equations 
$K_i |\Phi_{2}\rangle=|\Phi_{2}\rangle$, for  $i=1,..,N$, with 
$$K_i= Z_{i-2}X_{i-1}Z_{i}X_{i+1}Z_{i+2}, \quad i=3,..,N-2$$
in the bulk, and
$$K_1=X_2Z_3,  \quad K_{2}=X_1Z_2X_3Z_4 ,\quad K_{N-1}=Z_{N-3}X_{N-2}Z_{N-1}X_{N}, \quad K_{N}=Z_{N-2}X_{N-1}.
$$
at the boundary. The above constraints specify the state uniquely, up to a global phase.

 W.l.o.g we choose $N = 6k+4$. The stabilizer is an Abelian group, with a subgroup \begin{equation}
G_2= \mathbb{Z}_2^{4} \cong 
\left\langle 
\begin{array}{c} IZ (XIIIXI)\ldots XI,  ZX (IIIXIX)\ldots ZX, \\ XI(IIXIXI) \ldots  IZ ,   ZI (IXIXII) \ldots IX
\end{array}
\right\rangle . 
\end{equation}
$G_2$ is the symmetry group of interest. The $\tau=2$ automaton phase is the phase of $G_2$-symmetric states that contains the fixed point state.

The relevant string order parameters that capture the computational power of $G_2$ symmetric states are given by
\begin{align}{\label{QCAsop}}
\begin{split}
\sigma_1&=    \langle II\ldots I(IIIIIZ)(XIIIXI)\ldots XI \rangle,\\
\sigma_2&=   \langle II\ldots I(IIIIZX)(IIIXIX)\ldots ZX \rangle,\\
\sigma_3&=   \langle II\ldots I(IIIZXI)(IIXIXI)\ldots IZ \rangle,\\
\sigma_4&=  \langle II\ldots I(IIZXII)(IXIXII)\ldots IX \rangle,\\ 
\sigma_5&= \langle II\ldots I(IZXIII)(XIXIII)\ldots XZ \rangle,\\ 
\sigma_6&=\langle II\ldots I(ZXIIIX)(IXIIIX)\ldots ZI \rangle,
\end{split}   
\end{align} 
and expectation values of the all the non-trivial products of the operators involved in the first four string order parameters.
For illustration, we consider a one-dimensional line in the phase diagram of $\mathbb{Z}_2 ^{4}$-symmetric states. Namely, of our interest are the ground states $|\Phi_2(\alpha)\rangle$
 of the Hamiltonian, 
\begin{equation}\label{QCAHa}
H(\alpha) = -\cos \alpha \sum_{i=1}^{N} K_i -  \sin \alpha \sum_{i=3}^{N-2} X_i \; ,
\end{equation}
parametrized by an angle $\alpha$. 

\begin{figure}
\begin{center}
\includegraphics[width=10cm]{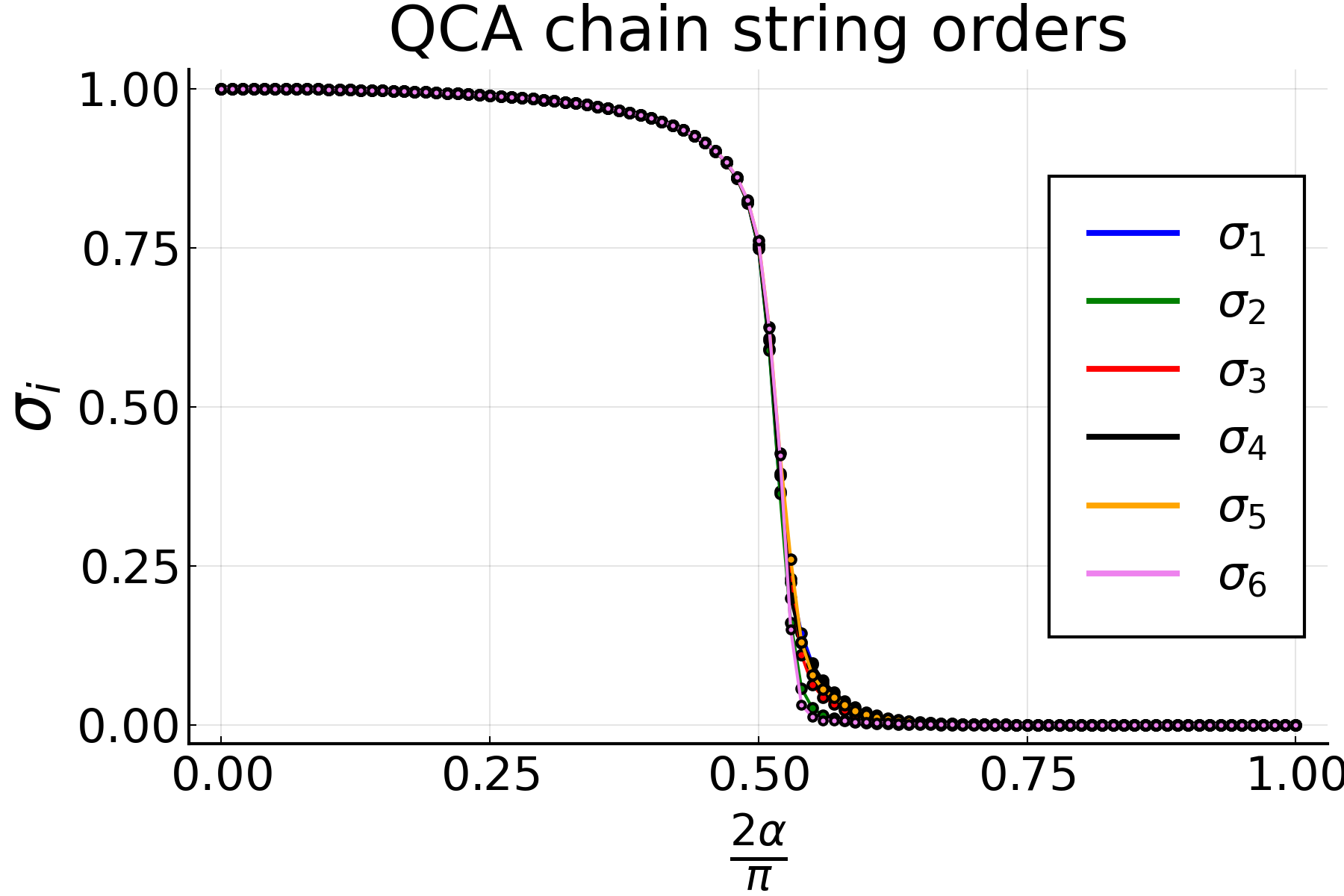}
\caption{\label{QCAComp} Order parameters $\sigma_{1}, \sigma_{2}, \sigma_{3}, \sigma_{4},\sigma_{5}, \sigma_{6}$ as a function $\alpha$ for the cellular automaton states $|\Phi_2(\alpha)\rangle$ on a chain length of $N=100$. As before, they all start near the site $N/4$ deep in the bulk.}
\end{center}
\end{figure}

\paragraph{Phase diagram.} When $\alpha = 0$, the ground state is the fixed point stabilizer state $|\Phi_2\rangle$. When $\alpha=\pi/2$, then the ground state is $2^4$-fold degenerate (due to the non-existence of magnetic fields at 4 boundary sites), $|\Phi_2(\pi/2)\rangle = |\pm\rangle_1 \pm\rangle_2 |+\rangle_3..|+\rangle_{N-2}|\pm\rangle_{N-1} |\pm\rangle_N$. At $\alpha=\pi/4$ occurs a {change-over} from QCA-like states to trivial states. This {change-over} is marked by the string order parameters $\sigma$ changing from non-zero to zero. See Fig.~\ref{QCAComp} for a plot of the order parameters in Eq.~(\ref{QCAsop}) as a function of $\alpha$. 

\subsubsection{The Ising chain}

From the perspective of symmetry protected topological order, the Ising chain with transverse magnetic field is an odd case. The second cohomology group of its symmetry group $\mathbb{Z}_2$ is the one-element group. Hence the only phase that exists in this model is the trivial phase, which has no computational power. The Ising model therefore is---from the quantum computational perspective---a non-example. The purpose of considering it here is to test how the new formalism handles it.\smallskip

We consider the Ising Hamiltonian in a transverse magnetic field,
\begin{equation}
H = -g \cos{\alpha} \sum_{i=1}^{N-1}Z_iZ_{i+1} -g \sin \alpha \sum_{i=1}^N X_i.
\end{equation}
This Hamiltonian is symmetric under the group $\mathbb{Z}_2$ generated by
\begin{equation}\label{SymmIsi}
U(g_1) = X_1X_2...X_{N-1}X_N.
\end{equation}
We are interested the ground state of this Hamiltonian, and if there is more than one, then we consider the eigenstates of the symmetry operator $g_1$ in the ground state manifold.

\section{SPT-MBQC in the MPS formalism}\label{MPSMBQC}

Here we review the existing formalism \cite{SPTO1,SPTO2} for MBQC in SPT phases which is based on matrix product states (MPS) \cite{MPSref}; see \cite{NovSchem} for the description of MBQC in terms of MPS. Previous results \cite{Bartl, MM2, SPTO1, SPTO2, 2Duniv, DW, QCA, MScDav, DAM} on computational phases of quantum matter use this framework. The purpose of this section is to create a reference point for the new formalism set up in Section~\ref{Res}. 

\begin{figure}
\begin{center}
\includegraphics[width=12cm]{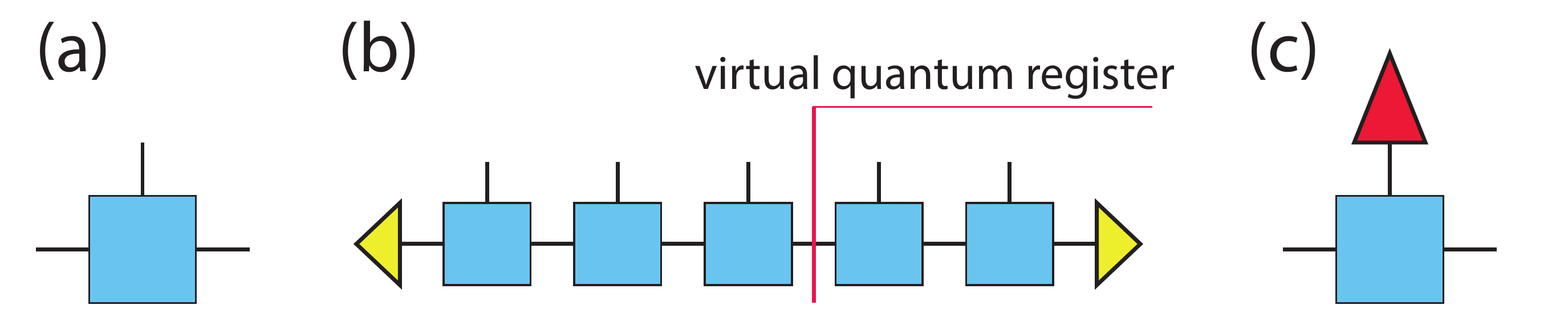}
\caption{\label{MPS}Basics of MBQC in the MPS framework. (a) Tensor representing one block of spins. The vertical leg represents the physical system, and the horizontal legs are virtual. (b) The virtual legs are contracted to represent the MBQC resource state. The logical quantum register simulated by MBQC is located on the virtual legs. (c) The elementary tensor with the physical leg contracted by a post-measurement quantum state represents a logical operation on the virtual quantum register.}
\end{center}
\end{figure}

The elementary physical unit is a block of spins on which the symmetry acts faithfully. (This creates some tension with MBQC, as such blocks typically contain more than a single spin, whereas the MBQC notion of locality is single-spin. More on that below.) Each block of spins is associated with an MPS tensor, with a `physical leg' for the block of spins, and `virtual legs' for the mediation of entanglement. The quantum register whose evolution is simulated by MBQC lives on the virtual legs \cite{NovSchem}, and each MPS tensor, with its physical leg contracted by measurement, represents a logical transformation of the virtual quantum register; see Fig.~\ref{MPS}.\medskip

The starting point for MBQC with uniform computational power is uniform wire \cite{Bartl}; i.e., the observation that if all blocks of spins are measured in the symmetry-respecting basis, then quantum information can be shuttled from one edge of a symmetry-protected spin chain to the other by local measurement, with perfect accuracy, for any ground state in a suitable SPT phase. 

The wire construction provides the following important technical ingredient. The Hilbert space associated with the virtual legs of the MPS tensor $A$ is a tensor product of the `logical subspace' and a `junk subspace'. The MPS tensors $A_s:=\langle s| A$, with the physical leg contracted by measurement {\em{in the symmetry-respecting basis}}, for all outcomes $s$, take the special form
\begin{equation}\label{TP}
A_s = C_s \otimes B_s.
\end{equation}
Therein, the operators $C_s$ act on the logical space, and the operators $B_s$ on the junk space. The $C_s$ are constant throughout the phase, and are determined by group cohomology. The $B_s$ are unknown and uncontrolled. This is the content of Theorem~1 in \cite{Bartl}. Random but heralded, perfectly accurate action occurs on the logical subsystem. All that happens is the accumulation of MBQC byproduct operators $C$. The evolution of the junk subsystem is unknown. As long as the two subsystems don't interact, the logical subsystem is fine. 

When applied to the $\mathbb{Z}_2\times \mathbb{Z}_2$ cluster phase, there are four measurement outcomes $s \in \mathbb{Z}_2\times \mathbb{Z}_2$, and the corresponding operators $C_s$ are in the Pauli group. The block consists of two spins-1/2.\medskip

To generalize from quantum wire to quantum computation, one has to tune the basis of block-local measurement away from the symmetry-respecting basis (the $X$-basis in the cluster case). But that creates a problem: The decomposition of the evolution operator $A_s$ into a tensor product, Eq.~(\ref{TP}), no longer holds. The resulting evolution on the virtual space makes the logical and the junk subsystem interact, effectively decohering the logical subsystem \cite{Bartl,Bartl3}. This is the core obstacle MBQC in SPT phases has to overcome.

A solution for this problem has been provided in \cite{SPTO1,SPTO2}. The first step is `oblivious wire'. Like standard wire, it operates by measuring a number of consecutive spins/blocks in the symmetry-respecting basis. The difference is that only the total accumulated byproduct operator is kept, and all other information provided by the measurement record is discarded. The resulting averaging procedure has the following effects (see Lemma~1 in \cite{SPTO1}): (i) the logical and the junk register become disentangled, and (ii) the state of the junk register is driven towards a fixed point $\rho_\text{fix}$. The fixed point state is a priori unknown, but reproducible conditions are achieved.

An elementary logical gate consists of  one block of spins measured away from the symmetry-respecting basis, followed by oblivious wire. It takes as input a state $\sigma_\text{log} \otimes \rho_\text{fix}$, and, to any desired degree of accuracy (determined by the length of oblivious wire), returns a state $\sigma'_\text{log} \otimes \rho_\text{fix}$. The separation of the logical and the junk subsystem is preserved, and the remaining question is after the resulting logical operation $\sigma_\text{log}' = {\cal{T}}(\sigma_\text{log})$.

For measurement angles $\beta$ away from the symmetry-respecting basis, in any direction $k$ of choice, the resulting operation ${\cal{T}}_\beta$ is, up to linear order in $\beta$, 
\begin{equation}\label{Tbeta}
{\cal{T}}_\beta(\sigma)  =  \sigma + i\beta \left[\frac{\nu_{k0} C(k) -\nu^*_{k0}C(k)^\dagger}{i}, \sigma \right] + {\cal{O}}(\beta^2),
\end{equation}
with $C(k)=C_0^{-1}C_k$. The important point to note is that deviations from the identity operation arise to linear order in $\beta$, whereas deviations from unitarity only arise to second order. This makes arbitrarily accurate computation possible, by splitting rotations about large angles into many rotations about small angles.

The parameters $\nu_{k0}$ above form the computational order parameter. For their definition, see Eq.~(20) of \cite{SPTO1}, or Section~\ref{StringComp} below. The gate constructions Eq.~(\ref{Tbeta}) introduce a constant multiplicative overhead dependent on the off-diagonal order parameter component $\nu_{k0}$. When  $|\nu_{k0}|$ is large, then computation is more efficient than when it is small. However, the value of $\nu_{k0}$ is irrelevant for what can be computed, as long as it is non-zero. 

So what can be computed?---This question is answered by Theorem~2 in \cite{SPTO1}: The byproduct operators $C_s$ span a Lie algebra of executable unitary gates, and Hermitian linear combinations of them can be measured.\medskip

In the 1D cluster phase (Example 1), the steps of identifying the computational power of MBQC return the following:
\begin{enumerate}
\item{{\em{Identifying the logical quantum register.}} With Lemma~2 of \cite{Bartl}, we find for the dimension of the logical quantum register $d_\text{log}=2$; i.e., one logical qubit is processed. With Lemma~1 of \cite{Bartl}, we identify the byproduct operators as $C_{00}=I$, $C_{01}=X$, $C_{10}=Z$, $C_{11}=Y$.}
\item{{\em{Figuring out what can be computed.}} With Theorem~2 of \cite{SPTO1}, we find that all gates in $SU(2)$ can be realized, and all Pauli observables be measured. Thus, MBQC in the entire 1D cluster phase is 1-qubit universal.}
\item{{\em{Efficiency.}} As described above, the computational order parameters $\nu_{k0}$ affect efficiency of computation, though not computability. Given a symmetric resource state, the parameters $\nu_{ij}$ can be obtained from its MPS representation.}
\end{enumerate}

{\em{Block vs. site locality:}} The existing formalism reviewed in this section is based on block locality whereas MBQC demands site locality. This creates a tension, and additional patches are required to move from block-locality to site locality in the MPS formalism. We describe the argument for the cluster chain below. It is indeed an advantage of the new formalism, to be introduced in the next section, that it can handle site-locality at the basic structural level.

For the cluster chain, the (symmetric) wire basis in each block is the simultaneous eigenbasis of $X\otimes I$ and $I\otimes X$. Closer analysis reveals that, in order to perform a rotation about the $x$-axis, the symmetric basis needs to be transformed by a unitary $\exp(i\beta\, Z\otimes I)$. For a $z$-rotation, it needs to be transformed by a unitary $\exp(i \beta\, X\otimes Z)$, and for a $y$-rotation by a unitary $\exp(i\beta \,Y\otimes Z)$. All these measurements are block-local. In addition, the measurements to implement $x$- and $z$-rotations are site-local, while the measurement to implement the $y$-rotation is not. The strategy is thus to leave out the $y$-rotations, in exchange for achieving site-locality. The Lie group generated by $X$ and $Z$ is still $SU(2)$; hence enforcing site-locality does not reduce computational power in this case.\medskip

Now returning to the general discussion, when setting up the new formalism in Section~\ref{Res}, we will address the following questions relating to the above review:
\begin{itemize}
\item[(i)]{What is the basic logical structure, i.e., the counterpart to the logical subsystem in Eq.~(\ref{TP})?}
\item[(ii)]{What is the statement of closure of logical operations; i.e., the counterpart to their action on the logical subsystem alone, cf. Eq.~(\ref{Tbeta})?}
\item[(iii)]{What is our statement of computational capability, i.e., the counterpart to Theorem~2 in \cite{SPTO1}?}
\item[(iv)]{How efficient is the computation?}
\end{itemize}

\section{MBQC on short-range entangled symmetric states}\label{Res}

In this section, we devise a new algebraic formalism for reasoning about computational phases of quantum matter. It contains our main result, Theorem~\ref{GT} on the relation between string order and MBQC computational power. In Section~\ref{Defs} we make the necessary definitions and introduce the constituents of MBQC; in Section~\ref{PP} we describe the circuit model evolution simulated by MBQC; and in Section~\ref{Sec:sreMBQC} we state the main theorem and explain how to use it. Section~\ref{GTproof} gives the proof of the main theorem; and Section~\ref{Exa2} applies it to the three examples introduced in Section~\ref{Exa}. The formalism takes some effort to set up, but once in place it is versatile and easy to use. 

The gist of the argument laid out in this section is as follows. We define a set of observables $T(g)$, $g\in G$, which can be understood as measurable properties of an MBQC-simulated quantum register; see Eq.~(\ref{T_enc}) below. Then, (i) We derive an evolution equation for the expectation values $\langle T(g)\rangle$, cf. Eq.~(\ref{LinRel})/ Lemma~\ref{LEMsreMBQC}; and (ii) We show that the evolved observables, at the output stage, can be {\em{locally}} measured (Lemma~\ref{Inf}). Both Lemmas combined yield our main result, Theorem~\ref{GT}. 

\subsection{The constituents of MBQC}\label{Defs}

Here we define the notions required to parse our main result, Theorem~\ref{GT} stated in Section~\ref{Sec:sreMBQC}. We begin by defining the pertaining representations of the symmetry group and their consistency conditions. After that, we define the MBQC measurement patterns and the string order parameters characterizing SPT in 1D systems, and specify  the gate operations MBQC on the symmetric, short-range entangled states $|\Phi\rangle$ can simulate.

\subsubsection{Representations of the symmetry group}\label{subsubsec:assumptions}

For the blocks $i=0,..,n+1$ that make up the spin chain we define three types of representations of the symmetry group $G$ (one linear, two projective). We require the following statement about projective representations of $\mathbb{Z}_2^n$.
\begin{Lemma}\label{cacLemma}
For all projective representations $v$ of a group $G=\mathbb{Z}_2^m$, $m\in \mathbb{N}$, it holds that $v(g)v(g') = \pm v(g')v(g)$, for all $g,g'\in G$.
\end{Lemma}
The proof of Lemma~\ref{cacLemma} is given in Appendix~\ref{cacAp}.\smallskip

The representations of interest satisfy consistency constraints. To prepare for the statement of these constraints, we introduce the subgroups ${\cal{G}}_i \subseteq G$ for all bulk sites $i=1,..,n$. The subgroups ${\cal{G}}_i$ comprise the generators of unitary transformations that can be effected through measurement on any given block $i$. 

The ${\cal{G}}_i$ become an important concept later on; and so we illustrate them here in an example. For the 1D cluster phase, as discussed at the end of Section~\ref{MPSMBQC}, there are two choices. (a) The block comprises two spins, such that the symmetry action is faithful. In this case ${\cal{G}}_i=G$, for all blocks $i$ in the bulk. Then, rotations about either of the three axes $x$, $y$, $z$ can be performed in each block. (b) The block comprises a single spin. Then, ${\cal{G}}_i=\langle g_{01}\rangle$ for $i$ odd, and ${\cal{G}}_i=\langle g_{10}\rangle$ for $i$ even. Correspondingly, only $x$-rotations can be performed on odd sites, and only $z$-rotations on even sites. This agrees with the standard treatment of MBQC on 1D cluster states.\medskip

We now introduce the relevant representations of $G$, separately for the three regions of the chain---right boundary, left boundary, bulk. 

{\em{Right boundary.}}  The right boundary, $i=n+1$, only carries the projective representation $v_{L,n+1}(G)$. It satisfies the commutation relations
\begin{equation}\label{KappaL}
v_{L,n+1}(g) v_{L,n+1}(g') = (-1)^{\kappa(g,g')} v_{L,n+1}(g') v_{L,n+1}(g), \;\;\;  \forall g,g'\in G,
\end{equation}
parametrized by a function $\kappa: G\times G \longrightarrow \mathbb{Z}_2$, in accordance with Lemma~\ref{cacLemma}. Eq.~(\ref{KappaL}) defines $\kappa$.\smallskip

{\em{Left boundary.}} On the left boundary, $i=0$, we have a linear representation $u_0(G)$, and a projective representation $v_{R,0}(G)$. They satisfy a mutual consistency condition, namely the symmetry group $G$ has a maximal subgroup $H$ with the  property 
\begin{equation}\label{u0}
[v_{R,0}(h),v_{R,0}(h')]=0,\; \forall h,h'\in H.
\end{equation}
For this subgroup it holds that
\begin{equation}
\label{H_spec}
u_0(h) = v_{R,0}(h),\;\forall h\in H.
\end{equation}
Furthermore, we match the commutation relations of $v_{L,n+1}(G)$ on the right boundary,
\begin{equation}\label{KappaR0}
    v_{R,0}(g)v_{R,0}(g') = (-1)^{\kappa(g,g')} v_{R,0}(g')v_{R,0}(g), \;\forall g,g'\in G.
\end{equation}

{\em{Bulk.}} For $i=1,..,n$, we have the linear representations $u_{i}(G)$ and the projective representations $v_{L,i}({\cal{G}}_i)$ and $v_{R,i}({\cal{G}}_i)$. Note that, in the bulk, those projective representations are only defined for the subgroups ${\cal{G}}_i$ of $G$, not a priori for $G$ itself.

We have the consistency constraints
\begin{equation}\label{KappaR}
    v_{R,i}(g)v_{R,i}(g') = (-1)^{\kappa(g,g')} v_{R,i}(g')v_{R,i}(g), \;\forall g\in G,\, g'\in {\cal{G}}_i,\; i=1,..,n.
\end{equation}
Furthermore
\begin{eqnarray}
\label{vLRcomm}
[v_{L,i}(g),v_{R,i}(g')] &= &0, \;\forall g \in G, g'\in {\cal{G}}_i,\;\;\; \forall i=1,..,n,\\
\label{url}
v_{L,i}(g)v_{R,i}(g) &= & u_i(g), \;\forall g\in G,\;\;\; \forall i=1,..,n.
\end{eqnarray}

{\em{Whole chain.}} The linear representation $U$ of $G$ on the entire spin chain is given by
\begin{equation}
\label{Udef}
U(g):=v_{R,0}(g)\left(\bigotimes_{i=1}^n u_i(g)\right) v_{L,n+1}(g),\;\; \forall g\in G.
\end{equation}
This is indeed a linear representation because the phase factors from Eqs.~(\ref{KappaL}) and (\ref{KappaR0}) cancel.
\medskip

\begin{figure}
\begin{center}
\includegraphics[width=10cm]{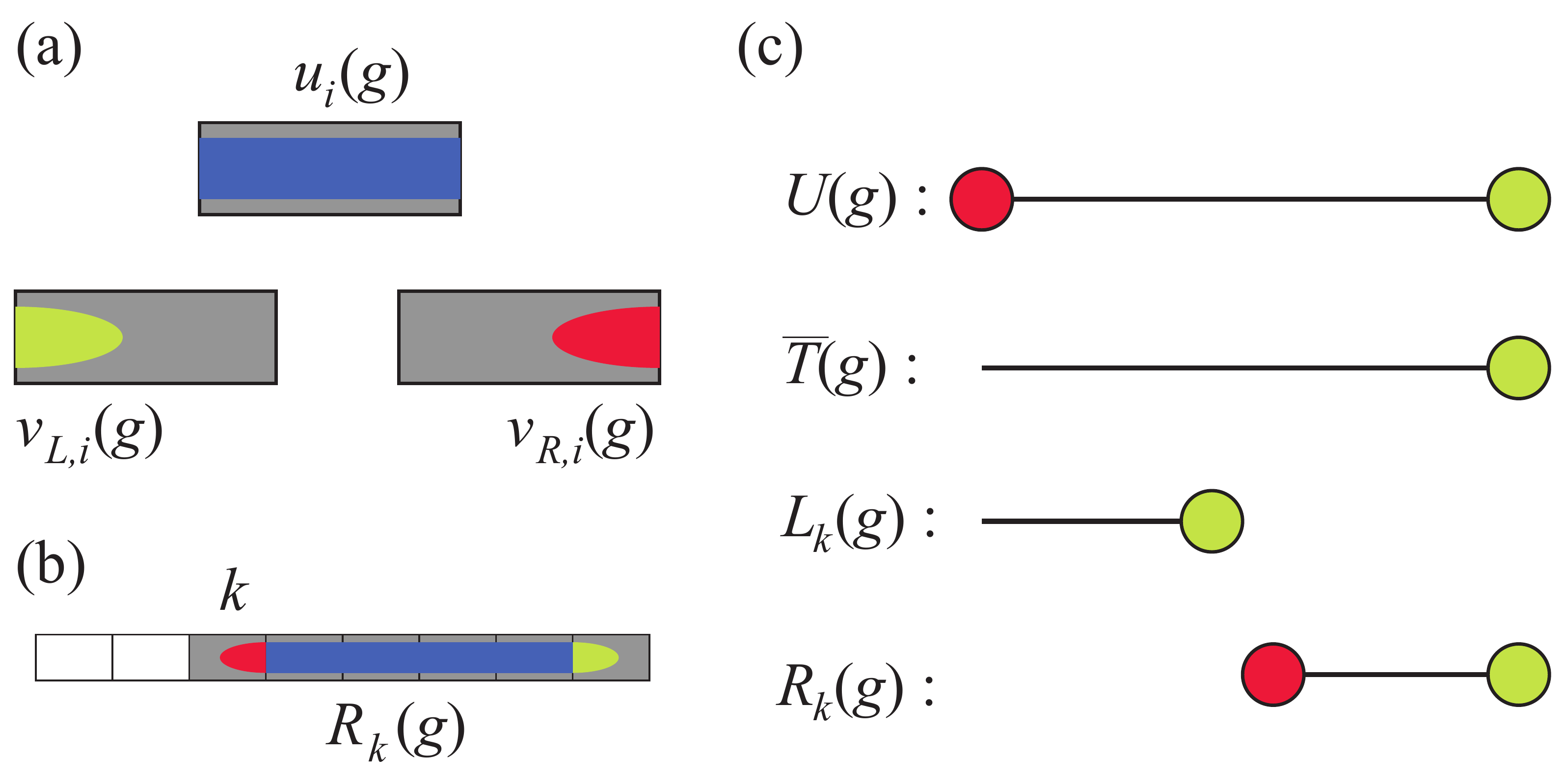}
\caption{\label{fig:Reps}Representations of the symmetry group $G$. (a) For each site in the bulk, we use a linear representation $u_i(G)$, and two projective representations $v_{L,i}(G)$, $v_{R,i}(G)$. The representation $v_L$ ``lives on the left'' of the block, and connects to other blocks to the left of it. $v_R$ connects to the blocks on the right. This is the reason for the naming. (b) An operator $R_k(g)$ giving rise to a string order parameter. (c) Simpler graphical notation for some operators of interest. The coloured dots at the end points represent the projective representations $v_L$ and $v_R$ (same colouring as in (a) and (b)), and the line connecting them a string of linear representations $u$. Shown are the action $U(G)$ of the symmetry $G$ on the whole spin chain, the logical operators $T(g)$, the operators $L_k(g)$ responsible for logical rotations evoked by measurement of the $k$-th block of spins, and $R_k(g)$, the operator giving rise to the string order parameter $\sigma_k(g) = \langle R_k(g) \rangle$. }
\end{center}
\end{figure}

This concludes the definition of the relevant linear and projective representations of $G$. See Fig.~\ref{fig:Reps} for a graphical display of the objects defined. The conditions  Eqs.~(\ref{u0}) -- (\ref{Udef}) need to be satisfied when applying the present formalism to examples. \smallskip

In a nutshell, the physical significance of the above constraints is as follows. (i) Eqs.~(\ref{u0}) and (\ref{H_spec}) define the initial state of the processed logical information. (ii) Given Eq.~(\ref{KappaL}), the definition of $\kappa$, Eq.~(\ref{KappaR0}) ensures that Eq.~(\ref{Udef}) indeed describes a linear representation of $G$, acting on the whole chain. (iii) Eqs.~(\ref{KappaR}), (\ref{vLRcomm}), (\ref{url}), ensure that the string order operators---to be defined in Eq.~(\ref{DefLR}), (\ref{string}) below---commute with the symmetry, hence can have non-zero expectation values. Eq.~(\ref{KappaR}) and (\ref{vLRcomm}) determine the sets ${\cal{G}}_i$, hence the executable gates.

\subsubsection{MBQC schemes and measurement patterns}\label{MP}

The main result of this section, Theorem~\ref{GT}, attributes computational power to certain symmetric quantum states---without explicitly mentioning the measurement pattern that unlocks this computational power. But the proof of the theorem is constructive, and the measurement patterns used are the ones described below. We introduce those measurement patterns now, because they are a first application of the definitions made in the previous section.

\paragraph{Independent constituents of MBQC measurement patterns.} All measurement patterns we consider have the same structure. They consist of various pieces of information, continuous and discrete. Some of those pieces are dependent, through the compatibility relations Eq.~(\ref{H_spec}) -- Eq.~(\ref{Udef}). We begin by listing the independent pieces. They are
\begin{enumerate}
\item{The symmetry group $G=\left(\mathbb{Z}_2\right)^m$, for some $m \in \mathbb{N}$, and the required linear and projective representations of $G$; namely
\begin{itemize}
\item{On the left boundary, i.e., block 0, the projective representation $v_{R,0}(G)$.}
\item{In the bulk, i.e., blocks $i=1,.., n$, the linear representations $u_i(G)$ and the projective representations $v_{L,i}(G)$.}
\item{On the right boundary, i.e., block $n+1$, the projective representation $v_{L,n+1}(G)$.}
\end{itemize}}
\item{The data that specifies any given quantum algorithm, namely
\begin{itemize}
\item{For each block $i=1,..,n$ in the bulk, the basis of measurement specified by: (i) a rotation plane $g_i \in {\cal{G}}_i$, and (ii) a rotation angle $\alpha_i \in [-\pi,\pi]$  (subject to the constraint that the angles can be non-zero only on blocks at least $2\Delta$ apart).}
\item{A subgroup $H \subset G$, specifying the logical initial state.}
\item{A subgroup $H' \subset G$, specifying the logical readout.}
\end{itemize}}
\item{The classical side processing relations to convert measurement record into computational output.  There is one bit worth of measurement adjustment $q_k$ for every block $k=1,..,n$, and one bit of output $o(h)$ for every group element $h\in H'$. The classical side-processing relations to obtain those from the measurement record $s_i(g)\in\mathbb{Z}_2$, $g\in G$, are
\begin{equation}\label{CPRgen}
\begin{array}{rcll}
q_k&=& \sum_{i=0}^{k-1}s_i(g_k) \mod 2,&\forall k=1,..,n,\vspace{1mm}\\
o(h) &=& \sum_{i=0}^{n+1} s_i(h) \mod 2, &\forall h \in H'.
\end{array}
\end{equation}}
\end{enumerate}
The three items listed above live at various levels of generality. The measurement angles, measurement planes, and subgroups $H$, $H'$ in item 2 describe a given quantum algorithm within a fixed MBQC scheme. They do not describe MBQC schemes themselves. Item 3, the classical side processing relations, is at the opposite end of the spectrum. As we will prove, the classical side processing relations are of the same form Eq.~(\ref{CPRgen}) for {\em{all}}  MBQC schemes in 1D. Hence they do not specify such  MBQC schemes. The remaining entry in the list, item 1, contains the only independent information that discriminates between and hence characterizes MBQC schemes in 1D. It is the basis for a future classification of MBQC schemes with $(\mathbb{Z}_2)^m$-symmetry in 1D.

\paragraph{Dependent constituents.} There are important parts of MBQC measurement patterns that are dependent through the constraints Eq.~(\ref{H_spec}) -- Eq.~(\ref{Udef}). Here we describe how to compute them.
\begin{enumerate}
\item{For the bulk blocks, $i=1,..,n$ we compute the projective representations $v_{R,i}(G)$ from $v_{L,i}(G)$ and $u_i(G)$ through Eq.~(\ref{url}).}
\item{On the left boundary, block 0, we compute $u_0(G)$ as follows. On $H\subset G$, $u_0(\cdot)$ is obtained from $v_{R,0}(G)$ through Eq.~(\ref{H_spec}). On $G\backslash H$, $u_0(\cdot)$ is free to choose, subject to the constraint that $u_0(G)$ is an Abelian group.}
\item{\label{GiComp}For the bulk blocks, $i=1,..,n$, the groups ${\cal{G}}_i\subset G$ of possible measurement planes are computed from the constraints Eq.~(\ref{KappaR}) and (\ref{vLRcomm}).} 
\item{The action of the symmetry group $G$ on the spin chain as a whole is given by Eq.~(\ref{Udef}).}
\end{enumerate}
Regarding item~\ref{GiComp} in the above list, we still need to show that the sets ${\cal{G}}_i$ resulting from this procedure are groups.

 \begin{Lemma}\label{CalG}
 For all blocks $i=1,.., n$, the maximal sets ${\cal{G}}_i$ are unique and are subgroups of $G$.
 \end{Lemma}

{\em{Proof of Lemma~\ref{CalG}.}} {\em{(i) Uniqueness.}} A set ${\cal{G}}_i \subset G$ is maximal in $G$ if it cannot be extended. The proof of uniqueness is by contradiction. Assume two distinct maximal sets exist, ${\cal{G}}_i \neq {\cal{G}}_i'$. Since all conditions on ${\cal{G}}_i$, ${\cal{G}}_i'$, namely Eq~(\ref{KappaR}), Eq.~(\ref{vLRcomm}) are element-wise, ${\cal{G}}_i\cup {\cal{G}}_i'$ is also a viable set ${\cal{G}}$. But ${\cal{G}}_i, {\cal{G}}_i' \subsetneq {\cal{G}}_i\cup {\cal{G}}_i'$, and hence ${\cal{G}}_i$, ${\cal{G}}_i'$ are not maximal -- contradiction.
Thus the maximal set is unique.

{\em{The maximal set ${\cal{G}}_i$ is a subgroup.}} It is easily verified that if $g',g'' \in G$ satisfy the constraints Eq.~(\ref{KappaR}) and (\ref{vLRcomm}) on $g'$, then so does $g'g''$. Further, $(g')^{-1}=g'$ for the present groups. Finally, $g'=I$ also satisfies Eqs.~(\ref{KappaR}), (\ref{vLRcomm}). $\Box$ 

\paragraph{Measurement procedure.} The measurements proceed from left to right, starting with block 0. The exception is block $n+1$ on the right boundary, whose basis is not adaptive and which therefore can be measured jointly with block 0 in the first measurement round. On the boundaries,  the measured observables are 
\begin{equation}\label{ObsBdy}
\begin{array}{rll}
O_0(g) =u_0(g),&\forall g\in G,& \text{for block 0},\\
O_{n+1}(h) = v_{L,n+1}(h),&\forall h \in  H',& \text{for block $n+1$}.
\end{array}
\end{equation}
In the bulk, the measured observables have a more complicated form. Namely, for any one $g_k \in {\cal{G}}_k$ chosen for block $k$,
\begin{equation}\label{ObsBlk}
O_k(g) =e^{i (-1)^{q_k} \frac{\alpha_k}{2} v_{L,k}(g_k)} u_k(g) e^{- i(-1)^{q_k} \frac{\alpha_k}{2} v_{L,k}(g_k)}, \forall g\in G,\;  k=1,..,n.
\end{equation}
Therein, $q_k\in \mathbb{Z}_2$ represents the adjustment of the measured observable according to measurement outcomes obtained elsewhere on the chain, as is usual in MBQC. Thus, in the bulk, the measurement in each block $k$ is specified by $q_k$, a measurement angle $\alpha_k$ and a logical rotation axis, given by the element $g_k$ of the symmetry group $G$. 

For any given $k\leq n$, the observables $O_k(g)$ pairwise commute for all $g\in G$, and can thus be measured simultaneously. We denote the corresponding measurement outcomes by $s_k(g)\in \mathbb{Z}_2$. Since, by construction, $O_k(g_1g_2)=O_k(g_1)O_k(g_2)$, it suffices to measure the observables corresponding to a set of generators of $G$. 
This completes the description of the measurement pattern. 

It remains to connect this procedure to the logical processing it affects. We do this below in Sections~\ref{LP} -- \ref{GTproof}.

\subsection{The logical observables}\label{PP} 

Here, we introduce computationally relevant quantities and operators defined on the large Hilbert space ${\cal{H}}$ in which the resource quantum state $|\Phi\rangle$ lives. These are the logical observables $T(g)$, $\forall g \in G$, the operators $L_k(g)$ upon which gate action is based, and the operators $R_k(g)$ that yield the string order parameters.

\subsubsection{Definition and properties}

We introduce the logical operators
\begin{equation}\label{T_enc}
T(g) := \left(\bigotimes_{i=0}^n u_i(g)\right)v_{L,n+1}(g),\;\; \forall g\in G.
\end{equation}
The observables $T(g)$ are encoded versions of $v_{L,n+1}(g)$, i.e., $T(g) = \overline{v_{L,n+1}(g)}$, $\forall g\in G$. They represent the basic logical structure in the present formalism, replacing the logical subsystem of the MPS-based formalism; see Eq.~(\ref{TP}). This addresses Question (i) of Section~\ref{MPSMBQC}. The logical subsystem is derived, whereas $T(G)$ is defined. The justification for the definition Eq.~(\ref{T_enc}) arises through the consequences for MBQC that it entails, specifically Theorem~\ref{GT} and Corollary~\ref{unit} below.\smallskip

Further, denote by $L_k(g)$ and $R_l(g)$ the operators
\begin{equation}\label{DefLR}
\begin{array}{rcll}
L_k(g)&:=&\left(\bigotimes_{i=0}^{k-1} u_i(g) \right)v_{L,k}(g),& \forall g\in {\cal{G}}_i,\;\;1\leq k\leq n,\\ 
R_l(g)&:=&v_{R,l}(g)\left(\bigotimes_{j=l+1}^n u_j(g)\right) v_{L,n+1}(g), & \forall g\in {\cal{G}}_i,\;\;1\leq l\leq n.
\end{array} 
\end{equation}
The expectation values
\begin{equation}\label{string}
\sigma_l(g):=\langle \Phi | R_l(g)|\Phi \rangle
\end{equation}
are string order parameters. We shall see later in Corollary~\ref{unit} that non-vanishing string order parameters provide computational power in MBQC. Eq.~(\ref{RUcomm}) is a precondition for the string order parameters to be non-zero. The operators $L_k(g)$ facilitate logical quantum operations, as discussed in Section~\ref{LP}.

\paragraph{Properties.} We now establish a few elementary properties that follow from the above definitions. A first  consequence of the above definitions is that $T$, $L$ and $R$ are related via
\begin{equation}
\label{Tdef2}
T(g) = L_i(g) R_i(g),\;\; \forall g\in {\cal{G}}_i,\;\;\;\forall i=1,..,n,
\end{equation}
With the definitions (\ref{Udef}), (\ref{T_enc}) it further holds that
\begin{equation}\label{TUrel}
u_0^\dagger(g) T(g) = v_{R,0}^\dagger(g)U(g),\;\;\forall g\in G,
\end{equation}
and specifically, with the condition Eq.~(\ref{H_spec}), 
\begin{equation}\label{TUrelH}
T(h) = U(h),\;\;\forall h\in H.
\end{equation}
With Eqs.~(\ref{PhiSymm}) and (\ref{TUrelH}), the resource state $|\Phi\rangle$ is an eigenstate of all logical operators $T(H)$, namely 
\begin{equation}\label{Hstab}
T(h)|\Phi\rangle =(-1)^{\chi(h)}|\Phi\rangle,\; \forall h\in H.
\end{equation}
This property describes the initialization of the logical quantum register; see Section~\ref{LogInit} below.\smallskip

The commutation relations among the basic representations $u$, $v_L$ and $v_R$ imply commutation relations among $U$, $T$, $L$ and $R$ which are of relevance for MBQC. In preparation for the proof of Theorem~\ref{GT}, we summarize these commutation relations in the following lemma.

\begin{Lemma}\label{cd}
(i) The following commutation relations hold.
\begin{eqnarray}
\label{RUcomm}
[R_k(g'),U(g)] &= &0,\;\;\forall g\in G, g'\in {\cal{G}}_k,\; k=1,..,n,\\
\label{RTcomm}
[R_k(g'),T(g)]&=&0, \;\; \forall g\in G, g'\in {\cal{G}}_k,\; k=1,..,n,\\
 \label{LTcomm}
 L_k(g')T(g)-(-1)^{\kappa(g,g')}T(g)L_k(g')&=& 0,\;\; \forall g\in G, g'\in {\cal{G}}_k,\; k=1,..,n,\\
 \label{TTcomm}
 T(g)T(g') - (-1)^{\kappa(g,g')} T(g')T(g) &=& 0, \;\; \forall g,g'\in G.
\end{eqnarray}
(ii) The operators $T(g)$, $L_k(g)$, $R_k(g)$, for all $g\in G$ and all $k$, can simultaneously be chosen Hermitian.
\end{Lemma}
Based on item (ii) of the lemma, henceforth we choose the operators $T(g)$, $L_k(g)$, $R_k(g)$, $\forall g\in G, \forall k$, to be Hermitian. As a first consequence, the string order parameters $\sigma_k(g)$ in Eq.~(\ref{string}) are all real.

Eq.~(\ref{RUcomm}) is a precondition for the string order parameters $\sigma_k(g)$ to be non-vanishing--which is a computational resource. Eq.~(\ref{RTcomm}) is a consistency condition. Eqs.~(\ref{LTcomm} and (\ref{TTcomm}) are used in the proof of Lemma~\ref{LEMsreMBQC}.\medskip

{\em{Proof of Lemma~\ref{cd}.}} (i) The logical operators $T(g)$ inherit their commutation relations from those of $v_{L,n+1}(g)$, cf. Eq.~(\ref{KappaL}). This establishes Eq.~(\ref{TTcomm}). For Eq.~(\ref{RTcomm}), with Eqs.~(\ref{KappaR}), (\ref{vLRcomm}) and (\ref{url}) we find that $R_k(g')T(g) =  (-1)^{\kappa(g,g')+\kappa(g,g')} T(g) R_k(g') = T(g) R_k(g')$.
The two contributions to the sign under commutation stem from blocks $k$ and $n+1$, and they cancel. The argument for Eq.~(\ref{RUcomm}) is the same.
For Eq.~(\ref{LTcomm}), $$L_k(g')T(g)=  T(g') R_k(g') T(g) = (-1)^{\kappa(g,g')}  T(g)  T(g') R_k(g') = (-1)^{\kappa(g,g')}  T(g) L(g').$$  Therein, we have used Eq.~(\ref{Tdef2}), and Eqs.~(\ref{RTcomm}), (\ref{TTcomm}) which are already established. \smallskip

(ii) We first show that $v_{L,k}(g)$ can be choosen Hermitian, $\forall g\in G$ and $k=1,..,n+1$. We have $v_{L,k}(g)^2\propto v_{L,k}(2g)=v_{L,k}(0)\propto I$, with the proportionality factor being a phase. Therefore, for any $k$ and $g$, we may adopt a phase convention such that $v_{L,k}(g)^2=I$. Then, the eigenvalues of the operator $v_{L,k}(g)$ are all $\pm 1$, hence all $v_{L,k}(g)$ are Hermitian.

By an analogous argument, $v_{R,k}(g)$ can be chosen Hermitian, $\forall g \in G$, and $k=0,..,n$. Likewise, for all $g\in G$ and $k=1,..,n$, it holds that $u_k(g)^2=u_k(2g)=u_k(0)=I$. Hence the eigenvalues of $u_k(g)$ are all $\pm1$, and all $u_k(g)$ are Hermitian.

We adopt the phase convention that makes $v_{L}$, $v_R$ Hermitian. With the definitions Eq.~(\ref{T_enc}) and (\ref{DefLR}) it then follows that the operators $T(g)$, $L_k(g)$, $R_k(g)$,  $\forall g\in G$, $\forall k$, are Hermitian. $\Box$
\medskip

\subsubsection{Logical initialization}\label{LogInit}

The MBQC resource states under consideration have the following property. 
\begin{Lemma}\label{L_init}
Consider a short-range entangled state $|\Phi\rangle$ of a spin-1/2 chain, symmetric under a group $G = (\mathbb{Z}_2)^m$ with the symmetry represented as described above. Then it holds that
\begin{equation}\label{InitEval}
\begin{array}{rcll}
\langle T(g)\rangle_\Phi &=& \displaystyle{(-1)^{\chi(g)}},& \text{if } g \in H,\\
\langle T(g)\rangle_\Phi &=& 0,& \text{if } g \in G\backslash H.
\end{array}
\end{equation}
\end{Lemma}
{\em{Proof of Lemma~\ref{L_init}.}} For $g\in H$, Eq.~(\ref{InitEval}) follows directly from Eq.~(\ref{Hstab}). For $g\in G\backslash H$, since by assumption $T(H)$ is a maximal Abelian subgroup of $T(G)$, exists an $h \in H$ for which $T(g)$ and $T(h)$ anti-commute. With Eq.~(\ref{Hstab}), $\langle \Phi| T(g) |\Phi \rangle = \langle \Phi| T(h)^\dagger T(g) T(h)|\Phi \rangle = - \langle \Phi| T(g) |\Phi \rangle$, hence $\langle \Phi| T(g) |\Phi \rangle =0$. $\Box$

\subsubsection{Logical Evolution}\label{LP}

To describe logical evolution in the present formalism, we introduce a sequence of `evolved' logical operators $T_0(g)=T(g), T_1(g), T_2(g), .. ,T_n(g)$, as in the Heisenberg picture,
\begin{equation}\label{ObsNL}
T_k(g) := V_{\leq n}^\dagger T(g) V_{\leq k},\;\; \forall g\in G,
\end{equation}
where
\begin{equation}\label{GateSeqs}
V_{\leq k}:= V_kV_{k-1}..V_2V_1,\;\; \forall k=1,..,n,
\end{equation}
and
\begin{equation}\label{Vdef}
V_k(\alpha_k):=\exp\left(-i \frac{\alpha_k}{2} L_k(g_k) \right), \;\; -\pi \leq \alpha_k \leq \pi,\;\; g_k \in {\cal{G}}_k.
\end{equation}
Why this sequence of logical operators represents an evolution is a priori not obvious. It is the content of the evolution equation~(\ref{LinRel}) and of Lemma~\ref{LEMsreMBQC} below. Of particular interest are the observables at the end of the evolution, $T_n(g)$; i.e. $k=n$.\medskip

In preparation for Lemma~\ref{LEMsreMBQC} below, we define the CPTP maps ${\cal{V}}_k$ which, as it turns out, are circuit model operations that can be simulated by MBQC on symmetric states $|\Phi\rangle$,
\begin{equation}\label{CPTP}
{\cal{V}}_k:= \frac{1+\sigma_k(g_k)}{2} [V_{\text{log},k}(\alpha)] + \frac{1-\sigma_k(g_k)}{2} [V_{\text{log},k}(\alpha)^\dagger]. 
\end{equation}
Herein, the brackets $[\cdot]$ denote superoperators, and
$$
V_{\text{log},k}(\alpha):=\exp\left(-i \frac{\alpha_k}{2} T(g_k) \right), \;\; -\pi \leq \alpha_k \leq \pi,\;g_k \in {\cal{G}}_k.
$$
As before in Eq.~(\ref{GateSeqs}), we define the concatenated operations ${\cal{V}}_{\leq k}:={\cal{V}}_k.. {\cal{V}}_2{\cal{V}}_1$ (The rotation angle $\alpha_k$ are suppressed, to simplify the notation). We have the following result.
\begin{Lemma}\label{LEMsreMBQC}
The measurement statistics resulting from action of the sequence of the logical CPTP maps ${\cal{V}}_n {\cal{V}}_{n-1} .. {\cal{V}}_2 {\cal{V}}_1$ on the resource state $|\Phi\rangle $ with expectation values given by Eq.~(\ref{InitEval}), followed by measurement of an Abelian subgroup of observables $T(H')$, $H'\subset G$, can be reproduced by measurement of the observables $T_n(h)$, $h\in H'$, on $|\Phi\rangle$, 
\begin{equation}\label{CPTPevol}
\langle \Phi |  T_n(h) |\Phi\rangle = \text{Tr}\left(T(h) {\cal{V}}_{\leq n}(|\Phi \rangle \langle \Phi| )\right).
\end{equation}
\end{Lemma}
Here, the l.h.s. represents the MBQC, and the r.h.s. represents the corresponding circuit simulation.\medskip

Now there are two important statements to make about the evolution of the above-introduced logical observables. First, the evolution of the expectation values $\{\langle T_t(g)\rangle_\Phi\}$ is closed. That is, the expectation values at any given time $t$ depend only on the same expectation values at time $t-1$, in a linear fashion. This is a significant simplification, since there are only $|G|$ operators $T_g(t)$, i.e., a constant number independent of the chain length. The evolution of these few observables decouples from the exponentially many other observables defined on the Hilbert space ${\cal{H}}$.

To manifest the property of closure, the following linear relations hold (as we prove below),
\begin{equation}\label{LinRel}
\left(\begin{array}{c}  \langle T_t(g_1) \rangle_\Phi \\ \langle T_t(g_2) \rangle_\Phi \\ : \\  \langle T_t(g_{|G|})\rangle_\Phi \end{array} \right)  =
\left[{\cal{V}}_t^\dagger(\alpha) \right] 
\left(\begin{array}{c}  \langle T_{t-1}(g_1) \rangle_\Phi \\ \langle T_{t-1}(g_2) \rangle_\Phi \\ : \\  \langle T_{t-1}(g_{|G|})\rangle_\Phi \end{array} \right),\;\; \forall t=1,..,n.
\end{equation}
Therein, the $|G|\times |G|$ matrix $\left[{\cal{V}}_t^\dagger(\alpha) \right]$ depends on one measurement angle $\alpha$ and a string order parameter as defined in Eq.~(\ref{string}).
Eq.~(\ref{LinRel}) is our fundamental evolution equation. It replaces Eq.~(\ref{Tbeta}) of the MPS-based formalism, answering Question~(ii) of Section~\ref{MPSMBQC}.
An added benefit is that  Eq.~(\ref{CPTP}) describes the exact dependence on the measurement angle, whereas Eq.~(\ref{Tbeta}) holds to linear order only.

Gaining closedness of the evolution, i.e., replacing $V_{\leq n}$ by ${\cal{V}}_{\leq n}$, has a flipside. Namely, we seem to lose unitarity. While the maps $V_t$ are unitary, the CPTP maps ${\cal{V}}_t$ are typically not. A priori, we are not interested in the noisy evolution afforded by CPTP maps; the goal is to implement unitary evolution. We will find, though, that this is not a problem---the unitary limit of interest can be recovered in a computationally efficient way; see Corollary~\ref{unit} in Section~\ref{HTU}.
\smallskip

The second important statement about the logical observables is that, although they are highly non-local objects, they can be measured in a block-local fashion. We have the following result.
\begin{Lemma}\label{Inf}
The measurement outcomes of the observables $T_n(h)$, $h\in H'$, with $H' \subset G$ and $v_{L,n+1}(H')$ an Abelian group, can be jointly inferred by block-local physical measurement and classical side processing. 
\end{Lemma}
Lemmas~\ref{LEMsreMBQC} and \ref{Inf} are proved in Section~\ref{GTproof}.

\subsection{MBQC in the presence of symmetry}\label{Sec:sreMBQC}

\subsubsection{Statement of the result}\label{StatRes}

With the above notions introduced, we can now state the main result. 
\begin{Theorem}\label{GT}
Consider a short-range entangled state $|\Phi\rangle$ of a spin-1/2 chain, symmetric under a group $G=(\mathbb{Z}_2)^m$; and representations $v_{L,i}$, $v_{R,i}$ (projective) and $u_i$ (linear) of $G$ satisfying Eqs.~(\ref{H_spec}) -- (\ref{Udef}). Then, MBQC using block-local measurements on $|\Phi\rangle$ can simulate quantum circuits consisting of (i) preparation of an initial state fully specified by the expectation values Eq.~(\ref{InitEval}); (ii) the sequence of CPTP maps ${\cal{V}}_{\leq n}={\cal{V}}_n..{\cal{V}}_1$ given by Eq.~(\ref{CPTP}), and (iii) final measurement of logical observables from an Abelian subgroup of $T(G)$. 
\end{Theorem}
This is the basic statement of computational capability of MBQC on symmetric states. We spell it out more in Corollary~\ref{unit} below.

\subsubsection{How to use Theorem~\ref{GT}}\label{HTU}

We now describe how to identify the computational power bestowed by the MBQC schemes described in Section~\ref{MP} on a given resource state $|\Phi\rangle$.

\paragraph{Computing the dependent constituents.} Proceed as described in Section~\ref{MP}. This yields in particular the definition of the string order parameters. Check which of the relevant string order parameters $\sigma_k(g_k)$, $g_k \in {\cal{G}}_k$, are non-zero.

\paragraph{Extracting the computational primitives.} For admissible measurement patterns, the computational primitives provided are identified through  the following procedure.
\begin{enumerate}
\item{The logical operators $T(g)$, $g \in G$, are computed through Eq.~(\ref{T_enc}).}
\item{{\em{Preparation:}} With Lemma~\ref{L_init}, the initial state $|\Phi\rangle$ has expectation values of the logical observables as specified by Eq.~(\ref{InitEval}).}
\item{{\em{Logical measurement:}} The final logical measurement is of the commuting observables $T(h)$, $h\in H'$. See Lemma~\ref{LEMsreMBQC}.}
\item{{\em{CPTP maps:}} The CPTP maps ${\cal{V}}_k$ appearing in Theorem~\ref{GT} are obtained as follows.
\begin{enumerate}
\item{Using Eq.~(\ref{KappaR}) and (\ref{vLRcomm}), the groups ${\cal{G}}_i$, for $i=1,.., n$ are computed.}
\item{There is one logical CPTP map for every element of any ${\cal{G}}_k$. Namely, if $g_k \in {\cal{G}}_k$ then the logical map
$$
{\cal{V}}_k(\alpha_k)= \frac{1 +\sigma_k(g)}{2} \left[e^{-i\frac{\alpha_k}{2}T(g_k)}\right] + \frac{1 -\sigma_k(g)}{2} \left[e^{i\frac{\alpha_k}{2}T(g_k)}\right] 
$$
can be realized by measuring the observables Eq.~(\ref{ObsBlk}) on block $k$, for all  $\alpha_k \in [0,\pi)$}.
\end{enumerate}
}
\end{enumerate}

\paragraph{Computational power.} In the theorem statement it is not made explicit how much computational power is provided with the CPTP maps ${\cal{V}}_k$, and under which circumstances. Specifically, one might want to know which unitaries are reachable as limits of the CPTP maps. This is clarified by the following corollary.

\begin{Cor}\label{unit}
Consider a short-range entangled state $|\Phi\rangle$ of a spin-1/2 chain, symmetric under a group $G=(\mathbb{Z}_2)^m$; and (projective) representations $v_{L,i}$, $v_{R,i}$ and $u_i$ of $G$ satisfying Eqs.~(\ref{H_spec}) -- (\ref{Udef}). Furthermore, assume that the string order parameters $\sigma_k(g)$ are bounded away from zero, $|\sigma_k(g)| \geq \sigma >0$ for all $k=1,..,n$. Then, 
\begin{itemize}
\item[(i)]{The group of realizable gates is $\mathcal{L}:=\exp (-i \mathcal{A})$, where $\mathcal{A}$ is the Lie algebra generated by $T(\mathcal{G})$, with ${\cal{G}}:=\bigcup_{k=1}^n {\cal{G}}_k$, under $[\;.,.]/i$ and linear combination.}
\item[(ii)]{The unitary gates 
\begin{equation}\label{Prim_Gates}
 U_g(\alpha) = \exp\left(-i\frac{\alpha}{2} T(g)\right), \;\;g \in \mathcal{G},
\end{equation}
when subdivided into $N$ parts each requiring one non-trivial measurement, can be implemented with error 
\begin{equation}\label{ApprErr}
\epsilon \leq \frac{\alpha^2}{N}\frac{1-\sigma^2}{\sigma^2}.
\end{equation}}
\end{itemize}
\end{Cor}
Corollary~\ref{unit} tells us that the fundamental criterion for MBQC power on short-range entangled and symmetric states is whether the string order parameters are zero or non-zero. Vanishing string order parameters lead to no computational power, and non-vanishing order parameter imply non-trivial computational power. Whatever the value of $\sigma$, as long as it is non-zero, the computational power is the same. Yet, the value of $\sigma$ determines how efficiently approximation errors can be suppressed. Any targeted approximation error $\epsilon$ can be reached by a sufficiently large number of steps $N$ into which the realization of $U_g(\alpha)$ is subdivided. The smaller $\sigma$, the larger $N$ needs to be. 

To summarize, item (i) of Corollary~\ref{unit} is the statement about computational power, and item (ii) the statement about computational efficiency. The former answers Question~(iii) of Section~\ref{MPSMBQC}, and the latter answers Question~(iv). \smallskip

\begin{figure}
\begin{center}
\includegraphics[width=9cm]{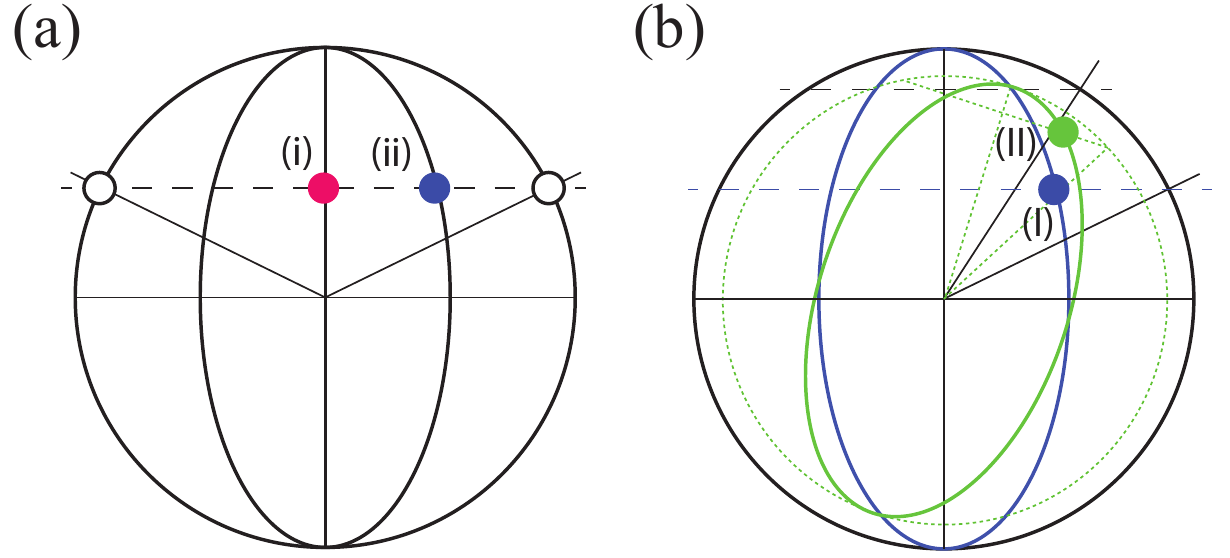}
\caption{\label{DivC}Splitting technique for MBQC-simulating unitary operations with high fidelity. The $X/Y$-equator of a Bloch sphere is shown. The implemented logical operation is a noisy rotation ${\cal{V}}(\alpha)$ about the $z$-axis. (a) With Eq.~(\ref{CPTP}), the nosy rotation ${\cal{V}}(\alpha)$ is a probabilistic mixture of a unitary rotation $U(\alpha)$ and $U(-\alpha)$. (i) When $\sigma=0$, the mixture is 50/50, and no rotation occurs at all. (ii) $\sigma=1/2$. The resulting operation amounts to rotation plus dephasing. (b) Splitting of one rotation into two of half the rotation angle. In the limit of small rotation angles, the total rotation angle is preserved, and the loss of purity is cut in half.}
\end{center}
\end{figure}

The proof of Corollary~\ref{unit} is given in Appendix~\ref{UnitProof}. The idea for part (ii) is to approximate the unitary $U(\alpha)$ by the $N$ fold iteration of the CPTP map  ${\cal{V}}(\alpha/\sigma N)$. Each such step incurs an error proportional to $\alpha^2/N^2$; hence the total error made in all $N$ steps combined is proportional to $1/N$. It can be made arbitrarily small by increasing $N$. The technique of splitting one large rotation into a number of small rotations is illustrated in Fig.~\ref{DivC}.

\subsection{Proof of Theorem~\ref{GT}}\label{GTproof}

The proof of Theorem~\ref{GT} is based on Lemmas~\ref{LEMsreMBQC} and \ref{Inf} which were stated in Section~\ref{LP}. We prove them first. Recall that Lemma~\ref{LEMsreMBQC} establishes the closedness of the evolution of the logical observables $T_t(g)$, $g\in G$.

{\em{Proof of Lemma~\ref{LEMsreMBQC}.}} There are two items to prove, namely (i) Eq.~(\ref{CPTPevol}), and (ii) that the initial expectation values are given by Eq.~(\ref{InitEval}).

{\em{(i).}} We prove Eq.~(\ref{CPTPevol}) by way of two intermediate steps. As per the assumptions of the theorem, we consider a $G$-symmetric short-range entangled state $|\Phi\rangle$ with entanglement range $\Delta$, and denote by $k$ and $l$ two blocks in the chain, such that $l-k > 2\Delta$. Further choose $\alpha_{k+1},..,\alpha_{l-1}=0$, i.e., $V_{k+1},..,V_{l-1}=I$. 
For easier book-keeping, we define sequences of states 
$$
|\Phi(k)\rangle:=V_{\leq k}|\Phi\rangle.
$$
Then it holds that 
\begin{equation}\label{TRfact}
\langle \Phi(k) |T(g) R_l(g')|\Phi(k)\rangle =  \sigma_l(g')  \langle \Phi(k) |  T(g)  |\Phi(k) \rangle.
\end{equation}
We use this to show that furthermore,
\begin{equation}\label{Step2}
\langle \Phi(l)|T(g)|\Phi(l)\rangle = \langle \Phi(k)|{\cal{V}}_l^\dagger(T(g))|\Phi(k)\rangle. 
\end{equation}
{\em{Proof of Eq.~(\ref{TRfact}).}} For all $g\in G$, it holds that $T(g) = \left. T(g)\right|_{\{\leq k\}} \otimes \left. T(g)\right|_{\{> k\}}$, $U(g) = \left.U(g)\right|_{\{\leq k\}} \otimes  \left.U(g)\right|_{\{> k\}}$, $\text{supp}(V_{\leq k}) \subseteq  \{\leq k\}$. Further, with Eq.~(\ref{TUrel}), $\left.U(g)\right|_{\{>k\}}=\left. T(g)\right|_{\{>k\}}$. Therefore,
\begin{equation}\label{LeftSupp}
\text{supp}\left( U(g)^\dagger V^\dagger_{\leq k} T(g) V_{\leq k} \right) \subseteq \{\leq k\}.
\end{equation}
Also,  $\text{supp}(R_l(g)) \subset  \{> k\}$, for all $l>k$, and therefore $[V_{\leq k}, R_l(g')]=0$, $\forall g' \in G$, whenever $l>k$. Thus we obtain
$$
\begin{array}{rcll}
\langle \Phi(k) |T(g) R_l(g')|\Phi(k)\rangle &=& \langle \Phi | V^\dagger_{\leq k}  T(g) R_l(g') V_{ \leq k} |\Phi \rangle\\
&=& \langle \Phi | V^\dagger_{\leq k}  T(g) V_{\leq k} R_l(g')  |\Phi \rangle\\
&=& (-1)^{\chi(g)} \langle \Phi | \left( U(g)^\dagger V^\dagger_{\leq k}  T(g) V_{\leq k}\right) R_l(g')  |\Phi \rangle\\
&=& (-1)^{\chi(g)} \langle \Phi | U(g)^\dagger V^\dagger_{\leq k}  T(g) V_{\leq k} |\Phi\rangle \langle \Phi| R_l(g')  |\Phi \rangle\\
&=& \sigma_l(g') \langle \Phi |  V^\dagger_{\leq k}  T(g) V_{\leq k} |\Phi\rangle\\ 
&=&  \sigma_l(g')  \langle \Phi(k) |  T(g)  |\Phi(k) \rangle.
\end{array}
$$ 
Therein, in the third equality we have used the symmetry property Eq.~(\ref{PhiSymm}) of $|\Phi\rangle$. In the fourth equality we have used Eq.~(\ref{LeftSupp}) and Lemma~\ref{Prod}, and in the fifth equality Eq.~(\ref{PhiSymm}) again. This proves Eq.~(\ref{TRfact}), completing the first step.\smallskip

{\em{Proof of Eq.~(\ref{Step2}).}} {\em{Case I:}} $[T(g_l),T(g)]=0$. With Lemma~\ref{cd} it follows that $[L(g_l),T(g)]=0$. The expectation value satisfies
$$
\langle \Phi(l)|T(g)|\Phi(l)\rangle = \langle \Phi(k) | V_l^\dagger T(g)V_l |\Phi(k)\rangle = \langle \Phi(k)|T(g)|\Phi(k)\rangle,
$$
and so the observable $T(g)$ is not evolving. By the case assumption, this is matched by the operation ${\cal{V}}^\dagger_l$ at the logical level, 
\begin{equation}
\label{comm}
 \langle \Phi(k)| {\cal{V}}_l^\dagger (T(g)) |\Phi(k)\rangle  =  \langle \Phi(k)| T(g)  |\Phi(k)\rangle.
\end{equation}
Eq.~(\ref{comm}) provides the matrix elements of $\left[{\cal{V}}^\dagger_l \right]$ in Eq.~(\ref{LinRel}), for the case where $T(g)$ and $T(g_l)$ commute. This concludes Case I.\smallskip

{\em{Case II:}} $[T(g_l),T(g)]\neq0$. With Lemma~\ref{cd} it follows that $L(g_l)$ and $T(g)$ anti-commute. In this case, the expectation value of interest is
$$
\begin{array}{rcl}
\langle \Phi(l)|T(g)|\Phi(l)\rangle &=& \langle \Phi(k)| V_l^\dagger T(g)V_l |\Phi(k)\rangle \\
&=& \cos(\alpha_l) \langle \Phi(k)| T(g) |\Phi(k)\rangle - i\sin(\alpha_l) \langle \Phi(k)|T(g) L_l(g_l)|\Phi(k)\rangle.
\end{array}
$$
We now focus on the expectation value in the term $\sim \sin(\alpha_l)$,
$$
\begin{array}{rcl}
\langle \Phi(k)|T(g) L_l(g_l)|\Phi(k)\rangle &=& \langle \Phi(k)|T(g) T(g_l) R_l(g_l)|\Phi(k)\rangle \\
&=&  \langle \Phi | V_{\leq k}^\dagger T(g) T(g_l) V_{\leq k} R_l(g_l)|\Phi \rangle \\
&=&  (-1)^{\chi(gg_l)}\langle \Phi |  \left( U(gg_l)^\dagger V_{\leq k}^\dagger T(g) T(g_l) V_{\leq k}\right) R_l(g_l)|\Phi \rangle \\
&=&  (-1)^{\chi(gg_l)}\langle \Phi |  U(gg_l)^\dagger V_{\leq k}^\dagger T(g) T(g_l) V_{\leq k} |\Phi\rangle \langle \Phi| R_l(g_l)|\Phi \rangle \\
&=& \sigma_l(g_l) \langle \Phi(k)| T(g)T(g_l)|\Phi(k)\rangle.
\end{array}
$$
Herein, in the fourth line we have used that $T(g)T(g_l) \propto T(gg_l)$, then Eq.~(\ref{LeftSupp}) for $gg_l\in G$, and finally Lemma~\ref{Prod}.
We thus arrive at
\begin{equation}\label{acomm}
\langle \Phi(l)|T(g)|\Phi(l)\rangle =  \cos(\alpha_l) \langle \Phi(k)| T(g) |\Phi(k)\rangle - i\sin(\alpha_l)  \sigma_l(g_l) \langle \Phi(k)| T(g)T(g_l)|\Phi(k)\rangle.
\end{equation}
Eq.~(\ref{acomm}) provides the matrix elements of $\left[{\cal{V}}^\dagger_l \right]$ in Eq.~(\ref{LinRel}), for the case where $T(g)$ and $T(g_l)$ anti-commute, and Eq.~(\ref{LinRel}) is thereby established. This concludes Case II.\smallskip

Eq.~(\ref{comm}) of the commuting and Eq.~(\ref{acomm}) of the anti-commuting case may jointly be written as 
\begin{equation}\label{cac}
\langle \Phi(l)|T(g)|\Phi(l)\rangle = \frac{1+\sigma_l(g_l)}{2}  \langle \Phi(k)| V_{\text{log},l}^\dagger  T(g) V_{\text{log},l} |\Phi(k)\rangle +  \frac{1-\sigma_l(g_l)}{2}  \langle \Phi(k)| V_{\text{log},l}  T(g) V_{\text{log},l}^\dagger |\Phi(k)\rangle.
\end{equation}
Recalling the definition of ${\cal{V}}_k$ from Eq.~(\ref{CPTP}), we thus find
$$
\langle \Phi(l)|T(g)|\Phi(l)\rangle = \langle \Phi(k)|{\cal{V}}_l^\dagger(T(g))|\Phi(k)\rangle,
$$
establishing Eq.~(\ref{Step2}). This completes the second step.\smallskip

We now apply Eq.~(\ref{Step2}) recursively. In accordance with the assumptions of the theorem, consider a sequence of unitaries $V_k$ where non-zero rotation angles $\alpha_k$ are sparse. Namely, they only occur at locations $1=k_1, k_2,..., k_\text{max}=n$, with the spatial separations $k_{i+1}-k_i>2\Delta$, $\forall i$. Under these conditions we can apply Eq.~(\ref{Step2}), and obtain
$$
\langle \Phi | V^\dagger_{\leq n} T(g) V_{\leq n} |\Phi\rangle = \langle \Phi| {\cal{V}}^\dagger_{\leq n} \left(T(g)\right)|\Phi\rangle,\;\;\forall g\in G.
$$
Using the cyclicity of trace on the r.h.s., we transform the above into
$$
\langle \Phi | V^\dagger_{\leq n} T(g) V_{\leq n} |\Phi\rangle = \text{Tr} \left( T(g) {\cal{V}}_{\leq n}(|\Phi\rangle \langle \Phi|) \right), \;\;\forall g\in G.
$$
This establishes Eq.~(\ref{CPTPevol}).\smallskip

{\em{(ii).}} Eq.~(\ref{CPTPevol}), which we have proved above, on the r.h.s. has the state $|\Phi\rangle \langle \Phi|$ as the initial state of the evolution. Its relevant expectation values have been provided by Eq.~(\ref{InitEval}) in Lemma~\ref{L_init}.  $\Box$\medskip

We recall that Lemma~\ref{Inf} states that the observables $T_{n}(g)$ can be measured in a local fashion. In preparation for the proof, we define the additional observables
\begin{equation}
T^{(\leq k)}(g):= V_{\leq k}^\dagger \left( \bigotimes_{j=0}^k u_j(g)\right)  V_{\leq k}, \;\; \forall k=0,..,n,\; g\in G.
\end{equation}
The observables $T^{(\leq k)}(g)$ have an intuitive interpretation, namely they represent the computational output aggregated up to block $k$. While those observables have random values, these values need to be known for properly adjusting measurement angles for the local observables $O_k(g)$ that drive the computation. That is, the observables $T^{(\leq k)}(g)$ are the quantum mechanical realization of the various time-instantiations of the information flow vector in MBQC \cite{CompMod}.

We observe that $T^{(\leq 0)}(g)=u_0(g)$, and 
\begin{equation}\label{Tng}
T_n(g) = T^{(\leq n)}(g) \otimes v_{L,n+1}(g), \;\; \forall g\in G.
\end{equation}
Key is the recursion relation 
\begin{equation}\label{RecRel}
T^{(\leq k+1)}(g) = T^{(\leq k)}(g) \left(e^{i\frac{\alpha_{k+1}}{2}   T^{(\leq k)}(g') \otimes v_{L,k+1}(g')} u_{k+1}(g) e^{-i\frac{\alpha_{k+1}}{2}   T^{(\leq k)}(g') \otimes v_{L,k+1}(g')} \right),\;\; \forall g\in G.
\end{equation}
Therein, $g':=g_{k+1} \in {\cal{G}}_{k+1}$ specifies the rotation axis for the logical operation associated with block $k+1$. Eq.~(\ref{RecRel}) follows from the commutation relations
\begin{equation}\label{LuCR}
[L_k(g_k),u_l(g)]=0,\;\; \forall g_k \in {\cal{G}}_k, g\in G,\;\text{whenever}\; k\neq l\leq n.
\end{equation}
Specifically, by direct calculation,
$$
\begin{array}{rcl}
T^{(k+1)}(g) &=& V_{\leq k+1}^\dagger \left( \bigotimes_{j=0}^k u_j(g) \right) \otimes u_{k+1}(g) V_{\leq k+1} \\
&=& \left(V_{\leq k}^\dagger V_{k+1}^\dagger \left( \bigotimes_{j=0}^k u_j(g) \right) V_{k+1} V_{\leq k} \right) \left( V_{\leq k}^\dagger V_{k+1}^\dagger  u_{k+1}(g) V_{k+1} V_{\leq k}\right)\\
&=& \left(V_{\leq k}^\dagger \left( \bigotimes_{j=0}^k u_j(g) \right) V_{\leq k} \right) \left( V_{\leq k}^\dagger V_{k+1}^\dagger  u_{k+1}(g) V_{k+1} V_{\leq k}\right)\\
&=& T^{(k)}(g)   \left( V_{\leq k}^\dagger V_{k+1}^\dagger  u_{k+1}(g) V_{k+1} V_{\leq k}\right)\\
&=& T^{(k)}(g)   \left(\left(V_{\leq k}^\dagger V_{k+1}^\dagger  V_{\leq k}\right)  u_{k+1}(g) \left(V_{\leq k}^\dagger V_{k+1} V_{\leq k}\right) \right).
\end{array}
$$
Therein, the second line is ordering of factors and insertions of identity, the third line follows with the commutation condition Eq.~(\ref{LuCR}). The fourth line is just the definition of $T^{(k)}(g)$, and the fifth line follows with Eq.~(\ref{LuCR}) again.

We now examine the unitaries $\left(V_{\leq k}^\dagger V_{k+1}^\dagger  V_{\leq k}\right)$, finding
$$
\begin{array}{rcl}
\left(V_{\leq k}^\dagger V_{k+1}^\dagger  V_{\leq k}\right) &=& V_{\leq k}^\dagger \exp\left(-i \frac{\alpha_{k+1}}{2} L_{k+1}(g') \right) V_{\leq k}\\
&=& \exp\left(i \frac{\alpha_{k+1}}{2} V_{\leq k}^\dagger  L_{k+1}(g') V_{\leq k}\right) \\
&=& \exp\left(i \frac{\alpha_{k+1}}{2}  T^{(\leq k)}(g') \otimes v_{L,k+1}(g') \right).
\end{array}
$$
Therein, we have used the short-hand $g':=g_{k+1} \in {\cal{G}}_{k+1}$ as before. Combining the respective last lines of the above two blocks of equations, we obtain Eq.~(\ref{RecRel}).\smallskip

{\em{Proof of Lemma~\ref{Inf}.}} 
We will show by induction that the outcomes $\lambda^{(\leq k)}(g)$ for all observables $T^{(\leq k)}(g)$ can be obtained by physical measurement of the local bulk and boundary observables Eqs.~(\ref{ObsBdy}), (\ref{ObsBlk}), and classical post-processing. Specifically, 
\begin{equation}\label{ValuRel}
\lambda^{(\leq k)}(g) = (-1)^{\sum_{j=0}^ks_k(g)},\;\; \forall k=0,..,n,\; g\in G,
\end{equation}
with $s_k(g)\in \mathbb{Z}_2$ the measurement outcome for the local observables Eq.~(\ref{ObsBdy}), (\ref{ObsBlk}).

{\em{Induction start.}} The induction begins with $k=0$. The observables $T^{(\leq 0)}(g)=u_0(g)$, $g\in G$, can indeed all be simultaneously measured. Also, Eq.~(\ref{ValuRel}) is valid for $k=0$.

{\em{Induction step.}} Now assume that the measurement outcomes for all observables $T^{(\leq k)}(g)$, $\forall g\in G$, are known, and that Eq.~(\ref{ValuRel}) holds at level $k$. Then, $\forall g\in G$, Eq.~(\ref{RecRel}) simplifies to
$$
\begin{array}{rcl}
T^{(\leq k+1)}(g) &\mapsto& T^{(\leq k)}(g) \otimes \left(e^{i\frac{\alpha_{k+1}}{2}   \lambda^{(\leq k)}(g') v_{L,k+1}(g')} u_{k+1}(g) e^{-i\frac{\alpha_{k+1}}{2}   \lambda^{(\leq k)}(g') \otimes v_{L,k+1}(g')} \right),\\
&=& T^{(\leq k)}(g) \otimes O_{k+1}(g),
\end{array}
$$
where $\lambda^{(\leq k)}(g')=\pm 1$ is the eigenvalue inferred for $T^{(\leq k)}(g')$, and $O_{k+1}(g)$ is a measured bulk observable, cf. Eq.~(\ref{ObsBlk}). Therein, the adaptation of the measurement angle agrees with Eq.~(\ref{ObsBlk}), because Eqs.~(\ref{CPRgen}) and (\ref{ValuRel}) show that $\lambda^{(\leq k)}(g') =(-1)^{q_{k+1}(g')}$.

Hence, for all $g\in G$, the eigenvalue  $\lambda^{(\leq k+1)}(g)=\pm 1$  for $T^{(\leq k+1)}(g)$ can be inferred from $\lambda^{(\leq k)}(g)$ and the value $(-1)^{s_{k+1}(g)}$  measured for $O_{k+1}(g)$, namely $\lambda^{(\leq k+1)}(g) = (-1)^{s_{k+1}(g)} \lambda^{(\leq k)}(g)$. Therefore, Eq.~(\ref{ValuRel}) also holds at level $k+1$. This completes the induction step. 

By induction, we can simultaneously infer the values $\lambda^{(\leq n)}(g)$ of $T^{(\leq n)}(g)$, $\forall g\in G$.

Finally, we need to measure the system at the right boundary. Since the representation $v_{L,n+1}$ is projective, we can in general only simultaneously measure a subgroup $v_{L,n+1}(H')$ of observables. With Eq.~(\ref{Tng}), the value of the observables of interest, $T_n(h)$, $h\in H'$ is then inferred from the values $\lambda^{(\leq n)}(h)$ and the values measured for $v_{L,n+1}(h)$, $\forall h\in H'$. $\Box$
\medskip

We are now ready to prove the main theorem.\smallskip

{\em{Proof of Theorem~\ref{GT}.}} Lemma~\ref{LEMsreMBQC} relates the evolution described in the theorem to the expectation values $\langle \Phi | T_n(h)|\Phi\rangle$, for all $h\in H'$, cf. Eq.~(\ref{CPTPevol}).  Lemma~\ref{Inf} shows how to measure the observables $T_n(h)$, $h\in H$, in a block-local fashion. $\Box$

\subsection{Examples}\label{Exa2}

In this first round of applying Theorem~\ref{GT} to our examples, we make the simplest choice for the representations, which will lead to ${\cal{G}}_i = G$, $\forall i=1,..,n$. For the cluster case, where previous results \cite{Bartl},\cite{SPTO1,SPTO2} exist, this will produce a block structure (blocks of size 2) compatible with those earlier results. We will return to the question of obtaining smaller blocks, specifically blocks of size one, in Section~\ref{BLtoSL}. 

\subsubsection{The cluster chain}\label{sec:ClusterB2}

We choose blocks $i=1,..,n$ of size 2, on the left boundary, for $i=0$ a block of size 2, and on the right boundary, for $i=n+1$, a block of size 1. In this way, we obtain a chain of odd length which is the default for the cluster chain.

As discussed in Section~\ref{HTU}, the measurement pattern is specified by the linear representations $u_i$ and the projective representations $v_{L,i}$. In the bulk, they are
\begin{equation}\label{Rep:bulk}
\begin{array}{rclrcl}
u_i(g_{01}) &=& IX,& u_i(g_{10}) &=& XI,\\
v_{L,i}(g_{01}) &=& ZI,&  v_{L,i}(g_{10}) &=& XZ, 
\end{array}
\end{equation}
The representations $u_0$, $v_{R,0}$ on the left boundary are

\begin{equation}\label{Rep:left}
\begin{array}{rclrcl}
u_0(g_{01}) &=& IX,& u_0(g_{10}) &=& XI,\\
v_{R,0}(g_{01}) &=& ZX,&  v_{R,0}(g_{10}) &=& XI, \\
\end{array}
\end{equation}
The representation $v_{L,n+1}$ on the right boundary is
\begin{equation}\label{Rep:right}
v_{L,n+1}(g_{01}) = Z,\; v_{L,n+1}(g_{10}) = X.
\end{equation}
This concludes the specification of the measurement pattern, and we now unpack it. \medskip

With Eq.~(\ref{url}) we find that in the bulk
\begin{equation}\label{RepR:Bulk}
v_{R,i}(g_{01}) = ZX,\;\; v_{R,i}(g_{10}) = IZ.
\end{equation}
With Eq.~(\ref{H_spec}) and Eq.~(\ref{Rep:left}), we find that $H=\langle g_{10}\rangle$. The representations are independent of the block label in the bulk, and therefore Eq.~(\ref{KappaR}) is satisfied for $i=1,..,n$. Comparing Eqs.~(\ref{Rep:bulk}) and (\ref{Rep:left}), we find that the commutation relations for $v_R$ are the same on block 0. 

From Eqs.~(\ref{KappaR}) and (\ref{vLRcomm}), we compute the sets ${\cal{G}}_i$, finding
\begin{equation}\label{ClusterSets}
{\cal{G}}_i = G,\;\;\forall i=1,..,n.
\end{equation}
Inserting Eqs.~(\ref{Rep:bulk}), (\ref{Rep:left}), (\ref{Rep:right}) into Eq.~(\ref{Udef}) produces 
\begin{equation}\label{ClusterGlobalSymm}
 g_{01} \cong ZXIXIXIXI...IXZ,\;\; g_{10} \cong XIXIXIX...IZ.
\end{equation}
All constraints are verified, and Theorem~\ref{GT} can be applied.\medskip

With Eq.~(\ref{T_enc}), (\ref{ClusterSets}) and Corollary~\ref{unit}, we find that we can implement rotations of form 
\begin{equation}\label{ZXrot}
e^{i\alpha X}, e^{i\beta Z}, e^{i\gamma Y},
\end{equation}
and hence all unitaries in $SU(2)$, by block-local measurements of block-size 2.

\subsubsection{The Kitaev-Gamma chain}
\label{subsec:MBQC_KG}

For the spin-1/2 bond-alternating Kitaev-Gamma chain, we choose blocks $i=1,2,...n$ of size $2$;
on the left boundary $i=0$ of block size $1$;
and on the right boundary $i=n+1$ of block size $1$.
Hence the chain length is chosen as $2n+2\in\mathbb{Z}$ in this case. 
We will work in the six-sublattice rotated frame  in this subsection. 

Next we specify the linear representations $u_i$ and the projective representations $v_{L,i},v_{R,i}$ of the $\mathbb{Z}_2\times \mathbb{Z}_2=\{1,R(\hat{x},\pi),R(\hat{y},\pi),R(\hat{z},\pi)\}$ group.
Since the group can be decomposed as $\langle R(\hat{x},\pi)\rangle\times \langle R(\hat{z},\pi)\rangle$,
it is enough to specify the representations of the two generators. 
In the bulk, $u_i$ and $v_{L,i}$ are ($i=1,...,n$)
\begin{equation}\label{eq:bulk_KG_rep}
\begin{array}{rclrcl}
u_i(R(\hat{x},\pi))&=&XX,&u_i(R(\hat{z},\pi))&=&ZZ,\\
v_{L,i}(R(\hat{x},\pi))&=&XI,&v_{L,i}(R(\hat{z},\pi))&=&ZI. 
\end{array}
\end{equation}
Using Eq. (\ref{url}), we find that in the bulk,
\begin{eqnarray}
v_{R,i}(R(\hat{x},\pi))=IX,&v_{R,i}(R(\hat{z},\pi))=IZ.
\label{eq:right_KG}
\end{eqnarray} 

On the left boundary, the projective representation $v_{R,0}$ is 
\begin{eqnarray}
v_{R,0}(R(\hat{x},\pi))=X,&v_{R,0}(R(\hat{z},\pi))=Z.
\end{eqnarray}
Since except the identity element, the other three elements in the $\mathbb{Z}_2\times\mathbb{Z}_2$ group anti-commute in $v_{R,0}$, 
the maximal abelian subgroup in $v_{R,0}$ can at most be $\mathbb{Z}_2$. 
As a result, $H$ in Eq. (\ref{u0}) can be chosen as $\langle R(\hat{x},\pi) \rangle$.
Defining the linear representation $u_0$ to be
\begin{equation}
\begin{array}{rclrcl}
u_0(R(\hat{x},\pi))&=&X,&u_0(R(\hat{z},\pi))&=&I,
\end{array}
\label{eq:left_KG}
\end{equation}
it can be verified that Eq. (\ref{H_spec}) is satisfied. 

Finally, on the right boundary, $v_{L,n+1}$ is
\begin{eqnarray}
v_{L,n+1}(R(\hat{x},\pi)) = X,&v_{L,n+1}(R(\hat{z},\pi)) = Z.
\end{eqnarray}

The linear representation $U$ of the $\mathbb{Z}_2\times \mathbb{Z}_2$ group on the whole chain can be obtained from Eq. (\ref{Udef}) as
\begin{eqnarray}
U(R(\hat{x},\pi))=\Pi_{i=0}^{n+1}X_i,&U(R(\hat{z},\pi))=\Pi_{i=0}^{n+1}Z_i,
\end{eqnarray}
which is the same as Eq. (\ref{eq:sym_Z2_Z2_KG}).
Furthermore, the group $\mathcal{G}_i$ ($i=1,2,...,n$) is just the full $\mathbb{Z}_2\times \mathbb{Z}_2$ group, 
since Eqs. (\ref{KappaR},\ref{vLRcomm}) can be verified using Eqs. (\ref{eq:bulk_KG_rep},\ref{eq:right_KG}). 
Then it can be checked that the assumptions in 
Sec. \ref{subsubsec:assumptions} are all satisfied. 
Therefore, from Corollary~\ref{unit}, we find that we can implement rotations of form 
$e^{i\alpha X}$, $e^{i\beta Z}$, $e^{i\gamma Y}$,  
and hence all unitary operations in SU(2), by block-local measurements of block-size 2.

Two comments are in order.
First, in addition to the Kitaev-Gamma model,
the MBQC procedure is applicable to 1D bond-alternating spin-1/2 XXZ and XYZ models as well, 
since these models are invariant under the $\{1,R(\hat{x},\pi),R(\hat{y},\pi),R(\hat{z},\pi)\}\cong\mathbb{Z}_2\times \mathbb{}{Z}_2$ symmetry group
and the constructions of the representations $u_i$, $v_{L,i}$ and $v_{R,i}$ are the same as the Kitaev-Gamma model. 
Second, the Hamiltonian of the bond-alternating Kitaev-Gamma model in the $U_6$ frame does not have a two-site translation invariance,
and instead, the periodicity is six.
However, this does not stop us from performing block-size-two measurements as translation invariance is not required in the present MBQC formalism. 

\subsubsection{Cellular automaton states}\label{QCA_block}

We choose blocks $i=1,\ldots,n$ of size $6$, for $i=0$ a block of size $2$, and on the right boundary, for $i=n+1$, a block of size $2$. Thus the natural chain length of choice for the $\tau=2$ automaton phase is $6n+4$.

As discussed in Section~\ref{HTU}, the measurement pattern is specified by the linear representations $u_i$ and the projective representations $v_{L,i}$. In the bulk, $u_i$ is generated by
\begin{equation}\label{Rep:QCAbulk_u}
\begin{array}{rclrcl}
u_i(g_{1}) &=& XIIIXI,&  u_i(g_{2})&=& IIIXIX,\\
u_i(g_{3}) &=& IIXIXI,& u_i(g_{4}) &=& IXIXII,
\end{array}
\end{equation}
and $v_i$ is generated by
\begin{equation}\label{Rep:QCAbulk_v}
\begin{array}{rclrcl}
v_{L,i}(g_{1}) &=& XIIIXZ, &  v_{L,i}(g_{2}) &=& IIIXZI,\\
v_{L,i}(g_{3}) &=& IIXZII, &  v_{L,i}(g_{4}) &=& IXZIII.
\end{array}
\end{equation}

The representation $u_0$ on the left boundary is given by
\begin{equation}\label{Rep:QCAleft}
 u_0(g_{1}) = IZ, \;\; u_0(g_{2}) =  II, \;\;
 u_0(g_{3}) = XI, \;\;  u_0(g_{4}) =  II .
\end{equation}

The representation $v_{L,n+1}$ on the right boundary is
\begin{equation}\label{Rep:QCAright}
v_{L,n+1}(g_{1}) = XI,\; v_{L,n+1}(g_{2}) = ZX,\;v_{L,n+1}(g_{3}) = IZ,v_{L,n+1}(g_{4}) = IX,
\end{equation}

and $v_{R,0}$ on the left boundary is 
\begin{equation}\label{RepR:QCAleft}
v_{R,0}(g_{1}) = IZ, \;\; v_{R,0}(g_{2}) = ZX, \;\;
 v_{R,0}(g_{3}) = XI, \;\;  v_{R,0}(g_{4}) = ZI.
\end{equation}
This concludes the specification of the measurement pattern, and we now unpack it. \medskip

With Eq.~(\ref{url}) we find that in the bulk
\begin{equation}\label{RepR:QCABulk}
v_{R,i}(g_{1}) = IIIIIZ,\;\; v_{R,i}(g_{2}) = IIIIZX,\;\; v_{R,i}(g_{3}) = IIIZXI,\;\; v_{R,i}(g_{4}) = IIZXII.
\end{equation}

With Eq.~(\ref{H_spec}) and Eq.~(\ref{Rep:QCAleft}), we find that $H=\langle g_1,g_3\rangle$. The representations are faithful and independent of the block label in the bulk.

From Eqs.~(\ref{KappaR}) and (\ref{vLRcomm}), we compute the sets ${\cal{G}}_i$, finding
\begin{equation}\label{QCASets}
{\cal{G}}_i = G,\;\;\forall i=1,..,n.
\end{equation}
Inserting Eqs.~(\ref{RepR:QCAleft}), (\ref{Rep:QCAbulk_u}), (\ref{Rep:QCAright}) into Eq.~(\ref{Udef}) produces 
\begin{align} \label{QCAGlobalSymm}
\begin{split}
 g_{1}  &\cong IZ (XIIIXI)\ldots XI ,\;\; g_{2} \cong ZX (IIIXIX)\ldots ZX, \\ 
 g_{3} &\cong XI(IIXIXI) \ldots  IZ,\;\; g_{4} \cong ZI (IXIXII) \ldots IX.
\end{split}  
\end{align}
All constraints are verified, and Theorem~\ref{GT} can be applied.\medskip

With Eq.~(\ref{T_enc}), (\ref{QCASets}) and Corollary~\ref{unit}, we find that we can implement rotations of form 
\begin{equation}\label{SU(4)}
e^{i\alpha_{ij} \sigma^{i} \sigma^{j}}, \text{ for } i,j=0,1,2,3
\end{equation}
where $\sigma^0 \equiv I, \sigma^1 \equiv X,\sigma^2 \equiv Y,\sigma^3 \equiv Z,$
and hence all unitaries in $SU(4)$, by block-local measurements of block-size 6.

\subsubsection{The Ising chain}

SPT analysis implies the absence of uniform MBQC computational power in the ground state of the infinite Ising chain with transverse magnetic field. We now show how the present formalism produces a matching result for all finite system sizes. In fact, the argument below applies to any system with $\mathbb{Z}_2$ symmetry, implemented in a manner consistent with Eqs.~(\ref{u0})--(\ref{url}). The Ising chain is only an example thereof, serving as illustration.\medskip

\noindent
There are two choices for each ${\cal{G}}_i$, (i) ${\cal{G}}_i=\{0\}$, and (ii) ${\cal{G}}_i= \mathbb{Z}_2$. 

Case (i), ${\cal{G}}_i=\{0\}$. There is no non-trivial computation. With Eq.~(\ref{ulZ2}), the measured observables $O_i(g)$, defined in Eq.~(\ref{ObsBlk}), remain $O_i(g) = u_i(g)$, irrespective of the measurement angle $\alpha_i$, and the resulting logical CPTP map, defined in Eq.~(\ref{CPTP}), is ${\cal{V}}_i = [I]$.\smallskip

Case (ii), ${\cal{G}}_i= \mathbb{Z}_2$. ${\cal{G}}_i$ now has one additional element, $g_1$. Inspecting Eq.~(\ref{vLRcomm}), we find 
\begin{equation}\label{vLRcommZ2}
[v_{L,i}(g),v_{R,i}(g')]=0,\;\; \forall g,g' = \mathbb{Z}_2.
\end{equation}
Further, $\mathbb{Z}_2$ has only linear representations, hence $[v_{L,i}(g),v_{L,i}(g')]=0,\;\; \forall g,g' = \mathbb{Z}_2$. Therefore, with Eq.~(\ref{url}), Eq.~(\ref{vLRcommZ2}) implies
\begin{equation}\label{ulZ2}
[v_{L,i}(g),u_{i}(g')]=0,\;\; \forall g,g' = \mathbb{Z}_2.
\end{equation}
With Eq.~(\ref{ulZ2}), the measured observables $O_i(g)$ given by Eq.~(\ref{ObsBlk}) are $O_i(g) \equiv u_i(g)$, irrespective of the measurement angle $\alpha_i$. Hence there is no way to imprint a non-trivial computation on the resource state.

Furthermore, for any block $i$ the only (potentially) non-trivial operation is
\begin{equation}\label{Z2evol}
{\cal{V}}_i(g_1) = \frac{1+ \sigma_i(g_1)}{2} [\exp(-i \alpha_i/2\, T(g_1))] + \frac{1- \sigma_i(g_1)}{2} [\exp(i \alpha_i/2\, T(g_1))]. 
\end{equation}
 The only non-trivial observable available for measurement is $T(g_1)$. Thus, all evolution according to Eq.~(\ref{Z2evol}) can be absorbed in the measurement, leaving it unchanged. 
 
 Further, with the maximality of $H$, we find $H=\mathbb{Z}_2$. Therefore, with Lemma~\ref{L_init}, the initial logical state satisfies $\langle T(g_1) \rangle = (-1)^{\chi(g_1)}$. Hence the only circuit that can be implemented is preparing an eigenstate of $T(g_1)$ and measuring the corresponding eigenvalue $(-1)^{\chi(g_1)}$. Again, no non-trivial computation arises.
 
\subsection{Approaching trivial SPT phases}

To conclude this section, we discuss an aspect common to the first three of the above examples, which interpolate between a computationally useful and a computationally trivial regime. Of our interest are parameter regimes where the thermodynamic limit is a trivial SPT phase. For such finite systems, the string order parameter may be non-zero for any given system size, providing some computational power. However, as we demonstrate below, the power decreases to zero with increasing system size.

If the string order parameter is zero in the thermodynamic limit,
it decays exponentially in finite size systems \cite{denNijs1989,Tasaki1990}, 
namely, there exist $D\in \mathbb{Z}$ and $\xi>0$, such that when $d>D$, we have $|\sigma_k(g)|\leq e^{-d/\xi}$,
where $d$ is the distance between block $k$ and the right boundary. 
Suppose the angle $\alpha$ is split into $N$ pieces as in Corollary \ref{unit}.
When $N>D/(2\Delta)$, there is an overflow of the split rotations into the exponentially decaying region. 
The MBQC rotation  angle that can be implemented in the exponentially decaying region is bounded from above  by 
\begin{eqnarray}
\sum_{n=0}^\infty \frac{\alpha}{N} e^{-(D+2n\Delta)/\xi}=\frac{1}{N}\frac{\alpha e^{-D/\xi}}{1-e^{-2\Delta/\xi}},
\end{eqnarray}
where the upper limit of the sum is set as infinity since we are only interested in an upper bound.
In the region within a $D$ distance from the right boundary,
the MBQC rotation angle is upper bounded by $\frac{\alpha D \sigma_{max}}{2 N\Delta}$,
where $\sigma_{max}=\text{sup}\{|\sigma_k(g)|\}_{k,g}$.
Combining the two regions, the overall implemented rotation angle is on order of $1/N$,
which approaches zero when $N$ becomes large,
equivalent to saying that the system size is increasingly large since it is bounded from below by $2N\Delta$.
Therefore, finite  unitary gate operations cannot be approximated to arbitrary high accuracies in this case,
meaning  MBQC power is lost in trivial SPT phases.

\section{From block-locality to site-locality}\label{BLtoSL}

As we highlighted in Section~\ref{AdvExpl}, an advantage of the present formalism is that in certain settings it permits a block size of one. That is, the notion of locality in MBQC, which is site-locality, is matched exactly by the notion of locality provided by the formalism. The prior formalism \cite{Bartl,SPTO1,SPTO2} based on matrix product states leads to a larger block size.

We establish block size one for the cluster chain and the cellular automation states, but find that the same cannot be obtained for the Kitaev Gamma chain.

\subsection{The cluster chain}\label{sec:ClusterB1}

We choose blocks $i=0,..,n+1$ of size 1. Consistent with the standard discussion of 1D cluster states as resources for MBQC, we take $n$ to be odd, such that the total chain length is odd as well. 

We can choose representations $u_i$, $v_{L,i}$ in the bulk and on the left boundary as follows
\begin{equation}\label{Rep:BulkB1}
\begin{array}{cc}
\text{even sites} & \text{odd sites}\vspace{2mm}\\ 
\begin{array}{rclrcl}
u_e(g_{01}) &=& I,& u_e(g_{10}) &=& X,\\
v_{L,e}(g_{01}) &=& Z,&  v_{L,e}(g_{10}) &=& I.
\end{array}
&
\begin{array}{rclrcl}
u_o(g_{01}) &=& X,& u_o(g_{10}) &=& I,\\
v_{L,o}(g_{01}) &=& I,&  v_{L,o}(g_{10}) &=& Z.
\end{array}
\end{array}
\end{equation}
The representation $v_{L,n+1}$  on the right  boundary is
\begin{equation}\label{Rep:rightB1}
v_{L,n+1}(g_{01}) = Z,\; v_{L,n+1}(g_{10}) = X.
\end{equation}
This concludes the definition of the measurement pattern.\medskip

With Eq.~(\ref{url}) we find the representations $u_{R,i}$ in the bulk and on the left boundary,
\begin{equation}\label{Rep:BL_R}
\begin{array}{rclrcl}
v_{R,e}(g_{01}) &=& Z,& v_{R,e}(g_{10}) &=& X,\\
v_{R,o}(g_{01}) &=& X,& v_{R,o}(g_{10}) &=& Z.
\end{array}
\end{equation}
With Eq.~(\ref{H_spec}) and Eqs.~(\ref{Rep:BulkB1}), (\ref{Rep:BL_R}) for $i=0$, we find that $H=\langle g_{10}\rangle$. 
From Eqs.~(\ref{KappaR}) and (\ref{vLRcomm}), we compute the sets ${\cal{G}}_i$, $1\leq i \leq n$, finding
\begin{equation}\label{ClusterSetsB1}
\begin{array}{rcll}
{\cal{G}}_i &=&\langle g_{01} \rangle,\;\;\forall i\;\text{even},\\
{\cal{G}}_i &=& \langle g_{10} \rangle,\;\;\forall i\;\text{odd}.
\end{array}
\end{equation}
Inserting Eqs.~(\ref{Rep:BulkB1}), (\ref{Rep:right}) into Eq.~(\ref{Udef}) reproduces the global symmetry action Eq.~(\ref{ClusterGlobalSymm}), 
$$
g_{01} \cong  ZXIXIXIX...IXZ,\;\; g_{10} \cong XIXIXIXI...IX.
$$
The preconditions all hold, and Theorem~\ref{GT} can be applied.\medskip

We now investigate which logical operations follow from the construction. With Eq.~(\ref{T_enc}), the logical operators are
$$
T(g_{01}) = IXI...IXZ =:\overline{Z},\;T(g_{10}) =XIX...XIX =:\overline{X},
$$
This labeling matches with the corresponding assignments in Eq.~(\ref{Rep:rightB1}).

We recall from Eq.~(\ref{DefLR}) that the operators $R_l(g)$ are defined only for $g \in {\cal{G}}_l$, and so are the corresponding string order parameters $\sigma_l(g)=\langle R_l(g)\rangle_\Phi$, cf. Eq.~(\ref{string}). With Eq.~(\ref{ClusterSetsB1}), there exists exactly one string order parameter per site. It is associated with $g_{01}$ for even and $g_{10}$ for odd sites. For example, the only non-trivial string order parameter for site $k=3$, with support on sites $l\geq 3$, is $\sigma_3(g_{10})=\langle IIIZXIX...XIX\rangle_\Phi$.

With Eq.~(\ref{CPTP}), the realizable logical operations ${\cal{V}}_k$ hinge on the string order parameters that are defined. Thus, we can perform the following logical CPTP maps by measurement on a single site,
$$
\begin{array}{rcll}
{\cal{V}}_k(g_{01}) &=& \frac{1+\sigma_k(g_{01})}{2} \left[ e^{-i\frac{\alpha_k}{2} \overline{Z}} \right] +\frac{1-\sigma_k(g_{01})}{2} \left[ e^{i\frac{\alpha_k}{2} \overline{Z}} \right], & \text{if}\; k\, \text{even},\vspace{2mm}\\
{\cal{V}}_k(g_{10}) &=& \frac{1+\sigma_k(g_{01})}{2} \left[ e^{-i\frac{\alpha_k}{2} \overline{X}} \right] +\frac{1-\sigma_k(g_{01})}{2} \left[ e^{i\frac{\alpha_k}{2} \overline{X}} \right], & \text{if}\; k\, \text{odd}.
\end{array}
$$
With Corollary~\ref{unit}, by splitting the realization of a logical operation over many sites, we can arbitrarily closely approximate the unitaries  $e^{i\alpha Z}, e^{i\beta X}$,
which generate the group $SU(2)$ as before.\smallskip

We may now compare to the preceding discussion of the cluster chain in Section~\ref{sec:ClusterB2}. In the present construction, we get one less elementary gate---rotations about the $y$-axis. However, from the perspective of computational power, that doesn't make a difference. The group of gates generated is $SU(2)$ in both cases. Yet, there is a gain in the present construction: the block size has been reduced from 2 to 1. That is, the computational scheme gets by with single-site measurements, which is the standard for MBQC. 

\subsection{The Kitaev-Gamma chain}

A site-local scheme is not available here. Any linear representation $u_i$ that brings the general expression  Eq.~(\ref{Udef}) of the symmetry action on the resource state in agreement with the specific symmetry action Eq.~(\ref{eq:sym_Z2_Z2_KG}) established for the Kitaev Gamma chain requires even block size.

\subsection{Cellular automaton states}

We choose blocks $i=1,..,n$ of size 1 and both the edge blocks $i=0 \text{ and } n+1$ of size 2. To be consistent with the framework set up in Section~\ref{QCA_block}, we choose $n$ to be divisible by $6$. Thus the total number of qubits in the chain is $6n+4$.

We start by choosing representations in the bulk. The site labels below are all  mod $6$. The projective representations $v_{L,i}$ are 
\begin{equation}\label{QCAsite_bulk_l}
   \begin{alignedat}{7}
     v_{L,1}(g_1)&= I,&\quad v_{L,1}(g_2)&= Z,&\quad v_{L,1}(g_3)&= I,&\quad v_{L,1}(g_4)&=I , \\ 
    v_{L,2}(g_1)&= Z,&\quad v_{L,2}(g_2)&= I,&\quad v_{L,2}(g_3)&= I,&\quad v_{L,2}(g_4)&=I , \\ 
    v_{L,3}(g_1)&= I,&\quad v_{L,3}(g_2)&= I,&\quad v_{L,3}(g_3)&= I,&\quad v_{L,3}(g_4)&=Z , \\ 
    v_{L,4}(g_1)&= I,&\quad v_{L,4}(g_2)&= I,&\quad v_{L,4}(g_3)&= Z,&\quad v_{L,4}(g_4)&=I  , \\
    v_{L,5}(g_1)&= I,&\quad v_{L,5}(g_2)&= Z,&\quad v_{L,5}(g_3)&= I,&\quad v_{L,5}(g_4)&=I , \\ 
    v_{L,6}(g_1)&= Z,&\quad v_{L,6}(g_2)&= I,&\quad v_{L,6}( g_3)&= I,&\quad v_{L,6}(g_4)&=I,    
   \end{alignedat} 
\end{equation}
and the linear representations $u_i $ are

\begin{equation}\label{QCAsite_bulk_u}
 \begin{alignedat}{7}
    u_{1}(g_1)&= X,&\quad u_{1}(g_2)&= I,&\quad u_{1}(g_3)&= I,&\quad u_{1}(g_4)&=I,  \\ 
    u_{2}(g_1)&= I,&\quad u_{2}(g_2)&= I,&\quad u_{2}(g_3)&= I,&\quad u_{2}(g_4)&=X,  \\ 
    u_{3}(g_1)&= I,&\quad u_{3}(g_2)&= I,&\quad u_{3}(g_3)&= X,&\quad u_{3}(g_4)&=I,  \\ 
    u_{4}(g_1)&= I,&\quad u_{4}(g_2)&= X,&\quad u_{4}(g_3)&= I,&\quad u_{4}(g_4)&=X , \\
    u_{5}(g_1)&= X,&\quad u_{5}(g_2)&= I,&\quad u_{5}(g_3)&= X,&\quad u_{5}(g_4)&=I , \\ 
    u_{6}(g_1)&= I,&\quad u_{6}(g_2)&= X,&\quad u_{6}(g_3)&= I,&\quad u_{6}(g_4)&=I .
\end{alignedat}   
\end{equation}

The representations at the left and right boundary are identical to ones mentioned in the Section~\ref{QCA_block} via Eqs.~\ref{Rep:QCAleft},\ref{Rep:QCAright}, as their block size remains unchanged.
This concludes the specification of the measurement pattern.

Using Eq.\ref{url}, we find the representations $v_{R,i}$ in the bulk as
\begin{equation}\label{QCAsite_bulk_r}
\begin{alignedat}{7}
v_{R,1}(g_1)&= X,&\quad v_{R,1}(g_2)&= Z,&\quad v_{R,1}(g_3)&= I,&\quad v_{R,1}(g_4)&=I , \\ 
v_{R,2}(g_1)&= Z,&\quad v_{R,2}(g_2)&= I,&\quad v_{R,2}(g_3)&= I,&\quad v_{R,2}(g_4)&=X , \\ 
v_{R,3}(g_1)&= I,&\quad v_{R,3}(g_2)&= I,&\quad v_{R,3}(g_3)&= X,&\quad v_{R,3}(g_4)&=Z  , \\ 
v_{R,4}(g_1)&= I,&\quad v_{R,4}(g_2)&= X,&\quad v_{R,4}(g_3)&= Z,&\quad v_{R,4}(g_4)&=X  , \\
 v_{R,5}(g_1)&= X,&\quad v_{R,5}(g_2)&= Z,&\quad v_{R,5}(g_3)&= X,&\quad v_{R,5}(g_4)&=I , \\ 
    v_{R,6}(g_1)&= Z,&\quad v_{R,6}(g_2)&= X,&\quad v_{R,6}(g_3)&= I,&\quad v_{R,6}(g_4)&=I .
\end{alignedat}
\end{equation}
From {Eqs.~(\ref{KappaR})} and (\ref{vLRcomm}), we compute the sets ${\cal{G}}_i$, $1\leq i \leq n$, finding
\begin{align}\label{QCABlockSets}
\begin{split}
\mathcal{G}_i= \begin{cases}
\langle g_2g_4 \rangle,& \, \forall i=1 \mod 6 \\
\langle g_1g_3 \rangle,& \, \forall i=2 \mod 6 \\
\langle g_4 \rangle,& \, \forall i=3 \mod 6\\
\langle g_3 \rangle,& \, \forall i=4 \mod 6\\
\langle g_2 \rangle,& \, \forall i=5 \mod 6\\
\langle g_1 \rangle,& \, \forall i=6 \mod 6
\end{cases}
\end{split}
\end{align}
Inserting Eqs.~(\ref{RepR:QCAleft}), (\ref{QCAsite_bulk_u}), (\ref{Rep:QCAright}) into Eq.~(\ref{Udef}) produces 
\begin{align} 
\begin{split}
 g_{1}  &\cong IZ (XIIIXI)\ldots XI ,\;\; g_{2} \cong ZX (IIIXIX)\ldots ZX, \\ 
 g_{3} &\cong XI(IIXIXI) \ldots  IZ,\;\; g_{4} \cong ZI (IXIXII) \ldots IX.
\end{split}  
\end{align}
All constraints are verified, and Theorem~\ref{GT} can be applied.\medskip

Just like the cluster case, in the site local measurement scheme, there exists exactly one string order parameter per site. For example, the only non-trivial string order for site $4$ (note that the first two sites are labelled as $-1$ and $0$), is given by
$\sigma_4(g_3)= \langle II IIIZXI (IIXIXI) \ldots IZ \rangle$.

Now, coming back to the question of the realizable gate sets with the site local scheme, with Eq.~(\ref{T_enc}), (\ref{QCABlockSets}) and Corollary~\ref{unit}, we find that we can implement rotations of form 
\begin{equation} \label{6_set_SU(4)}
e^{i\alpha ZI}, e^{i\beta XZ}, e^{i\gamma IX}, e^{i\delta IZ}, e^{i\tau ZX}, e^{i\eta XI}.
\end{equation}
Fortunately, these unitaries are enough to generate the whole $SU(4)$ group. 

Comparing the present construction to the block local measurement scheme in Section ~\ref{QCA_block}, we see that we get only $6$ elementary gates in the site local case compared to the $15$ earlier. But, in the end, it doesn't make any difference as the group of gates generated is $SU(4)$ in both cases. However, the analysis in this section is better-suited to standard MBQC discussions as we get by with site-local measurements.

\section{String vs. computational order parameters}\label{StringComp}

Ref. \cite{SPTO1} introduced computational order parameters $\nu_{ij}$ that  can be extracted from the MPS description of the resource state. They govern the effectiveness of MBQC in 1D SPT phases. In the present formalism, precisely the same role is played  by the string order parameters $\sigma_k(g_k)$, $g_k\in {\cal{G}}_k$. Comparing Eq.~(\ref{CPTP}) with the corresponding Equations~(24), (26) of \cite{SPTO1}, we anticipate a linear relation between $\sigma(g_k)$ and the $\nu_{ij}$.  In Theorem~\ref{SO}  below we confirm this. 

Ref. \cite{SPTO1} focusses on symmetry groups of type $\mathbb{Z}_d\times \mathbb{Z}_d$, $d\in \mathbb{N}$, and the present work on $\mathbb{Z}_2^k$, $k\in \mathbb{N}$. We can only compare in the intersection, i.e., the group $\mathbb{Z}_2 \times \mathbb{Z}_2$. We also restrict to translation-invariant systems (in the bulk), since Ref. \cite{SPTO1} makes that assumption.\medskip

{\em{Background.}} In Section~\ref{MPSMBQC} we recalled a basic result from \cite{Bartl}, for the MPS tensors of $\mathbb{Z}_d \times \mathbb{Z}_d$ symmetric states, cf. Eq.~(\ref{TP}). Namely,  in the maximally non-commuting phase, the components of these tensors  (in the symmetric basis) factorize as $A_s = C_s \otimes B_s$, with $C_s$ constant and determined by symmetry, and $B_s$ varying across the phase in potentially arbitrary ways. The definition of $\nu$ is based on the `junk' matrices $B_s$. Namely, we define the channel ${\cal{L}}(\rho):=\sum_s B_s (\rho) B_s^\dagger$. ${\cal{L}}$ has a unique attractive fixed point $\rho_\text{fix}$. The $\nu_{ij}$ are specified by the relation (cf. Eq.~(20) of \cite{SPTO1}), 
\begin{equation}\label{nuDef}
\lim_{n\rightarrow \infty} {\cal{L}}^n \left(B_i \rho_\text{fix} B_j^\dagger\right)  = \nu_{ij}\, \rho_\text{fix}.
\end{equation}
The coefficients $\nu_{ij}$ are conveniently arranged in matrix form, $[\nu_{ij}]=:\nu$. Somewhat surprisingly, $\nu$ satisfies the constraints of a density matrix, i.e., it has unit trace, and is Hermitian and positive. An interpretation of what the `state' $\nu$ represents is given in Section~VIII B of \cite{SPTO1}. \smallskip

We recall that for the 1D cluster state, which is inside the maximally non-commuting phase with $\mathbb{Z}_2\times \mathbb{Z}_2$-symmetry, the junk system has dimension 1, and the junk matrices may be set
\begin{equation}\label{signConv}
B_s(x)=+1,\; \forall s\in  \mathbb{Z}_2\times \mathbb{Z}_2,\; \forall \;\text{blocks} \; x\;\;\; \text{(1D cluster state)}.
\end{equation} 
We observe that Eq.~(\ref{signConv}) contains a sign convention, since the transformation (*) $C_k \longrightarrow -C_k$, $B_k\longrightarrow - B_k$, for any given $k\in \mathbb{Z}_2\times \mathbb{Z}_2$,  doesn't change the MPS tensor $A$, hence not the quantum state described.

With the above notions and conventions introduced, we have the following result.
\begin{Theorem}\label{SO}
In the maximally non-commuting phase with $\mathbb{Z}_2 \times \mathbb{Z}_2$-symmetry, subject to the sign convention Eq.~(\ref{signConv}), it holds that
\begin{equation}\label{OPrel}
\sigma_x(g) = \sum_i \nu_{i,i+w_g},
\end{equation}
where $w_g$ is such that $|i+w_g\rangle \propto v_{R,x}(g)|i\rangle$. 
\end{Theorem}
The above theorem demonstrates the anticipated linear dependence between the string order parameters $\sigma_x(g)$ and the computational order parameters $\nu_{ij}$. 

It remains to be clarified why the sign convention Eq.~(\ref{signConv}) is enforced. The reason is that, while the transformations (*) don't change the state hence not the string order parameters $\sigma(g)$, they change $\nu$ (cf. definition Eq.~(\ref{nuDef})). Therefore, Eq.~(\ref{OPrel}) is not invariant under (*), and a sign convention must be picked. As the proof of Theorem~\ref{SO} reveals, Eq.~(\ref{signConv}) is a convenient choice.

Additionally, as a consistency check, we observe that both sides of Eq.~(\ref{OPrel}) are real-valued. The lhs is because $R_x(g)$ is Hermitian, cf. Eqs.~\ref{Rep:bulk}), (\ref{RepR:Bulk}). The rhs is, because 
$$
\sum_i \nu_{i,i+w_g} = \frac{1}{2} \left( \sum_i \nu_{i,i+w_g}  + \sum_i \nu_{i+w_g,i} \right) = \frac{1}{2} \left( \sum_i \nu_{i,i+w_g} + \nu_{i,i+w_g}^* \right).
$$
In the first step, we have split the sum into two halves and reorganized the summation in the second half. In the second step, we used Hermiticity of $\nu$, cf. Eq.~(21) of \cite{SPTO1}.
\medskip

\begin{figure}
\begin{center}
\includegraphics[width=15cm]{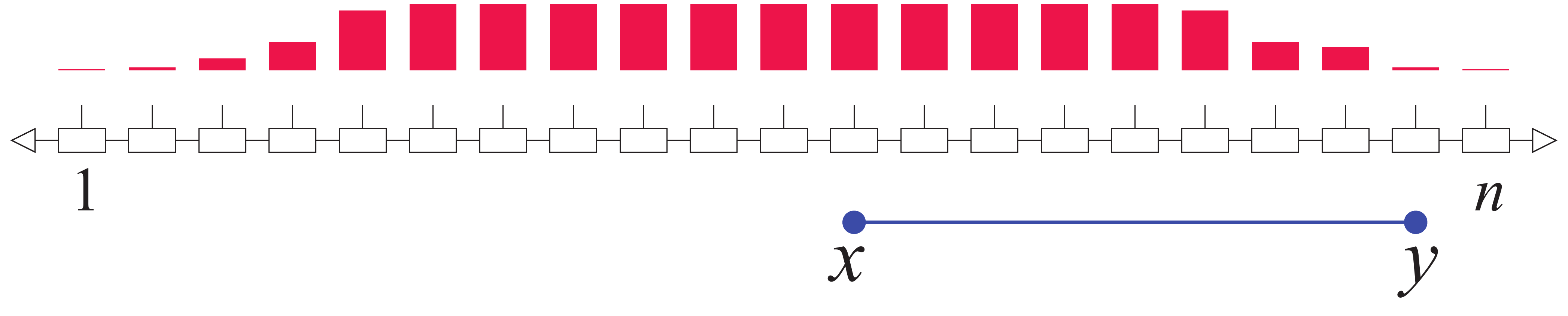}
\caption{\label{SigmaXY}Conventions for the resource state $|\Phi\rangle$ considered. The symmetric perturbation, indicated by red bars, fades off at the end of the chain, and is constant in the bulk. The string order operators $\sigma_{xy}(g)$ are such that the end point $x$ is in the bulk, and the end point $y$ is near the right edge, where the perturbation is tuned off.}
\end{center}
\end{figure}

{\em{Proof of Theorem~\ref{SO}.}} The proof is in the MPS formalism, within which the parameters $\nu$ are defined (see \cite{Bartl3} for the application of similar techniques). Of interest are the string order parameters $\sigma_x(g)=\langle \Phi |R_x(g)|\Phi\rangle$. For technical reasons, we consider the string order parameters with two end points, $\sigma_{xy}(g):= \langle \Phi |R_x(g) R_y(g)|\Phi\rangle$, with the site $x$ deep in the bulk, and the site $y$ near the right boundary of the chain where the perturbation is turned off. Near the end points, the chain looks like the cluster chain, and therefore $R_y(g)|\Phi\rangle = |\Phi\rangle$, for all $g\in \mathbb{Z}_2 \times \mathbb{Z}_2$ and all  block locations $y$ near the right boundary $n$. Hence,
\begin{equation}\label{xyx}
\sigma_{xy}(g) = \sigma_x(g),\; \forall g\in \mathbb{Z}_2 \times \mathbb{Z}_2, \; y\approx n.
\end{equation}
In the graphical calculus of tensor networks, we have
\begin{equation}\label{sig2}
\sigma_{xy}(g) = \frac{1}{\text{Norm}} \times \left(\parbox{6cm}{\includegraphics[width=6cm]{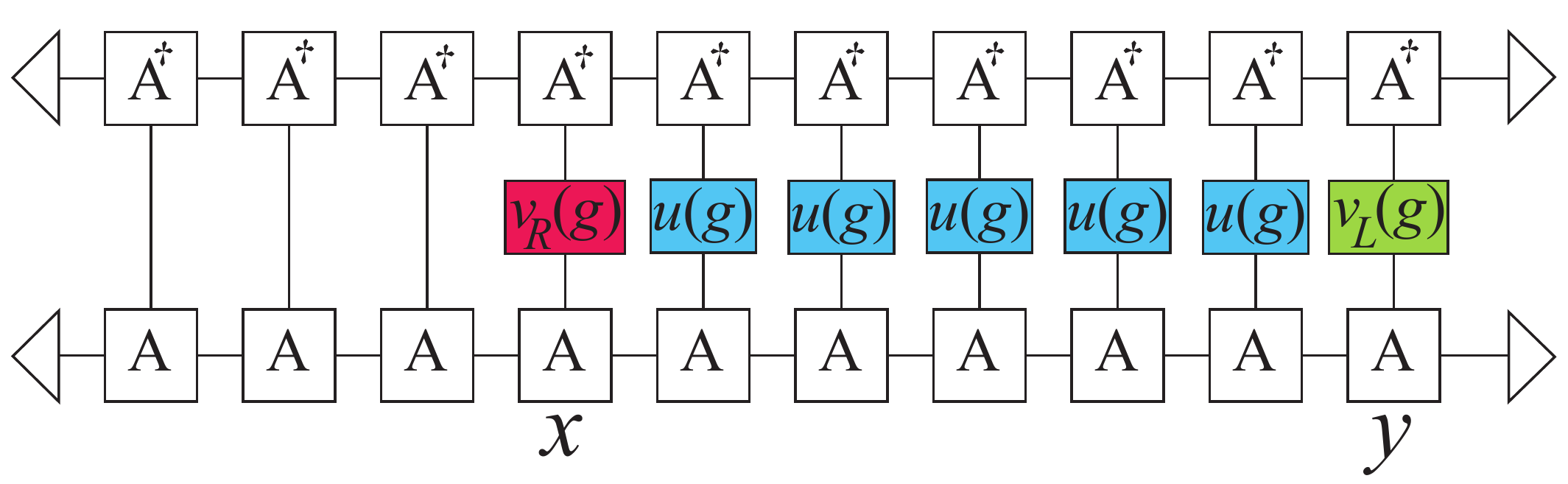}} \right),
\end{equation}
where the (projective and linear) representations $v_L$, $v_R$ and $u$ are as defined in Section~\ref{subsubsec:assumptions}, and
\begin{equation}\label{sig3}
\text{Norm} = \left(\parbox{6cm}{\includegraphics[width=6cm]{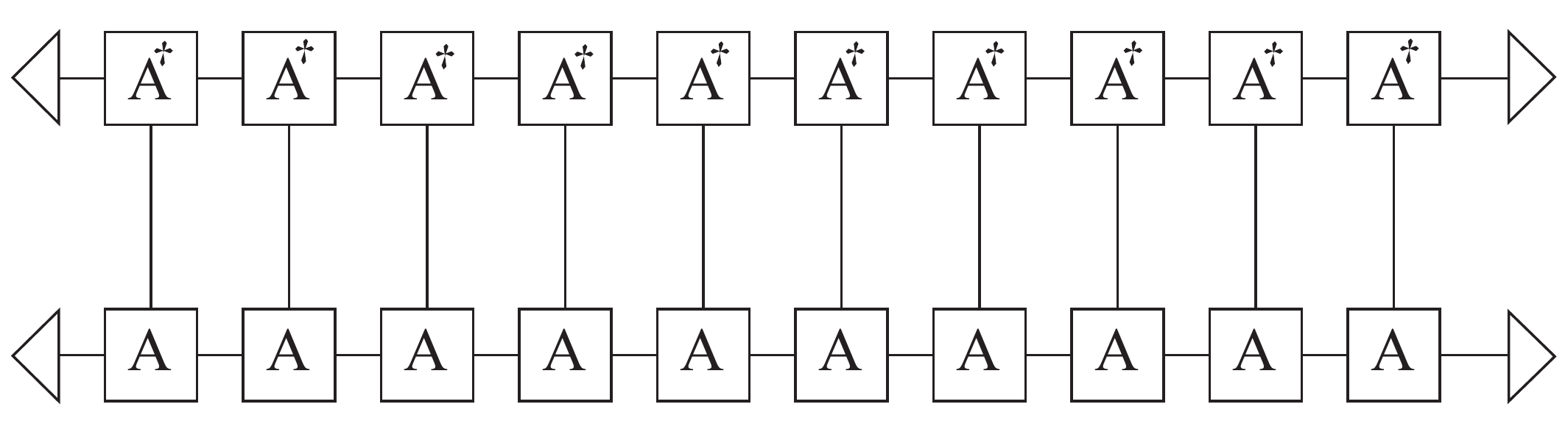}} \right).
\end{equation}
The first step in manipulating the tensor network on the rhs of Eq.~(\ref{sig2}) is to use the symmetries of the tensors $A$ between blocks $x$ and $y$,
\begin{equation}\label{sig4}
\sigma_{xy}(g) = \frac{1}{\text{Norm}} \times \left(\parbox{7.5cm}{\includegraphics[width=7.5cm]{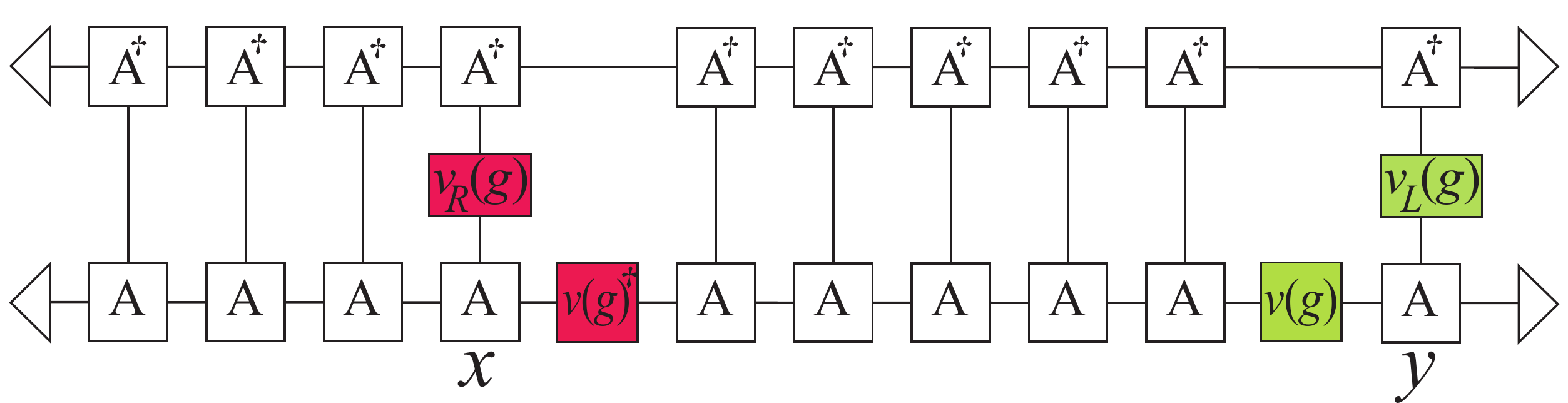}} \right),
\end{equation}
where $v(g)$ the projective representation of the symmetry acting on the virtual legs of $A$; see \cite{Bartl}.

Next we turn to the network of tensors near $x$,
$$
{\tt{TNW}}(x):= \parbox{1.5cm}{\includegraphics[width=1.5cm]{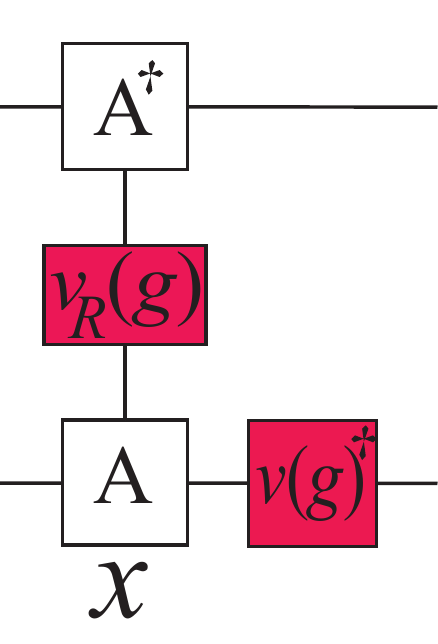}} = \sum_{ij} (B_i \otimes C_i v_g) \otimes (B_j^\dagger \otimes C_j^\dagger) \langle j| v_L(g)|i\rangle.
$$
Therein, with Eq.~(\ref{RepR:Bulk}), $\langle j| v_L(g)|i\rangle =\delta_{j,i+w_g}e^{i\phi_{j,w_g}}$, for some phases $\phi_{j,w_g}$. Thus,
$$
\begin{array}{rcl}
{\tt{TNW}}(x)&=& \sum_{j}   (B_{j+{w_g}} \otimes C_{j+{w_g}}v(g)) \otimes (B_j^\dagger \otimes C_j^\dagger) e^{i\phi_{j,w_g}}\\
&=& \sum_{j}   (B_{j+{w_g}} \otimes C_jC_{w_g}v(g)) \otimes (B_j^\dagger \otimes C_j^\dagger) e^{i\phi_{j,w_g}+i\phi'_{j,w_g}}
\end{array}
$$
Now specializing to the cluster case, with Eq.~(\ref{signConv}) we obtain
\begin{equation}\label{sig5}
\left(\parbox{1.5cm}{\includegraphics[width=1.5cm]{sigmaXY_5}} \right)_\text{cluster} = \sum_{j}   (C_jC_{w_g}v(g)) \otimes C_j^\dagger e^{i\phi_{j,w_g}+i\phi'_{j,w_g}}.
\end{equation}
We further employ the fact that, at the cluster point, the tensor $A$ has the additional symmetry
$$
\parbox{3.7cm}{\includegraphics[width=3.7cm]{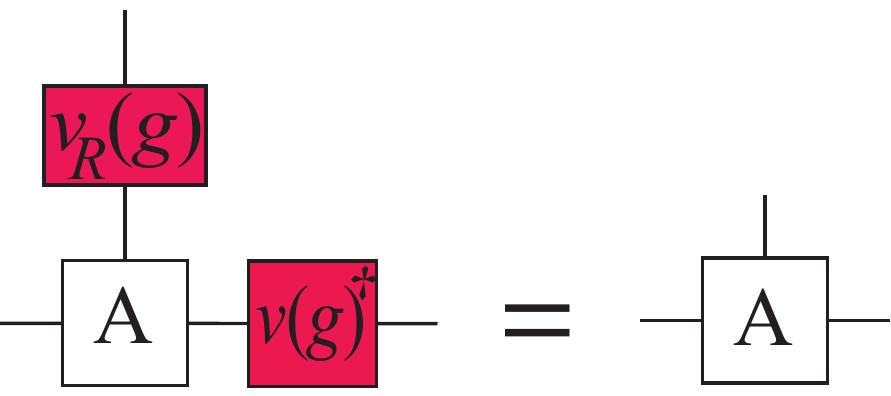}}.
$$
Therefore,
\begin{equation}\label{sig7}
\left(\parbox{1.5cm}{\includegraphics[height=2.2cm]{sigmaXY_5}} \right)_\text{cluster} = \left(\parbox{0.9cm}{\includegraphics[height=2.2cm]{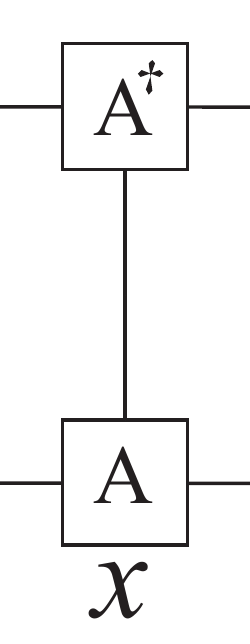}} \right)_\text{cluster} = \sum_{j}   C_j \otimes C_j^\dagger.
\end{equation}
Comparing Eqs.~(\ref{sig5}) and (\ref{sig7}), we find that $C_{w_g}v(g)e^{i\phi_{j,w_g}+i\phi'_{j,w_g}}=I$. Since the matrices $C_j$ and the phase factors are constant across the cluster phase, this relation holds in the entire  phase. The tensor network near $x$ therefore simplifies to
\begin{equation}
{\tt{TNW}}_x = \sum_{j}   (B_{j+{w_g}} \otimes C_j) \otimes (B_j^\dagger \otimes C_j^\dagger). 
\end{equation} 
Analogously, with additional simplification arising through Eq.~(\ref{signConv}),
$${\tt{TNW}}_y =   \sum_{j}   C_j \otimes C_j^\dagger = \sum_{j}   A_j \otimes A_j^\dagger.$$ 
Summarizing so far, we have
\begin{equation}\label{sig8}
\sigma_{xy}(g) = \frac{1}{\text{Norm}} \times \left(\parbox{6cm}{\includegraphics[width=6cm]{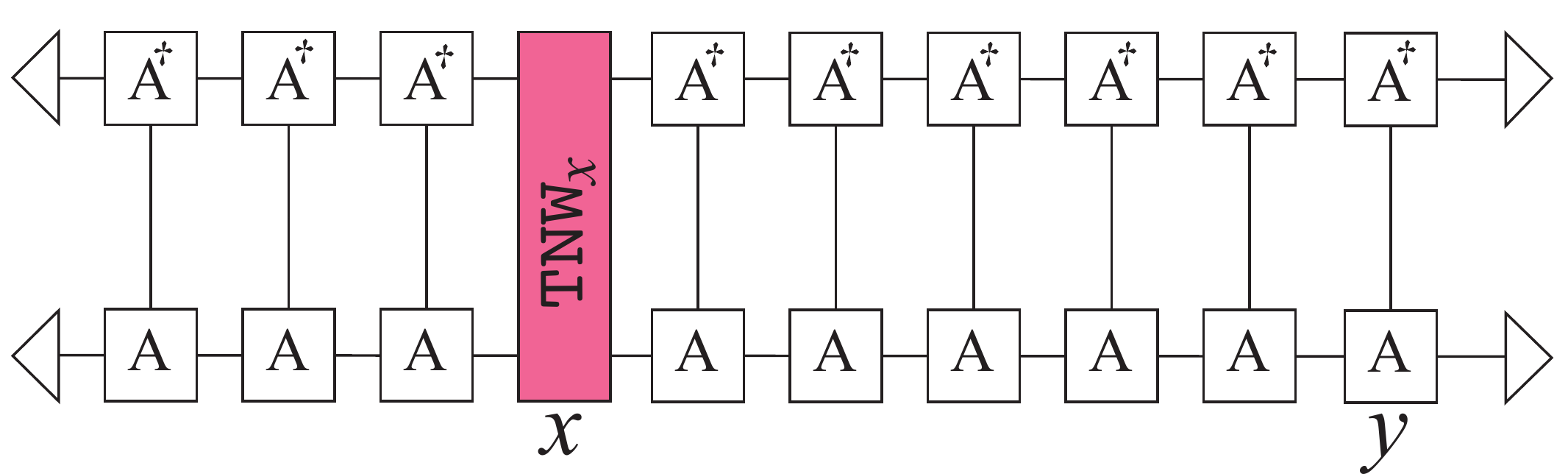}} \right).
\end{equation}
We now propagate forward all byproduct operators $C_{i_k}$, applying them to the right boundary condition. Due to the tuning-off of the perturbation near the boundaries, the virtual  boundary states $|L\rangle$, $|R\rangle$ are those of the cluster state, $|L\rangle = |0\rangle$, $|R\rangle = |+\rangle$. Denoting the overall byproduct $\Sigma_\textbf{i} = C_{i_1}C_{i_2}...C_{i_{n-1}}C_{i_n}$, the configuration $\textbf{i}$ makes a contribution $\propto |\langle 0|\Sigma_\textbf{i}|+\rangle|^2=1/2$. Therefore,
\begin{equation}\label{sig9}
\sigma_{xy}(g) = \frac{1}{2^n}\frac{1}{\text{Norm}} \times \left(\parbox{6cm}{\includegraphics[width=6cm]{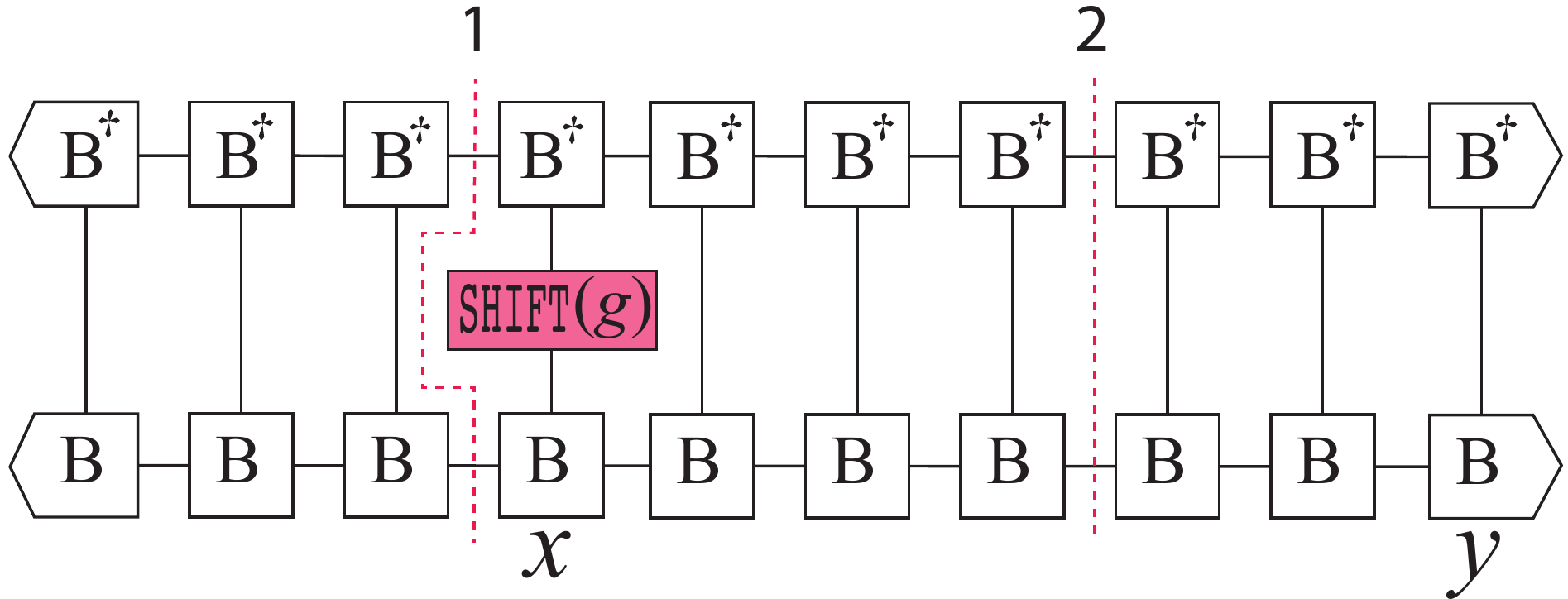}} \right),
\end{equation}
where 
\begin{equation}\label{TNW'}
{\tt{SHIFT}}(g):=\sum_j|j+w_g\rangle \langle j|.
\end{equation}
We now interpret the tensor network on the rhs of Eq.~(\ref{sig9}). The left part until the vertical cut 1 represents the creation of some state, followed by repeated application of the channel ${\cal{L}}$. The result is the fixed point state $\rho_\text{fix}$. Then comes the map ${\tt{SHIFT}}(g)$, and thereafter further applications of the channel ${\cal{L}}$. With Eqs.~(\ref{nuDef}) and (\ref{TNW'}), the operator resulting from the left part of the tensor network up to cut 2 is $\left( \sum_j \nu_{j+w_g,j}\right) \rho_\text{fix}$. The complementary part, i.e. the network to the right of cut 2, prepares a corresponding effect $E_\text{fix}$. Therefore, the entire expression becomes
$$
\sigma_{xy}(g) = \frac{1}{2} \frac{1}{\text{Norm}} \times   \left[ \sum_j \nu_{j+w_g,j}\right] \text{Tr}\left(\rho_\text{fix}E_\text{fix}\right).
$$
The norm factor Eq.~(\ref{sig3}) corresponds to the tensor network in Eq.~(\ref{sig2}), for the choice $g=0$. Therefore, 
$$
\text{Norm} = \frac{1}{2} \times \text{Tr}\left(\rho_\text{fix}E_\text{fix}\right).
$$
Combining the last two equations gives
$$
\sigma_{xy}(g) =  \sum_j \nu_{j+w_g,j}.
$$
Invoking Eq.~(\ref{xyx}) and rewriting the sum yields Eq.~(\ref{OPrel}). $\Box$\medskip

{\em{Remark:}} Of independent interest, we observe in Eq.~(\ref{sig9}) that the string order parameters $\sigma_x(g)$, $\forall g\in \mathbb{Z}_2\times \mathbb{Z}_2$ , can be evaluated as expectation values of local operators ${\tt{SHIFT}}(g)$ on a `junk state' (recall that $\sigma_x(g)=\sigma_{xy}(g)$, for $y\approx n$). Also see \cite{Bartl3}, \cite{SPTO1}.

\section{An application: string order and contextuality} \label{COeSO}

Ref.~\cite{DM} established a connection between string order in symmetry protected topological phases and non-local games \cite{NLgames}---an area at the foundations of quantum physics and information theory. Inspired by this, here we establish a link between string order and quantum contextuality, a subject closely related to non-local games. Our result is stronger in the sense that it applies to an entire symmetry protected phase, not just a sub-region thereof; yet our setting is more permissive.\smallskip 

To begin, we review the notion of quantum contextality, namely the inviability of non-contextual hidden-variable models \cite{Merm}. In a non-contextual hidden variable model (nc HVM), pre-determined outcomes $\lambda_A$ are assigned to observables $A \in {\cal{O}}$ in such a way that the following conditions are met: (i) $\lambda_A$ is an eigenvalue of $A$; (ii) $\lambda_A$ is a function of $A$ only, and in particular does not depend on what compatible observables $B,C,..$ are measured jointly with $A$ (context-independence); and (iii) if jointly measurable observables satisfy an algebraic relation, $X=f(A,B,..)$, the corresponding values satisfy the same relation, $\lambda_X=f(\lambda_A,\lambda_B,..)$.

If the set ${\cal{O}}$ is such that the above constraints admit no solutions $\lambda: {\cal{O}} \longrightarrow \mathbb{R}$, then ${\cal{O}}$ is called contextual. The Kochen-Specker theorem \cite{KS} says that contextual sets ${\cal{O}}$ exist whenever the Hilbert space dimension is $\geq 3$. If a given set ${\cal{O}}$ does admit solutions $\lambda: {\cal{O}} \longrightarrow \mathbb{R}$, then one may attempt to mimick the randomness of quantum measurement by a probability distribution $p$ over the admissible value assignments $\lambda$. The notion of state-dependent contextuality then applies:

\begin{Def} Consider a set ${\cal{O}}$ of observables for which the set $\Omega$ of non-contextual value assignments $\lambda: {\cal{O}} \longrightarrow \mathbb{R}$ is non-empty.
A state $|\Psi\rangle$ is contextual wrt. ${\cal{O}}$ if no probability function $p: \Omega \longrightarrow \mathbb{R}_+$ reproduces the measurement statistics of compatible observables from ${\cal{O}}$ on $|\Psi\rangle$.
\end{Def}
An immediate consequence of this definition is that if ${\cal{O}} \subset {\cal{O}}'$ and $|\Psi\rangle$ is contextual wrt. ${\cal{O}}$ then $|\Psi\rangle$ is contextual wrt. ${\cal{O}}'$.
With those notions introduced, we have the following result.

\begin{Theorem}\label{CT}
Consider a family ${\cal{F}}_{\Delta,\sigma}$ of short-range entangled states with symmetry group  $\mathbb{Z}_2\times \mathbb{Z}_2$  acting as in Eq.~(\ref{symm}). All states $|\Psi\rangle \in {\cal{F}}_{\Delta,\sigma}$ have an entanglement range of $\Delta\leq 0$, and for all string order parameters it holds that $|\sigma_k(g)|\geq \sigma>0$, $g\in {\cal{G}}_k$, for all sites $k$. Further, ${\cal{F}}_{\Delta,\sigma}$ contains a member of any finite size. Then, the family ${\cal{F}}_{\Delta,\sigma}$ contains quantum states that are contextual wrt. the set of site-local observables. 
\end{Theorem}
\medskip

\noindent
{\em{Proof of Theorem~\ref{CT}.}} The overall strategy of the proof is to employ MBQC as a {\em{witness}} of contextuality, and then invoke Corollary~\ref{unit} to demonstrate that the witness can be implemented.

(i) {\em{Construction of the contextuality witness.}} We derive the witness from the simplest contextual MBQC \cite{AB}, based on Mermin's star \cite{Merm}, with two bits of classical input. To accommodate the error that applying Corollary~\ref{unit} will introduce, we use Theorem~3 of \cite{RR13}, which handles probabilistic MBQCs. For the case of two input bits it says that any MBQC which computes a non-linear Boolean function with a success probability greater than $3/4$ is contextual. 

Concretely, we MBQC-simulate the following set of four quantum circuits, depending on two input bits $a$ and $b$, and yielding one output bit $s$,
\begin{equation}\label{BoolCirc}
\frac{I+(-1)^s X}{2}\exp(\left((-1)^{a+b} i \frac{\pi}{8} Z \right)\exp(\left((-1)^{b} i \frac{\pi}{8} Z \right)\exp(\left((-1)^{a} i \frac{\pi}{8} Z \right)\exp(\left( i \frac{\pi}{8} Z \right)|+\rangle.
\end{equation}
It is easily verified that this circuit family computes an OR-gate, $s=a\cup b$, with unit success probability. The OR-gate is a non-linear Boolean function.\smallskip

(ii) {\em{Implementation of the contextuality witness.}} We choose the one-site blocking discussed in Section~\ref{sec:ClusterB1}. With Corollary~\ref{unit} applied to the present case of $\mathbb{Z}_2\times \mathbb{Z}_2$ symmetry (the example is completely worked out in Section~\ref{sec:ClusterB1}) we find that the circuits of Eq.~(\ref{BoolCirc}) can be realized, with an error that decreases towards zero with increasing size of the resource state, cf. Eq.~(\ref{ApprErr}). Since the family ${\cal{F}}_{\Delta,\sigma}$ contains members of any size, one state is large enough to surpass the contextuality  threshold in the success probability of 3/4. $\Box$\medskip

 Comparing with Theorem~2 in \cite{DM}, we find that the above Theorem~\ref{CT} is stronger in the sense that contextuality is established for any non-vanishing value of the string order parameters, not only beyond a threshold value ($\sigma>1/3$ in Theorem~2 of \cite{DM}). On the other hand, the present contextuality setting is more permissive than the non-local game setting in \cite{DM}. Specifically, in the non-local game setting, the sites/blocks on the chain cannot communicate with one another. The contextuality setting above places no such constraint. The measurable observables are local, but which ones among the local observables are measured, and under which conditionings---e.g. on earlier measurement outcomes---is not restricted.
 
\section{Conclusion and outlook}\label{Concl}

We have devised a new framework for MBQC on short-range entangled symmetric resource states. It requires fewer assumptions than previously known. Specifically, we can handle finitely extended systems and do not require translation-invariance. Further, our formalism matches the site-locality of measurements in MBQC, as opposed to interpreting larger blocks of spins as the local unit. 

We strengthen the existing connection \cite{DB1,CBD} between string order and MBQC computational power in one-dimensional systems. Our physical insight is that whenever the string order parameters are non-zero, a constant group of unitary gates can be implemented arbitrarily accurately. The larger the values of the order parameters, the more efficient the implementation. This is the content of Theorem~\ref{GT} and Corollary~\ref{unit} in Section~\ref{Sec:sreMBQC}.

Our conceptual insight is that, once we move to finite systems such that SPT phases can no longer form the basis for classification (because they no longer exist), we observe a reversal of importance: the measurement scheme becomes the central object, the object suitable for classification; and the resource state becomes the accessory. The latter has to be symmetric, short-range entangled, and possess string order matching the symmetries; and nothing else about it needs to be known. In the thermodynamic limit discussed in previous works, the essential property of an SPT ordered resource state is the SPT phase it belongs to, characterized by group cohomology. For finite size, this cohomological information reappears in the description of the measurement scheme.\smallskip

We conclude with the following open questions:
\begin{itemize}
    \item{{\em{Can we classify the MBQC schemes with $(\mathbb{Z}_2)^m$-symmetry in spatial dimension one?}} We remark that in Section ~\ref{MP}, item 1 of the list of independent constituents of MBQC measurement patterns, we provided the raw material for the classification, i.e. the data to be classified. This data contains projective representations of the symmetry group, as in the classification of SPT phases  in 1D. However, because the linear representations---also part of the classification---don't need to be faithful on individual blocks (see the examples of  Section~\ref{BLtoSL}),  the classification of MBQC schemes in 1D is more complex.}
    \item{{\em{Can the present construction be generalized to other symmetry groups?}} Likely, the non-Abelian case will differ more from the present treatment than the general Abelian. In this regard, it shall be noted that the first computational phase of quantum matter identified was for a non-Abelian symmetry group, $S_4$ \cite{MM2}, and that a Wigner-Eckart theorem for MBQC---applicable to both Abelian and non-Abelian symmetry groups---has been established in \cite{AbhiTCW}.}

    \item{{\em{Can the present computational scheme be generalized to higher spatial dimension?}}}
\end{itemize}

\paragraph{Acknowledgements.} WY, AA and RR are funded from the Canada First Research Excellence Fund, Quantum Materials and Future Technologies Program. AA and RR are funded by USARO (W911NF2010013). WY and RR were funded by NSERC, through the European-Canadian joint project FoQaCiA (funding reference number 569582-2021). RR is funded by the Humboldt foundation. WY and AA thank A. Nocera for discussions, and AA and RR thank P. Feldmann, D. Bondarenko, and D.T. Stephen for discussions. The numerical simulations in this work were performed using the software package ITensor\cite{itensor}, and are available \cite{Programs}. AA thanks E.M. Stoudenmire for software support and discussions. 

\appendix

\section{Proof of Lemma~\ref{cacLemma}}\label{cacAp}

{\em{Proof of Lemma~\ref{cacLemma}.}} Since $v$ is a projective representation, for any pair $g,g'\in G$ it holds that $v(g)v(g') = e^{i\varphi_{gg'}}v(g')v(g)$, for some $\varphi_{gg'}\in \mathbb{R}$. Further, using the fact that $g'\in G=\mathbb{Z}_2^m$, $I \propto v(0) = v(g'+g') \propto v(g')^2$. Hence, $v(g')^{-1} \propto v(g')$, for all $g'\in G$. Since proportionality constants don't affect commutation, we have $v(g)v(g')^{-1}=e^{i\varphi_{gg'}}{v(g')}^{-1}v(g)$. Combining the two commutation relations,
$$
v(g) = v(g)\,v(g')^{-1}v(g') = e^{2i\varphi_{gg'}}v(g')^{-1}v(g')\, v(g) = e^{2i\varphi_{gg'}} v(g),
$$
hence $e^{i\varphi_{gg'}}=\pm1$, for all $g,g' \in \mathbb{Z}_2^m$. $\Box$

\section{Proof of Corollary~\ref{unit}}\label{UnitProof}

{\em{Proof of Corollary~\ref{unit}.}} {\em{Item (i):}} If we have two gates $\exp(-id\alpha T(g))$ and $\exp(-id\alpha T(g'))$ at our disposal, we can also implement the unitary gate
$$\exp(-id\alpha T(g)) \exp(-id\alpha T(g')) \exp(id\alpha T(g)) \exp(id\alpha T(g')) \approx \exp(-(d\alpha)^2 [T(g),T(g')]).$$
Iterating, if necessary, we find that all gates generated by $T(\mathcal{G})$ can be implemented. 

We further observe that 
$$
\exp\left(i\sum_{i|\, g_i \in {\cal{G}}} \alpha_i T(g_i) \right) = \prod_{i|\, g_i \in {\cal{G}}} \exp\left(i\alpha_iT(g_i)\right) + O(\alpha_i\alpha_j).
$$
Thus, for small rotation angles, rotations generated by any element in the Lie algebra $\mathcal{A}$ can be reduced to gates generated by $T({\cal{G}})$, with only higher-order approximation error.

{\em{Item (ii):}} We compare the unitary $U(\alpha)$ to its approximation by the $N$-fold concatenation of the CPTP map ${\cal{V}}$ with rotation angle $\propto 1/N$. We find that the total approximation error is proportional to $1/N$; hence it is of advantage for accuracy to split the rotation into many small parts. 

To quantify the error $\epsilon$ of the total gate operation, we use the diamond norm $\|\cdot\|_\diamond$ \cite{Diamond}, with the properties stated in Lemma 12 therein. 
In the following we denote the CPTP maps of interest by ${\cal{V}}_g$; it is the group label that matters, not the site label. 

For small rotation angles $\beta$, that is whenever
$$
\frac{\beta}{\sigma(g)}\ll 1,
$$
the unitary $U_g(\beta)$ is best approximated by the CPTP map ${\cal{V}}_g(\beta/\sigma(g))$. The approximation error then is
\begin{equation}\label{E1}
\begin{array}{rcl}
\left\|[U_g(\beta)] - {\cal{V}}_g(\beta/\sigma(g))\right\|_\diamond  & \leq& \frac{3\beta^2}{4}  \frac{1-\sigma(g)^2}{\sigma(g)^2} + O\left( (\beta/\sigma(g))^3 \right)\\
&\leq& \beta^2 \frac{1-\sigma^2}{\sigma^2}.
\end{array} 
\end{equation}
In the last line, we have used the assumption that $\sigma_k(g)\geq \sigma$, for all $k$, $g\in G$, and furthermore have bounded the contribution of the trailing orders to 1/3 of the leading order contribution to the error. This will always be satisfied for sufficiently small $\beta$.

We now bound the error of the $N$-fold iteration ${\cal{V}}\left(\frac{\alpha}{\sigma(g)N}\right)$. For brevity, we will write ${\cal{V}}_g[U_g]^{-1}$ for ${\cal{V}}_g(\alpha/\sigma(g) N) [U_g]^{-1}(\alpha/N)$. We have that
$$
\begin{array}{rcl}
\epsilon &=& \|[U_g(\alpha)] - {\cal{V}}_g(\alpha/\sigma(g)N)^N\|_\diamond \\
&=& \|[U_g(\alpha/N)]^N - {\cal{V}}_g(\alpha/N)^N\|_\diamond \\
&=&  \|I -\left({\cal{V}}_g[U_g]^{-1}(\alpha/N)\right)^N\|_\diamond \\
&=&  \left\|\left(I -\left({\cal{V}}_g[U_g]^{-1}\right)\right)\left(I + {\cal{V}}_g[U_g]^{-1}+ \left({\cal{V}}_g[U_g]^{-1}\right)^2 +... + ({\cal{V}}_g[U_g]^{-1})^{N-1} \right)\right \|_\diamond \\
&\leq&  \left\| I -\left({\cal{V}}_g[U_g]^{-1}\right)\right\|_\diamond \; \left\| I + {\cal{V}}_g[U_g]^{-1}+ \left({\cal{V}}_g[U_g]^{-1}\right)^2 +... + ({\cal{V}}_g[U_g]^{-1})^{N-1} \right\|_\diamond \\
&\leq&  \left\| I -\left({\cal{V}}_g[U_g]^{-1}\right)\right\|_\diamond \sum_{i=0}^{N-1} \left\| \left({\cal{V}}_g[U_g]^{-1}\right)^i \right\|_\diamond \\
&=& N \left\| I -\left({\cal{V}}_g[U_g]^{-1}\right)\right\|_\diamond\\
&=& N  \|[U_g(\alpha/N)] - {\cal{V}}_g(\alpha/\sigma(g)N)\|_\diamond.
\end{array}
$$
Now combining the above with Eq.~(\ref{E1}) for the angle $\beta = \alpha/N$, we obtain Eq.~(\ref{ApprErr}). $\Box$

\section{More on the Kitaev-Gamma chain}
\label{app:KG}

\subsection{Hamiltonian in the unrotated frame}

\begin{figure}
\begin{center}
\includegraphics[width=11cm]{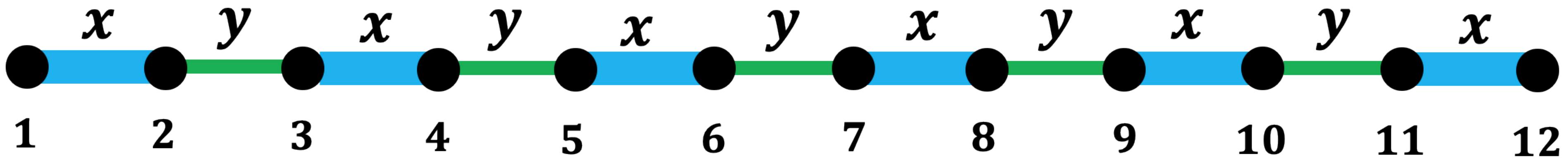} 
\caption{
Bond pattern for the 1D bond-alternating Kitaev-Gamma model in the unrotated frame.
The thick and thin lines represent the alternating pattern of the bond strengths. 
\label{fig:bond_original} 
}
\end{center}
\end{figure}

The 1D spin-1/2 bond-alternating Kitaev-Gamma model is defined by the following Hamiltonian \cite{Luo2021,Yang2022a},
\begin{eqnarray}
H_{K\Gamma}=\sum_{\gamma=<ij>} g_\gamma [K S_i^\gamma S_j^\gamma+\Gamma (S_i^\alpha S_j^\beta+S_i^\beta S_j^\alpha)],
\label{eq:KG}
\end{eqnarray}
in which
$\gamma\in\{x,y\}$ is the spin direction associated with the bond connecting the nearest neighboring sites $i$ and $j$ as shown in Fig. \ref{fig:bond_original}, 
and $(\gamma,\alpha,\beta)$ form a local right-handed coordinate system in spin space. 
A useful unitary transformation for studying the 1D Kitaev-Gamma model is the 
rotation $U_6$ \cite{Yang2019}, 
which acts as $I$, $R(\frac{1}{\sqrt{2}}(0,1,-1),\pi)$,  $R(\frac{1}{\sqrt{3}}(1,1,1),\frac{2\pi}{3})$,  $R(\frac{1}{\sqrt{2}}(-1,1,0),\pi)$,  $R(\frac{1}{\sqrt{3}}(1,1,1),-\frac{2\pi}{3})$,  $R(\frac{1}{\sqrt{2}}(1,0,-1),\pi)$
on sites $1+6m$, $2+6m$, $3+6m$, $4+6m$, $5+6m$, $6+6m$ ($m\in \mathbb{Z}$), respectively.
After the $U_6$ transformation, the Hamiltonian $H^\prime_{K\Gamma}=U_6H_{K\Gamma}(U_6)^{-1}$ acquires the form in Eq. (\ref{eq:KG_U6}) \cite{Luo2021,Yang2022a}.
In this appendix, the spin coordinate systems before and after the $U_6$ transformation will be termed as unrotated and rotated frames, respectively.

There are two duality transformations for the system which are mostly easily seen using $H_{K\Gamma}$ in Eq. (\ref{eq:KG}) in the unrotated frame \cite{Luo2021}: 
$U(R(\hat{z}),\pi)$ maps the parameters $(K,\Gamma,g_x,g_y)$ to $(K,-\Gamma,g_x,g_y)$,
and $U(R(\hat{z}),\pi/2)T_a$ maps $(K,\Gamma,g_x,g_y)$ to $(K,\Gamma,g_y,g_x)$. 
As a result, 
it is enough to consider the parameter  region $\phi\in(-\pi/2,\pi/2)$, $g\in (0,1)$;
and the EH$^\prime$, OH, and OH$^\prime$ phases in Fig. \ref{fig:phase_KG} (a) can be obtained from the EH phase by applying $U(R(\hat{z},\pi))$, $U(R(\hat{z},\pi/2))T_a$, and $U(R(\hat{z},-\pi/2))T_a$ in the unrotated frame, respectively,
which is the reason why only the EH phase is discussed in Sec. \ref{subsec:KG}.

\subsection{The odd-Haldane phase}

\begin{figure}
\begin{center}
\includegraphics[width=9cm]{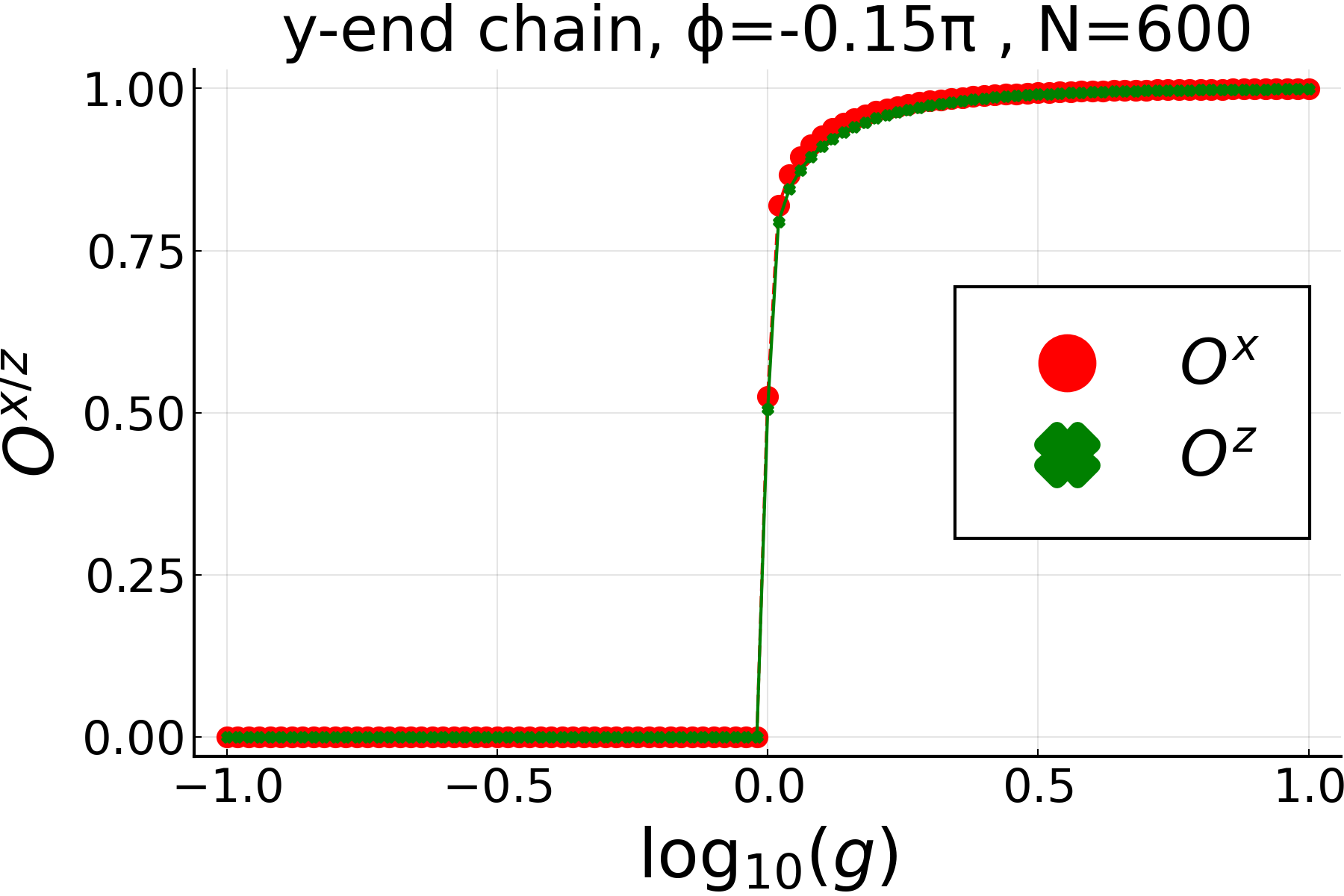}
\caption{\label{fig:y_end} 
String order parameters $O^\alpha(\frac{N}{2}+1,N)$ ($\alpha=x,z$) as functions of $\log(g)$ at $\phi=-0.15\pi$ for  $y$-end chains.
DMRG simulations are performed on open chains with system size $N=600$.
}
\end{center}
\end{figure}

In what follows, an open chain with even length $N$ will be named as an $x$-end (or a $y$-end) chain 
if the bond between the last two sites $N-1$ and $N$ is an $x$- (or $y$-) bond.
As discussed in Sec. \ref{subsec:KG}, for $x$-end chains, the EH phase (and also the EH$^\prime$ phase) has   nonvanishing string order parameters  $O_e^\alpha$ ($\alpha=x,y,z$) defined in Eq. (\ref{eq:def_Oe}). 
On the other hand, since the odd-Haldane phases (i.e., the OH and OH$^\prime$ phases) are related to the even Haldane phases by $U(R(\pm \hat{z},\pi/2))T_a$,
it is expected that the string order parameters characterizing the odd-Haldane phases 
should be defined in $y$-end chains with the same expression in Eq. (\ref{eq:def_Oe}).

Fig. \ref{fig:y_end}  shows the numerical values of the string order parameters $O^\alpha(\frac{N}{2}+1,N)$, for $\alpha=x,z$, in the rotated frame as a function of $\log (g)$ 
for an even-length $y$-end chain,
where $\phi$  is fixed to be $-0.15\pi$. 
As is clear in Fig. \ref{fig:y_end},
the string order parameters are nonzero and zero in the $g>1$ and $g<1$ regions, respectively,
which identifies the OH phase in the $g>1$ region, in contrast to the EH phase in the $g<1$ region.
For MBQC purposes, $y$-end chains must be used in the OH and OH$^\prime$ phases.

\end{document}